\documentclass[review]{elsarticle}

\usepackage{etex}
\usepackage{microtype}%if unwanted, comment out or use option "draft"
\usepackage{amsmath}
\usepackage{amssymb}
\usepackage{xspace}
\usepackage{proof-dashed}
\usepackage{tikz}
\usepackage[arrow,matrix,curve]{xy}%\CompileMatrices

\usepackage{lineno,hyperref}
\modulolinenumbers[5]

\journal{Journal of \LaTeX\ Templates}

\begin{document}

\begin{frontmatter}

\title{On the Boundary between Decidability and
  Undecidability of Asynchronous Session Subtyping
% Decidable and Undecidable Fragments of Asynchronous Subtyping
%   for Session Types
  % \footnote{}
}

\author{Mario Bravetti}
%\ead{mario.bravetti@unibo.it}
\address{University of Bologna, Department of Computer Science and Engineering / INRIA FOCUS
%\\ 
%Mura A. Zamboni 7, 40127 Bologna, Italy
}

\author{Marco Carbone}
%\ead{carbonem@itu.dk}
\address{Department of Computer Science, IT University of Copenhagen
%\\
%  Rued Langgaards Vej 7, 2300 Copenhagen, Denmark
}

\author{Gianluigi Zavattaro}
%\ead{gianluigi.zavattaro@unibo.it}
\address{University of Bologna, Department of Computer Science and Engineering / INRIA FOCUS
%\\ 
%Mura A. Zamboni 7, 40127 Bologna, Italy
}

\begin{abstract}
  Session types are behavioural types for guaranteeing that concurrent
  programs are free from basic communication errors.  Recent work has
  shown that asynchronous session subtyping is undecidable. 
However, since session types have become popular in mainstream programming
languages in which asynchronous communication is the
norm rather than the exception, it is crucial to detect significant
decidable subtyping relations.
  Previous work considered extremely restrictive fragments in which limitations were imposed to the size of communication buffer (at most 1)
  or to the possibility to express multiple choices (disallowing them completely in
  one of the compared types).
  In this work, 
  for the first time, we show decidability of a
  fragment that does not impose any limitation on communication buffers
  and allows both the compared types to include multiple choices
  for either input or output, 
  thus yielding a fragment which is more significant
  from an applicability viewpoint.
In general, we study the boundary between decidability and undecidability
by considering several fragments of subtyping. Notably,
we show that subtyping remains
  undecidable even if restricted to not using output covariance
  and input contravariance. 
%  
%  
% 
%  Additionally, we further refine the existing
%  undecidability results by showing 
%  In general, the contribution of this work is
%%two-fold: we detect some kind of restrictions 
%to analyze % forms of
%restrictions 
%of % the
%asynchronous subtyping,
%concerning both limitations to the choice capability
%and to the communication buffer,
% and classifying 
% them into decidable and undecidable fragments.
  \end{abstract}


\end{frontmatter}

%\linenumbers

%%%% TEXT
\newcommand{\selection}{single-choice\xspace}
\newcommand{\Selection}{Single-Choice\xspace}

\newcommand{\myparagraph}[1]{{\bigbreak\noindent\bf #1}}

\newcommand{\calt}{{\cal T}}

 %%% inference rule %%%
\newcommand{\infr}[2]
        {\renewcommand{\arraystretch}{1.5}
        \begin{array}{c}
        #1\\
        \hline
        #2
        \end{array}}

\newcommand{\auxarrow}
        {\mathop{\longrightarrow}}

 %%% transition relation %%%
\newcommand{\arrow}[1]
        {\, \auxarrow\limits^{#1} \,}

%%% THESE ARE ALL MACROS -- PLEASE BE CONSISTENT WITH PLACING THEM IN
%%% THE RIGHT PLACE

%% THEOREM ENVs
%\theoremstyle{plain}
        \newtheorem{proposition}{Proposition}[section]
        \newtheorem{definition}[proposition]{Definition}
        \newtheorem{corollary}[proposition]{Corollary}
        \newtheorem{theorem}[proposition]{Theorem}
        \newtheorem{lemma}[proposition]{Lemma}
        \newcommand{\proof}{{\em Proof.}\ }
        \newtheorem{example}[proposition]{Example}

\newcommand{\restateTHM}[1]
{
  \noindent
  {\bfseries
    Theorem~\ref{#1}.}
}
\newcommand{\restateLEM}[1]
{
  \noindent
  {\bfseries
    Lemma~\ref{#1}.}
}
\newcommand{\restatePRO}[1]
{
  \noindent
  {\bfseries
    Proposition~\ref{#1}.}
}

%% TYPE LANGUAGE NAMES
\newcommand{\fulllang}{\mathcal S_{\text{full}}}
\newcommand{\opluslang}{\mathcal S_{\oplus}}
\newcommand{\ampersandlang}{\mathcal S_{\&}}

%% TYPES SYNTAX
%
\newcommand{\Tout}[1]{!\langle {#1}\rangle}
\newcommand{\Tin}[1]{?({#1})}

\newcommand{\ToutK}[1]{k!\langle {#1}\rangle}
\newcommand{\TinK}[1]{k?({#1})}
\newcommand{\Tbranchindex}[4]{\&\{{#1}_{#3}\!:{#2}_{#3}\}_{#3\in #4}}
\newcommand{\Tbranch}[2]{\Tbranchindex{#1}{#2}{i}{I}}
\newcommand{\Tbranchsingledec}[3]{\&^{#3}\{{#1}\!:{#2}\}}
\newcommand{\Tbranchsingle}[2]{\&\{{#1}\!:{#2}\}}

\newcommand{\Tbranchset}[3]{\&\{{#1}\!:\!{#2}\}_{#1\in #3}}

\newcommand{\TbranchK}[2]{k\Tbranchindex{#1}{#2}{i}{I}}

\newcommand{\TbranchindexM}[5]{\&\{{#1}_{#3}(#5_{#3})\!:{#2}_{#3}\}_{#3\in #4}}
\newcommand{\TbranchM}[3]{\TbranchindexM{#1}{#2}{i}{I}{#3}}

\newcommand{\Tbranchsimple}[2]{\&\{{#1}\!:{#2}\}}

\newcommand{\Tselectindex}[4]{\oplus\{{#1}_{#3}\!:{#2}_{#3}\}_{#3\in #4}}
\newcommand{\Tselect}[2]{\Tselectindex{#1}{#2}{i}{I}}
\newcommand{\Tselectsingle}[2]{\oplus\{{#1}\!:{#2}\}}
\newcommand{\Tselectset}[3]{\oplus\{{#1}\!:{#2}\}_{#1\in #3}}

\newcommand{\TselectK}[2]{k\Tselectindex{#1}{#2}{i}{I}}

\newcommand{\TselectindexM}[5]{\oplus\{{#1}_{#3}\langle #5_{#3}\rangle:{#2}_{#3}\}_{#3\in #4}}
\newcommand{\TselectM}[3]{\TselectindexM{#1}{#2}{i}{I}{#3}}

\newcommand{\Tselectsimple}[2]{\oplus\{{#1}\!:{#2}\}}

\newcommand{\Trec}[1]{\mu \mathbf{#1}}
\newcommand{\Tvar}[1]{\mathbf{#1}}
\newcommand{\Tend}{\mathbf{end}}
\newcommand{\context}[3]{\mathcal{#1}[{#2}]^{#3}}

\newcommand{\colorcontext}[4]{\mathcal{#1}^{#4}[{#2}]^{#3}}

\newcommand{\rderiv}{\rightarrow}

\newcommand{\notrderiv}{\rightarrow_{\mathsf{err}}}
\newcommand{\success}{\rightarrow_{\mathsf{ok}}}

\newcommand{\blocking}{\rightarrow_{\mathsf{blk}}}
\newcommand{\terminating}{\rightarrow_{\mathsf{ter}}}

\newcommand{\iostep}[1]{\, {\auxarrow\limits^{#1}}_{\mathsf{io}} \,}

%% SYNTACTIC FUNCTIONS
\newcommand{\unfold}[2]{\mathsf{unfold}^{#1}(#2)}
\newcommand{\unfoldNOPAR}[1]{\mathsf{unfold}^{#1}}
\newcommand{\outunfold}[1]{\mathsf{outUnf}(#1)}
\newcommand{\depth}{\mathsf{outDepth}}
\newcommand{\antSet}[1]{\, \mathsf{antSet}_{#1} \,}
\newcommand{\antEq}[1]{\, \mathsf{antEq}_{#1} \,}
\newcommand{\extAntEq}[1]{\, \mathsf{extAntEq}_{#1} \,}
\newcommand{\extAntSet}[1]{\, \mathsf{extAntSet}_{#1} \,}

%% SYMBOLS

\newcommand*{\DashedArrow}[1][]{\mathbin{\tikz
    [baseline=-0.25ex,-latex, dashed,#1] \draw [#1] (0pt,0.5ex) --
    (1.3em,0.5ex);}}%

\newcommand{\subtypedc}{\subtype_{\mathsf{DC}}}
\newcommand{\subtypes}{\subtype_{\mathsf{s}}}
\newcommand{\subtypeo}{\subtype} %_{\mathsf{o}}} %TRICK!!!!!!!!!!!!!
\newcommand{\subtypetin}{\subtype_{\mathsf{tin}}}
\newcommand{\subtypetout}{\subtype_{\mathsf{tout}}}
\newcommand{\subtypetintout}{\subtype_{\mathsf{tin,tout}}}
\newcommand{\subtypesin}{\subtype_{\mathsf{sin}}}
\newcommand{\subtypesout}{\subtype_{\mathsf{sout}}}
\newcommand{\subtypesinsout}{\subtype_{\mathsf{sin,sout}}}
\newcommand{\subtypesintout}{\subtype_{\mathsf{sin,tout}}}
\newcommand{\subtypetinsout}{\subtype_{\mathsf{tin,sout}}}
\newcommand{\selsubtypesin}{\selsubtype_{\mathsf{sin}}}
\newcommand{\selsubtypesout}{\selsubtype_{\mathsf{sout}}}
\newcommand{\selsubtype}{{<\!\!<}}
\newcommand{\subtype}{{\leq}}
\newcommand{\subtypea}{\;{\leq_{\mathsf{a}}}\;}
\newcommand{\subtypebound}{{\leq_{\mathsf{bound}}}}
\newcommand{\subtypeak}{\;{\leq_{\mathsf{a}}^{\mathsf k}}\;}
\newcommand{\subtypet}{\;{\leq_{\mathsf{t}}}\;}
\newcommand{\subtypesa}{\;{\leq_{\mathsf{sa}}}\;}
\newcommand{\semT}[2]{[\![{#1}]\!]^{#2}}
\newcommand{\BsemT}[2]{{[\![\![{#1}]\!]\!]}^{#2}}
\newcommand{\semS}[1]{[\![{#1}]\!]}
\newcommand{\BsemS}[1]{[\![\![{#1}]\!]\!]}
\newcommand{\semTcont}[1]{{\{\!\!\{{#1}\}\!\!\}}}

\newcommand{\dual}[1]{\overline{#1}}

\newcommand{\decore}[1]{\mathsf{ann}(#1)}
\newcommand{\undecore}[1]{\mathsf{unann}(#1)}
\newcommand{\mincycle}[2]{\mathsf{minCycle}(#1,#2)}

\newcommand{\leafset}[1]{\mathsf{leafSet}(#1)}
\newcommand{\noIn}[1]{\mathsf{noIn}(#1)}
\newcommand{\antOut}[2]{\mathsf{antOut}(#1,#2)}
\newcommand{\antOutInf}[1]{\mathsf{antOutInf}(#1)}
\newcommand{\reach}[1]{\mathsf{reach}(#1)}

\newcommand{\finclose}[1]{\mathsf{fin}(#1)}
\newcommand{\recclosure}[1]{\mathsf{close}(#1)}
\newcommand{\closure}[1]{\mathsf{subterms}(#1)}

\newcommand{\labelunion}{\uplus}

\newcommand{\encodeSC}[1]{[\![{#1}]\!]}
\newcommand{\unencodeSC}[1]{\{\!\!\{{#1}\}\!\!\}}

% %%% Local Variables: 
% %%% mode: latex
% %%% TeX-master: "main"
% %%% End: 

\section{Introduction}\label{sec:introduction}
Session types~\cite{HVK98,HYC08} are types for controlling the
communication behaviour of processes over channels. In a very simple
but effective way, they express the pattern of sends and receives that
a process must perform. Since they can guarantee freedom from some
basic programming errors, session types are becoming popular with many
main stream language implementations, e.g., Haskell~\cite{LM16},
Go~\cite{NY16} or Rust~\cite{JML15}.

As an example, consider a client that invokes service operations by
following the protocol expressed by the session type
$$
\oplus\{op1\!: \& \{resp1\!: \Tend\},\ op2\!: \& \{resp2\!: \Tend\} \}
$$
indicating that the client decides whether to call operation $op1$ or
$op2$ and then waits for receiving the corresponding response ($resp1$
or $resp2$, respectively).  For the sake of simplicity we consider
session types where (the type of) communicated data is abstracted
away.  The symmetric behaviour of the service is represented by the
complementary (so-called {\it dual}) session type
$$\&\{op1\!: \oplus \{resp1\!: \Tend\},\ op2\!: \oplus \{resp2\!: \Tend\} \}
$$
indicating that the server receives the call to operation $op1$ or
$op2$ and then sends the corresponding response ($resp1$ or $resp2$,
respectively).

We call \emph{output selection}
the construct $\oplus\{l_1:T_1, \dots, l_n:T_n\}$. It is used to
denote a point of choice in the communication protocol:
each choice has a label $l_i$ and a continuation $T_i$. 
In communication protocols,
when there is a point of choice, there is usually a peer that
internally takes the decision and the other involved peers receive
communication of the selected branch.
Output selection is used to describe the behaviour of the peer that takes the decision:
indeed, in our example it is the client that decides which operation to call.
Symmetrically, we call \emph{input branching} the construct $\&\{l_1:T_1, \dots, l_n:T_n\}$. 
It is used to describe the behaviour
of a peer that receives communication of the selection
done by some other peers.  In the example, indeed, the service receives
from the client the decision about the selected operation.\footnote{In session type terminology \cite{HVK98,HYC16}, 
the output selection/input branching constructs are usually simply called \emph{selection}/\emph{branching}; we call them output selection/input branching
because we consider a simplified syntax for session types in which there
is no specific separate construct for sending one output/receiving one input. 
Anyway, such output/input types 
%(without carried type,
%that are not considered in our simplified setting) 
can be seen as an output selection/input branching with only one choice.}

When composing systems whose interaction protocols have been specified
with session types, it is % particularly
significant to consider variants of their specifications that still
preserve safety properties.  In the above example, the client can be
safely replaced by another one with session type
$$\oplus\{op1\!: \& \{resp1\!: \Tend\} \}
$$
indicating that it can call only one specific service operation.  But
also the service can be safely replaced by another one accepting an
additional operation:
$$\&\{op1\!: \oplus \{resp1\!: \Tend\},\ op2\!: \oplus \{resp2\!:
\Tend \},\ op3\!: \oplus \{resp3\!: \Tend\} \} $$
Subtyping relations %$\preceq$
have been formally defined for session types, e.g., by Gay and
Hole~\cite{GH05} %ESOP09,
and Chen et al.~\cite{MariangiolaPreciness}, in order to precisely
capture this safe replacement notion. For instance, a subtyping
relation like that of Gay and Hole~\cite{GH05} (denoted by
$\subtypes$), where processes are assumed to simply communicate via
synchronous channels\footnote{Here, we focus on the so-called
  process-oriented subtyping, as opposed to channel-based
  subtyping~\cite{G16}.}, would imply, for the client, that:
%\vspace*{-2mm}
%$$\oplus\{\text{add\_to\_cart}: \oplus \{\text{pay}: \Tend\}\}\ \subtypes\ \Trec t.\oplus\{\text{add\_to\_cart}: \Tvar t\ ,\ \text{pay}: \Tend\}
%$$
$$\oplus\{op1\!: \& \{resp1\!: \Tend\} \}\ \subtypes\ 
\oplus\{op1\!: \& \{resp1\!: \Tend\},\ op2\!: \& \{resp2\!: \Tend\}
\}$$
%\vspace*{-2mm}
according to the so-called \emph{output covariant} property, while, for the server
%\vspace*{-2mm}
$$
\begin{array}{l}
  \&\{op1\!: \oplus \{resp1\!: \Tend\},\ op2\!: \oplus \{resp2\!: \Tend \},\ op3\!: \oplus \{resp3\!: \Tend\} \}\\
  \hspace{2cm}\ \subtypes\; \&\{op1\!: \oplus \{resp1\!: \Tend\},\ op2\!: \oplus \{resp2\!: \Tend\} \}
\end{array}
$$
%$$\Trec t.\& \{\text{add\_to\_cart}: \Tvar t,\ \text{remove}: \Tvar t,\ \text{pay}: \Tend\}\ \subtypes\ \Trec t.\& \{\text{add\_to\_cart}: \Tvar t,\ \text{pay}: \Tend\}
%%\vspace*{-2mm}
%$$
according to the so-called \emph{input contravariant} property.

%.
%The above examples are instances of the so-called \emph{output covariant} and 
%\emph{input contravariant} subtyping, according to which a subtype can have 
%less internal nondeterminism and more external nondeterminism.

When processes communicate via asynchronous channels, a more generous
notion of subtyping $\subtype$ like that of Chen et al.~\cite{%ESOP09,
  MariangiolaPreciness} can be considered. E.g., a process using an
asynchronous channel to call a service operation, that receives the
corresponding response and then sends a huge amount of data (requiring
heavy computation), could be safely replaced by a more efficient one
that computes and sends all the data immediately without waiting for
the response:
%\vspace*{-2mm}
$$\oplus \{op\!: \oplus\{huge\_data\!: \&\{resp\!: \Tend\}\}\} \
\subtype \ \oplus \{op\!: \&\{resp\!: \oplus \{huge\_data\!:
\Tend\}\}\}
%\vspace*{-2mm}
$$
Intuitively, this form of asynchronous subtyping reflects
the possibility to \emph{anticipate the output} of the huge data 
w.r.t.\ to the input of the response because such data 
are stored in a buffer waiting for their reader to consume them.

%Traditionally, types for programming languages make use of the notion
%of subtyping, a relation between types that allows to safely replace
%programs with other programs: a given program $P$ with type $S$ can
%always be safely replaced by another program $Q$ with type $T$
%whenever $T\subtype S$ ($T$ is a subtype of $S$). In the literature,
%subtyping has been thoroughly studied for both binary session types
%(types expressing communication patterns between two entitities) and
%multiparty session types~\cite{HYC16} (types expressing communication
%patterns among many entitities). In particular, it is possible to
%compute whether two types are related by the subtyping relation
%defined for synchronous communication~\cite{GH05}. 

\subsection{Previous Results}

Recently, Bravetti
et al.~\cite{BCZ16} and Lange and Yoshida~\cite{LY17} have
independently shown that asynchronous subtyping 
%relations for asynchronous communication 
(the subtyping relation with output anticipation) is
%instead 
undecidable. 
In particular, in Bravetti
et al., this is done by showing undecidability of the much simpler {\it single-choice relation}
$\selsubtype$ that is defined as a restriction of asynchronous subtyping $\subtype$ where
related $T \, \selsubtype \, S$ types are such that: all output selections in $T$ have a single choice (output
selections are covariant, thus $S$ is allowed have output selections with multiple choices) and all input branchings in $S$ have a single choice (input branchings are contravariant, thus $T$ is allowed to have input branchings with multiple
choices).
Moreover, those papers prove decidability for very small
fragments of the asynchronous subtyping relation: the most significant
one basically requires one of the two compared types to be
such that {\it all its input branchings and output selections have a single choice}.
%non-branching, i.e., not to perform choices neither for inputs nor for
%outputs. 
In particular, in Bravetti et al. this is done by showing decidability of the two relations $\selsubtypesin$ ({\it single-choice input} $\selsubtype$) and $\selsubtypesout$ ({\it single-choice output} $\selsubtype$), both
defined as {\it further} restrictions of $\selsubtype$ where for related $T \, \selsubtypesin \, S$ ($T \, \selsubtypesout \, S$, resp.) types we {\it additionally} require that: 
all input branchings (output selections, resp.) in $T$ and $S$ have a single choice.
Other decidable fragments, considered by Lange and Yoshida, pose %strong 
limitations on the communication behaviour that causes communication buffers
%i.e., 
to store at most one message,
%
%
%hat their 
%to ha at most
%size %is just 
%$1$ 
or they are used in half duplex
modality (messages can be sent in one direction, only if the buffer
for the opposite direction has been emptied).
% that only the buffer for one of the two communication directions of the binary session can be in use.
% for a relation when communication is asynchronous, i.e., outputs
% are.
%If $T\subtype S$, type $S$ must be able to simulate every input or
%output that $T$ does. However, in the asynchronous setting, $S$ is
%allowed to anticipate outputs that are prefixed by inputs, i.e.,
%replacing a program of type $S$ with one of type $T$ that performs an
%output earlier is safe.
Although asynchronous subtyping is undecidable, it is important
to reason about more significant cases for which such a relation is decidable.
%, which are more significant w.r.t. those considered in ~\cite{BCZ16} and~\cite{LY17}. 
This
because session types have become popular in mainstream programming
languages, and, in such cases, asynchronous communications are the
norm rather than the exception.

\subsection{Contributed Results} 

%%%%% DRAWING WITH VARIOUS RESULTS
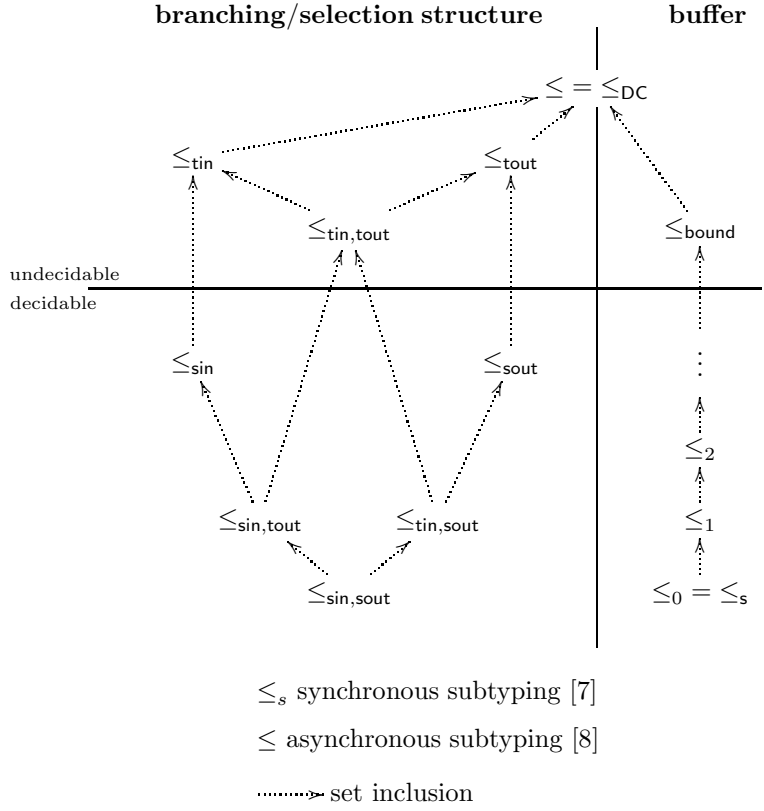
\begin{figure}[t]
  \begin{center}
%    \begin{tabular}{ll}
      % \begin{tabular}{lllll}
      %   $\subtype_s$  & standard binary \cite{IandCmostrous}\\
      %   $\subtype_o$  & orphan-message-free \cite{MariangiolaPreciness}\\
      %   $\subtype_m$  & multiparty \cite{ESOP09}\\        
      %   $\longrightarrow$ & reduces to\\[3cm]
      % \end{tabular}\hspace{-3cm}
      % &
        \begin{tabular}{c}
          \xymatrixcolsep{0pc}
          \xymatrix@C=-0.5em@R=1.2em{
&&          & &  &  & \hspace{-2cm}{\bf branching/selection \; structure} \hspace{-2cm} & &  &  &  & \hspace{0.2cm}{\bf buffer}  &  & \\
&&          & &  &  &  &  &  &  & \subtype=\subtypedc\ar@{-}[dddddddd]\ar@{-}[u]  &  &  & \\
&&          & & \subtypetin\ar@{.>}^{}[urrrrrr] &  &  &  &
                                                             \subtypetout\ar@{.>}_{}[urr]&
                                            & &
                                                %\subtypebound\ar@{.>}_{}[ul] 
& & \\
&&          & & &  &
                   \subtypetintout\ar@{.>}^{}[ull]\ar@{.>}^{}[urr]&
                                      & & %\selsubtype \ar@{.>}[ruu] 
& & \subtypebound\ar@{.>}_{}[uul]   \\
          \ar@{-}[rrrrrrrrrrrrrrrrrrr]^{\hspace{-9.6cm}\text{undecidable}}_{\hspace{-9.9cm}\text{decidable}}
&\qquad & \quad         &&&&&&&&&&&&&&&&&&&\  \\
&&          & & \subtypesin\ar@{.>}^{}[uuu] & & & &
                                                     \subtypesout\ar@{.>}_{}[uuu] & & &
                                                                                        \begin{tabular}{l}$\vdots$\end{tabular}\ar@{.>}^{}[uu] &&\\
&&          & & &  &  & &  &  &  &  \subtype_2\ar@{.>}^{}[u] &  & \\
&&          & %\selsubtypesin\ar@{.>}[ruu]\ar@{.>}[rrrrrruuuu] 
& &
                                               \subtypesintout\ar@{.>}^{}[luu]\ar@{.>}^{}[ruuuu]
                  &
                                    &\subtypetinsout\ar@{.>}^{}[ruu]\ar@{.>}^{}[luuuu]
                                      & &
                                          %\selsubtypesout\ar@{.>}[uul]\ar@{.>}[uuuu] 
& & \subtype_1\ar@{.>}^{}[u] & & \\
&&          & & & &
                \subtypesinsout\ar@{.>}^{}[ul]\ar@{.>}^{}[ur]
                                    & & & & &
                                             \subtype_0=\subtypes\ar@{.>}^{}[u]  & & \\
&&          & & & & & & & & \ & & & \\
          }
        \end{tabular}
   % \end{tabular}
%\mbox{}      \\\\
      % $\DashedArrow[->,dotted            ]$\\
      \begin{tabular}{
%l@{\qquad \quad %\qquad
%}l@{\ }
l}
        $\subtype_s$  synchronous subtyping~\cite{GH05}
        %\quad
%&%\quad
%               $\selsubtype$ & single-choice relation~\cite{BCZ16}
        \\
        $\subtype$  %orphan-message-free
asynchronous subtyping~\cite{MariangiolaPreciness}
        %\quad
%&%\quad
%               $\selsubtypesin$ & single-choice input $\selsubtype$~\cite{BCZ16}
        \\
        \!\!\!\!\!\xymatrix{\ \ar@{.>}[r] & \text{set inclusion}}
        %\quad
%&%\quad
%               $\selsubtypesout$ & single-choice output $\selsubtype$~\cite{BCZ16}
      \end{tabular}
  \end{center}
  \caption{Lattice of the asynchronous subtyping relations considered in this paper.}
  \label{fig:lattice}
\end{figure}

The aim of this paper 
is to detect significant decidable fragments of asynchronous session subtyping and to establish % somehow
a more precise boundary between
decidability and undecidability.
In particular, concerning decidability, as discussed above, the few
  decidable fragments of asynchronous subtyping known so far are
  extremely restrictive: our relations $\selsubtypesin$, $\selsubtypesout$ \cite{BCZ16}
  and the decidable relations considered by Lange and Yoshida \cite{LY17}. Here, for the first time, we show decidability of a
  fragment that does not impose any limitation on communication buffers
  and allows both the subtype and the supertype to include multiple choices
  (either for input branchings or for output selections), thus opening the possibility for some practical applicability in restricted specific scenarios (e.g. session types for clients/services in web-service systems, see below). More precisely, while $\selsubtypesin$ ($\selsubtypesout$, resp.), being it defined as a restriction of $\selsubtype$, admits multiple choices only for output selections in the supertype  (input branchings in the subtype, resp.), here we consider and show decidability for a much larger relation, we denote by $\subtypesin$ ($\subtypesout$, resp.). Such a relation is defined as the {\it restriction of the whole $\subtype$ relation} (instead of the $\selsubtype$ relation),
where, in related types, all input branchings (output selections, resp.) must have a single choice.
Therefore, differently from $\selsubtypesin$ ($\selsubtypesout$, resp.), in $\subtypesin$ ($\subtypesout$, resp.)
{\it both the subtype and the supertype can include multiple choices} for output selections
  (for input branchings, resp.).
  The combination of non restricted buffers and presence of multiple choices
  on both related types requires a {\it totally new approach} for guaranteeing
  the termination of the subtyping algorithm (both for the termination condition itself and for the related decidability proof).
  For instance, if multiple choices are admitted for input branchings,
  the termination condition has to deal
%  (and related decidability proof technique)
  with complex recurrent patterns to be checked on the leafs of trees representing input branchings 
  (with multiple choices), 
  instead of detecting simple repetitions on strings representing sequences of single-choice inputs 
  (as in previous work \cite{BCZ16,LY17}). 

Concerning undecidability, all previous results \cite{BCZ16,LY17} exploit the capability of asynchronous subtyping
of matching input branchings/output selections by means of covariance/contravariance. We here show that
asynchronous subtyping remains undecidable even if we restrict it by disallowing this feature. %such a feature is not considered.
As asynchronous subtyping is based on the combination of output covariance/input
contravariance and output anticipation deriving from asynchronous communication, 
our result means that (provided that the syntax of types is not constrained) the source of undecidability is to be precisely 
localized into the output anticipation capability. 
The undecidability proof has a structure similar to that of Bravetti et al.\ \cite{BCZ16},
where the termination problem for queue machines, a well-known
Turing-equivalent formalism, is encoded into asynchronous session subtyping. 
However, differently from Bravetti et al.\ \cite{BCZ16}, not having covariance/contravariance makes it impossible
to encode queue machines deterministically. 
As we will see, the need to cope with nondeterminism makes it necessary to restrict the class of encodable
queue machines to a new ad-hoc fragment 
that we introduce in this paper and we prove being Turing-equivalent:
single consuming queue machines. Moreover a much more complex encoding, that uses nondeterminism, must be adopted. 
%
%
%
%For this subclass of queue machines we present an encoding 
%for which 
%
%
%We show how to encode this class of queue machines nondeterministically.
%
%Hence, we thus need to deal with 
%
%
%deterministically simulating the behaviour of queue machines. 
%
%
%
%
%
%
% 
%deal with a non deterministic encoding of queue machines
%
%
%nput contrvariance/output covariance
%
%
%
%
%With respect to \cite{BCZ16} here we however need to devise an ad-hoc
%Turing complete fragment of queue machines (they 

In general, the contribution of this work is
%two-fold: we detect some kind of restrictions 
to analyze % forms of
restrictions 
of % the
asynchronous subtyping and classifying them into decidable and undecidable fragments.
%: % relation
%%or of
%%the syntax of types
%those that make subtyping decidable (but which are much more permissive w.r.t.\ those in~\cite{BCZ16} and~\cite{LY17}, thus requiring a more involved proof of decidability); and those
%for which, instead, subtyping is still undecidable (however, due to the restrictions considered, requiring a more involved proof of undecidability w.r.t.~\cite{BCZ16} and~\cite{LY17}).
%
More precisely, as detailed in the following, we focus on two kinds of restrictions of asynchronous subtyping: limitations to the branching/selection structure and to the communication buffer, giving rise to the numerous relations shown in the lefthand part and righthand part of Figure~\ref{fig:lattice}, respectively (see Section \ref{sec:preliminaries} for formal definitions of all the relations). 
The relations are depicted as a lattice according to their inclusion as sets of pairs. Notice that decidability/undecidability is not logically related to set inclusion (e.g.\ the emptyset and the set of all pairs are both decidable and are the bottom and the top of the lattice).
%(such a figure also includes the  $\subtype$, $\selsubtype$,
%$\selsubtypesin$, $\selsubtypesout$ and $\subtypes$ relations we already discussed). 

Concerning asynchronous subtyping $\subtype$ itself, we consider the {\it orphan message free} notion of subtyping introduced in Chen et al.~\cite{MariangiolaPreciness}: %. This because
it is commonly recognized that a convenient notion of asynchronous subtyping should prevent existence of messages that remain ``orphan'', i.e.\ that are never consumed from the communication buffer. Operationally, this implies that inputs in the supertype cannot be indefinitely delayed by output anticipation: eventually such inputs must be performed by the subtype so to correspondingly consume messages from the buffer.
As a side result, in this paper we introduce a new, elegant, way
  of defining orphan message free asynchronous subtyping.
 The new definition is based on just adding a
  constraint about closure under duality to the standard (non orphan message free) coinductive
  definition of asynchronous subtyping~\cite{MY15}. 
%VA BENE???????????????
We thus use such a novel approach based on dual closeness to give a concise definition of asynchronous subtyping: we call $\subtypedc$ the obtained relation. We then show $\subtypedc$ (also included in Figure~\ref{fig:lattice}) to be equal to orphan message free subtyping $\subtype$ of Chen et  al.~\cite{MariangiolaPreciness}.
We also show that the most significant decidable and undecidable fragments of $\subtype$, i.e.\ the $\subtypesin \cup \subtypesout$ relation and subtyping without covariance/contravariance, are dual closed subtyping relations according to our new definition.

%show 
%%refine the
%undecidability of subtyping to hold even for a 
%
%a more fine-grained relation. % , where
%% covariance and contravariance no longer are permitted.

\myparagraph{Limitations to the branching/selection structure.} We limit the asynchronous subtyping $\subtype$ capability 
of managing input branchings/output selections, giving rise to the subtyping relations shown in the lefthand part of Figure \ref{fig:lattice}, as follows:
\begin{itemize}
\item requiring that in both the subtype and the supertype output selections (input branchings, resp.) have a {\it single choice}: in this case the $\mathsf{sout}$ ($\mathsf{sin}$, resp.) subscript is added to $\subtype$;
\item requiring that each output selection (input branching, resp.) performed by the subtype is matched by an  
 output selection (input branching, resp.) performed by the supertype with the exactly the same {\it total} set of labels, i.e.\ output covariance (input contravariance, resp.) is not admitted: in this case the $\mathsf{tout}$ ($\mathsf{tin}$, resp.) subscript is added to $\subtype$.
\end{itemize}
We now summarize our decidability/undecidability results for these relations.

%\myparagraph{Summary of our Main Contributions.} pippo
\begin{itemize}

\item {\em Decidability of asynchronous subtyping for single-in types ($\subtypesin$)
or single-out types ($\subtypesout$).}  
%To be precise, 
We consider the class of single-in (single-out, resp.) session types, i.e.\
types where all output selections (input branchings, resp.) have a single choice.
%%As discussed above,
%%  session types make it possible to use branching for both
%%  inputs (externally chosen) and outputs (internally chosen). 
%%  In the case of inputs, this results in an
%%  external choice: a process exposes different labels that the sender
%%  of a message can pick. On the other hand, branching outputs
%%  implement internal choices, i.e., when a system decides to call a
%%  particular operation or another. 
%  Single-out (resp. single-in) session types are session types where
%  outputs (resp. inputs) are always singleton, hence there is no
%  output (resp. input) branching. 
We present and prove correct an
  algorithm for deciding whether two single-out (resp. single-in, by
  exploiting the closure under duality property) types are in the
  subtyping relation.  From a modeling viewpoint, assuming binary
  sessions to happen between a single-in and a single-out party, this
  entails that internal decisions are taken by the single-in party,
  while the single-out one passively accepts them.  This kind of
  behaviour can occur in, e.g., %find applicability
% stateless 
  web-service systems where
%\cite{XXX}
  %as well as micro services 
%  where 
  a client internally chooses a request-response operation \cite{WSDL}
  % to call
  and then waits for a corresponding (non branching) input,
  while the server accepts invocations on several operations %requests
  but then it reacts by answering on a related response channel
  (independently of the actual returned
  data/error%and that no internal choice in the server can modify the set of requests it accepts
  ); see the examples at the beginning of this Introduction section. Our algorithms for subtyping of single-out/single-in types could
  thus be used in typing systems for server/client code.
With minor variants to the machinery introduced to show decidability of $\subtypesin$ and $\subtypesout$, we also show
that $\subtypesintout$, $\subtypetinsout$ and $\subtypesinsout$ are decidable.

\item {\em Undecidability of Asynchronous Subtyping without Output
    Covariance and Input Contravariance ($\subtypetintout$).
    } As discussed above, subtyping 
    for session types makes
  use of output covariance and input contravariance: an output
  $\oplus\{l_i:T_i\}_{i\in I}$ is a subtype of an output with more
 labels $\oplus\{l_j:T_j\}_{j\in J}$, for $I\subset J$; and an
  input $\&\{l_i:T_i\}_{i\in I}$ is a subtype of an input with less
  labels $\&\{l_j:T_j\}_{j\in J}$, for $J\subset I$. 
  Existing
  results on the undecidability of asynchronous subtyping exploit its capability
 of relating types with a different number of branches.  
We consider a restricted form of subtyping, the $\subtypetintout$ relation, which disallows this feature, i.e. which does not
use output covariance and input contravariance.
We show that, also with such a restriction, subtyping remains undecidable by 
encoding the termination problem for single consuming queue machines, a Turing-equivalent formalism that (as already explained) we introduce on purpose, into $\subtypetintout$.
The same encoding we use for $\subtypetintout$ shows also undecidability of $\subtypetin$, $\subtypetout$ and $\subtype$ (thus also providing an alternative proof for the undecidability of $\subtype$ with respect to those by Bravetti
et al.~\cite{BCZ16} and Lange and Yoshida~\cite{LY17}).
\end{itemize}

\myparagraph{Limitations to the communication buffer.}
We limit the communication buffer capability, giving rise to the subtyping relations shown in the righthand part of Figure \ref{fig:lattice}, by restricting the capability of $\subtype$ to anticipate outputs: this is equivalent to putting an upper limit to communication buffers between two parties, a common fact in practice.
In this context our decidability/undecidability results are the following ones.

\begin{itemize}

\item {\em Decidability of ${\mathsf k}$-bounded Subtyping ($\subtype_{\mathsf k}$), with ${\mathsf k} \geq 0$.}  
%Our first decidability result focuses on a variant of subtyping, called 
In ${\mathsf k}$-bounded asynchronous
  subtyping we restrict the capability of $\subtype$ to anticipate
  outputs: they can only be anticipated w.r.t.\ a number of inputs that is less or equal to ${\mathsf k}$.
  % number of inputs that can precede an output that is to be
  % anticipated while establishing subtyping.
%  This is equivalent to putting an upper limit to communication
%  buffers between two parties, a common fact in practice.  
We give and
  prove correct an algorithm for deciding whether any two session
  types are in a ${\mathsf k}$-bounded subtyping relation. Notice that, in the case ${\mathsf k}=0$ we obtain synchronous subtyping 
 $\subtypes$~\cite{GH05}.
 Moreover, if we consider ${\mathsf k}=1$ we have a notion of subtyping along the lines of that, 
 we already mentioned,
 obtained by Lange and Yoshida \cite{LY17} imposing restrictions on the communication 
 behaviour.

\item {\em Undecidability of Bounded Asynchronous Subtyping
    ($\subtypebound$).}  We say that a pair of session types is in
  bounded asynchronous subtyping relation if there exists a
  ${\mathsf k}$ such that such pair is in ${\mathsf k}$-bounded
  subtyping relation.  Bounded asynchronous subtyping relates types
  that do not unboundedly put messages in a buffer. For instance, the
  types
%  \vspace*{-2mm}
  $$\Trec t.\oplus \{huge\_data:\oplus \{huge\_data: \& \{ack:\Tvar
  t\}\}\}
% \vspace*{-2mm}
% $$
% and \vspace*{-2mm} $$
$$
$$
\qquad\qquad \Trec t.\& \{ack: \oplus \{huge\_data:\Tvar t\}\}
%\vspace*{-2mm}
$$ 
  are related by %standard 
  asynchronous subtyping but not by \emph{bounded} 
  asynchronous subtyping: the augmented data production frequency
  of the subtype requires to store an unbounded amount of huge data.
%   that are anticipated. 
%  We show that bounded 
%  asynchronous subtyping
%  deciding whether there exists a
%  $k$ for which two types are in the $k$-bounded subtyping relation 
%  is undecidable.
  Since in practice buffers are bounded, this could have been
  an acceptable candidate notion for replacing standard asynchronous
  subtyping, however we prove that it is undecidable as well.
  We do this by showing undecidability of a property for queue machines: 
  bounded non termination.
  
\end{itemize}

%(\S~\ref{subsec:singleout}).
%
%(\S~\ref{subsec:undec_first}).
%
%(\S~\ref{subsec:kbounded})
%
%(\S~\ref{subsec:undec_bounded})

\myparagraph{Outline.} 
In Section~\ref{sec:preliminaries}
we present session types and definition of asynchronous subtyping $\subtype$, the novel dual closed reformulation $\subtypedc$,
the fragments of $\subtype$ shown in Figure \ref{fig:lattice} and a discussion about their properties.
Section~\ref{sec:algorithm} presents all
decidability results, notably for 
 ${\mathsf k}$-bounded subtyping ($\subtype_{\mathsf k}$) and for subtyping over single-in types ($\subtypesin$) and
over single-out types ($\subtypesout$).
%relations $\subtype_k$ in the righthand bottom part of Figure \ref{fig:lattice}, limiting the communication buffer capability to be $k$ bounded, and for the relations $\subtypesin$, $\subtypesout$ and the other relations in the lefthand bottom part of Figure \ref{fig:lattice}, limiting the branching/selection structure to either single-in or single-out types only.
Section~\ref{sec:undecidability} presents all
undecidability results, notably for 
bounded subtyping ($\subtypebound$) and for subtyping without output
    covariance and input contravariance ($\subtypetintout$).
Section~\ref{sec:relatedwork} discusses 
related work and
Section~\ref{sec:conclusions}
presents concluding remarks.
Detailed proofs of theorems, lemmas and
propositions can be found in the Appendix. We chose to put proof technicalities, that often include additional definitions and intermediate results, in the Appendix so not to disrupt the paper prose.

% %%% Local Variables: 
% %%% mode: latex
% %%% TeX-master: "main"
% %%% End: 

\section{Session Types and Asynchronous Subtyping}\label{sec:preliminaries}
We begin by formally introducing the various ingredients needed for
our technical development.
%
% \myparagraph{Session Types.} 

We start with the formal syntax % of a variant
of binary session types.
%, firstly introduced by Honda et al.~\cite{HVK98}. 
Similarly to Chen et
al.~\cite{MariangiolaPreciness} we %use a variant that merges outputs
%with internal choice and inputs with external choice.
do not use a dedicated construct for sending an output/receiving an input, we  
instead represent outputs and inputs directly inside choices.
More precisely, we consider output selection $\Tselect{l}{T}$, 
expressing an internal choice among outputs,
and input branching $\Tbranch{l}{T}$, expressing an external choice among inputs. 
Each possible choice % in a choice
is labeled by a label
$l_i$, taken from a global set of labels $L$, followed by a session
continuation $T_i$. Labels in a branching/selection are assumed to be pairwise distinct.

%we consider {\em output} type $\Tselect{l}{T}$ denoting the type of a
%channel over which we can send one of the labels $l_i$ (for $i\in I$).
%% If all output subterms occurring in a session type are singleton, we
%% call such type single-out. 
%Dually, the {\em input} type $\Tbranch{l}{T}$ denotes the type of a
%channel ready to receive one of the labels $l_i$. 
%
\begin{definition}[Session Types]\label{def:sessiontypes}
  Given a set of labels $L$, ranged over by $l$, the syntax of binary
  session types is given by the following grammar:
  \begin{displaymath}
    \begin{array}{lrl}
      T &::=&   %\Tout U. T
      % \mid \Tin U. T
      % \mid 
                                          \Tselect{l}{T} 
              \quad \mid\quad  \Tbranch{l}{T} 
              \quad \mid\quad  \Trec t.T
              \quad \mid\quad \Tvar t
              \quad \mid\quad \Tend
              % \\\\
              % U,V & ::= & S\mid \mathsf{bool}\mid\mathsf{nat}\mid\ldots 
    \end{array}
  \end{displaymath}
%  $T \neq \Tvar t$ in
  % $\Trec t.T$.  
%.
  % We assume guarded recursion by imposing $T \neq \Tvar t$ in
  % $\Trec t.T$.  
  
  \noindent A session type is \emph{single-out} if, for all of its
  subterms $\Tselect{l}{T}$, $|I|=1$. Similarly, a session type is
  \emph{single-in} if, for all of its subterms $\Tbranch{l}{T}$,
  $|I|=1$.
\end{definition}
% Above, the {\em output} type $\Tselect{l}{T}$ denotes the type of a
% channel over which we can send one of the labels $l_i$ (for $i\in I$).
% If all output subterms occurring in a session type are singleton, we
% call such type single-out.  Dually, the {\em input} type
% $\Tbranch{l}{T}$ denotes the type of a channel ready to receive one of
% the labels $l_i$.
In the sequel, we leave implicit the index set $i \in I$ in input branchings and
output selections when it is already clear from the denotation of the
types. Note also that we abstract from the type of the message that
could be sent over the channel, since this is orthogonal to our
theory. Types $\Trec t.T$ and $\Tvar t$ denote standard tail recursion
for recursive types.  We assume recursion to be guarded: in
$\Trec t.T$, the recursion variable $\Tvar t$ occurs within the scope
of an output or an input type. In the
following, %, excluding where explicitly stated,
we will consider closed terms only, i.e., types with all recursion
variables $\Tvar t$ occurring under the scope of a corresponding
definition $\Trec t.T$.  Type $\Tend$ denotes the type of a channel
that can no longer be used.

In our development, it is crucial to count the number of times we need
to unfold a recursion $\Trec t.T$. This is formalised by the following
function:
\begin{definition}[$n$-unfolding]\label{def:unfolding}
  \begin{displaymath}
    \begin{array}{l@{\qquad}l}
      \unfold 0T = T %\qquad\text{(for all T)}
      &
%      \unfold 1{\Tout U. T} = \Tout U. \unfold 1{T}
%      \\
%      \unfold 1{\Tin U. T } = \Tin U. \unfold 1{T }
%      \\
      \unfold 1{\Tselect{l}{T}} =  \oplus\{{l}_{i}:{\unfold 1 {T_i}}\}_{i\in I}
      \\
      \unfold 1{\Trec t.T} = T\{\Trec t.T/\Tvar t\}
      &
      \unfold 1{\Tbranch{l}{T} } = \&\{{l}_{i}:{\unfold 1 {T_i}}\}_{i\in I}      
      \\
%      \multicolumn{2}{l}{
%      \unfold 1{\Tvar t} = \Tvar t
%      \qquad
      \unfold 1{\Tend} = \Tend
%      \qquad
&
      \unfold n{T} = \unfold 1{\unfold {n-1}T}
%      }
    \end{array}
  \end{displaymath}
\end{definition}

The definition of asynchronous subtyping uses the notion of input
context, a type context consisting of a sequence of inputs preceding
holes where types can be placed:
\begin{definition}[Input Context]\label{def:context}
  An input context $\mathcal A$ is a session type with multiple holes
  defined by the syntax:\quad 
  $\mathcal A\ ::=\ \quad[\,]^n
  \quad\mid\qquad %\Tin U.\mathcal A\quad\mid\quad
  \Tbranch{l}{\mathcal A}$.\\
  An input context $\mathcal A$ is well-formed whenever all its holes
  $[\,]^n$, with $n \in \mathbb{N}^+$, are consistently enumerated, i.e.\ there exists $m \geq 1$ such that $\mathcal A$ includes one and only one $[\,]^n$ for each $n \leq m$. % with a natural number.
  Given a well-formed input context $\mathcal A$ with holes indexed 
  over $\{1,\ldots,m\}$ and types $T_1$,\dots, $T_m$, we use $\context {A} {T_k} {k\in
        \{1,\ldots, m\}}$ to denote the type obtained by filling
  each hole $k$ in $\mathcal A$ with the corresponding term $T_k$.
\end{definition}

From now on, whenever using input contexts we will assume them to be
well-formed, unless otherwise specified.

For example, consider the input context 
$$\mathcal A=\&\{l_1: []^1,\ l_2: []^2\}$$
we have:
%Thus, in the subtyping example above, the well-formed input context considered to express the output $l$ anticipation is
$$\context {A} {\oplus\{l:T_i \}} {i\in
        \{1,2\}} = \&\!\big\{l_1:\oplus\{l:T_1 \}, l_2:\oplus\{l:T_2 \} \big\}$$

We start by considering the standard notion of
asynchronous subtyping $\subtype$ given by Chen et 
al.~\cite{MariangiolaPreciness}. We choose it because of its orphan message free property that is commonly recognized to be convenient: only subtypes are allowed that do not cause incoming messages to remain ``orphan'' (because they are never consumed from the communication buffer). 
In the definition of asynchronous subtyping given by Chen et 
al., orphan message freedom causes a specific dedicated constraint to be included
(which is, e.g., instead not present in the asynchronous subtyping definition
by Mostrous and Yoshida~\cite{MY15}).
We now formally present the asynchronous subtyping relation $\subtype$, rephrased w.r.t. that of Chen et 
al.~\cite{MariangiolaPreciness} in a technical format that is convenient for showing our results, which follows
a coinductive simulation-like definition.

\begin{definition}[Asynchronous Subtyping, $\subtype$]\label{def:subtyping} \label{subtype}
  $\mathcal R$ is an asynchronous subtyping relation if %whenever it is \emph{dual closed} and 
  $(T,S)\in\mathcal R$ implies that:
  \begin{enumerate}

    % \item if $T\in\{\mathsf{bool}\mid\mathsf{nat}\mid\ldots \}$ then
    %   $S=T$;

  \item %\label{item:end}
  
    if $T=\Tend$ then $\exists n\geq 0$ such that
    $\unfold nS = \Tend$;

  \item %\label{item:internal}
  
    if $T=\Tselect{l}{T}$ then $\exists n\geq 0,\mathcal A$ such that
    \begin{itemize}
      % \item $T'$ input/branch-free implies $\mathcal A=[]^1$,
    \item
      $\unfold nS = \context {A} {\Tselectindex l{S_k}{j}{J_k}} {k\in
        \{1,\ldots, m\}}$, %$ \qquad for some $J_k$,
    \item $\forall k\in\{1,\ldots, m\}.  \, I\subseteq J_k$,
    \item
      $\forall i\in I,%k\in\{1,\ldots, m\}. 
      (T_i,\context {A} {{S_{ki}}} {k\in \{1,\ldots, m\}})\in\mathcal R$ and
 
   \item if $\mathcal A\neq[\,]^1$ then $\forall i\in I. \; \& \in T_i$ (no orphan message constraint);

    \end{itemize}

  \item %\label{item:external}
  
  if $T=\Tbranch{l}{T}$ then $\exists n\geq 0$ such that
    $\unfold nS = \Tbranchindex lSjJ$, $J\subseteq I$ and
    $\forall j\in J. (T_j,S_j)\in\mathcal R$;

  \item %\label{item:rec}
  
  if $T= \Trec t.{T'} $ then $(T'\{T/\Tvar{t}\}, S)\in\mathcal R$.

  \end{enumerate}
 where with ``$\& \in T_i$'' we mean that $T_i$ contains at least an input branching.
   $T$ is an asynchronous subtype of $S$, written $T \subtype S$, if there 
  is an asynchronous subtyping relation $\mathcal R$ such that $(T,S) \in \mathcal R$.
\end{definition}

Intuitively, two types $T$ and $S$ are related by $\subtype$, whenever
$S$ is able to simulate $T$, but with a few twists: type $S$ is
allowed to anticipate outputs nested in its syntax tree (asynchrony);
and, output and input types enjoy covariance and contravariance,
respectively.
Moreover, the above definition includes the no orphan message constraint~\cite{MariangiolaPreciness},
namely: we allow the supertype inputs to be delayed only if also the subtype contains 
some input.
%it requires that if the smaller type contains no inputs, then
%there must be no input that could be delayed in the bigger type (no
%orphan messages~\cite{MariangiolaPreciness}). 

A {\it synchronous} subtyping 
relation $\subtypes$ like that of Gay and Hole~\cite{GH05} is obtained by requiring that, in item 2.\ of the above Definition \ref{subtype}, it always holds $\mathcal A=[\,]^1$.
%The obtained relation is here denoted by .

%In the case, in Definition \ref{subtype}, we assume that, in item 2., it always holds $\mathcal A=[\,]^1$ we get a {\it synchronous} subtyping 
%relation, we call $\subtypes$, like that of Gay and Hole~\cite{GH05}.

\begin{example}
Consider 
$T= \Trec t.\oplus\{l:\& \{l_1:  \Tvar t, l_2:  \Tvar t\}\}$
and 
$S= \Trec t.\& \{ l_1: \& \{l_2: \oplus\{l: \Tvar t\} \} \}$.
We have $T \, \subtype \, S$ because the 
following is an asynchronous subtyping relation:
$$
\begin{array}{ll}
\{& (T,S)\ ,\ 
   (\oplus\{l:\& \{l_1:  T, l_2: T\}\},S)\ ,\
   (\& \{l_1:  T, l_2:  T\} , 
    \& \{ l_1: \& \{l_2: S \} \})\ ,\ \\
&   (T,\& \{l_2: S \})\ ,\
   (\oplus\{l:\& \{l_1:  T, l_2: T\}\},\& \{l_2: S\})\ ,\\
& \qquad   (\& \{l_1:  T, l_2:  T\} , 
    \& \{l_2:\& \{ l_1: \& \{l_2: S \} \} \}), \\          
&   (T,\& \{ l_1: \& \{ l_2: S \}\})\ ,\
    (\oplus\{l:\& \{l_1:  T, l_2: T\}\},\& \{ l_1: \& \{ l_2: S \}\})\ ,\\           
& \qquad   (\& \{l_1:  T, l_2:  T\} , 
    \& \{ l_1:\& \{l_2:\& \{ l_1: \& \{l_2: S \} \} \}\})\ , \\
&   (T,\& \{l_2:\& \{ l_1: \& \{l_2: S \} \} \})\ ,\ \ldots\qquad
 \}
\end{array}
$$
Note that the relation contains infinitely many pairs that differ in the 
sequence of inputs, alternatively labeled with $l_1$ and $l_2$,
that are accumulated at the beginning of the r.h.s. type.
\end{example}

We now introduce an alternative way of defining orphan message free asynchronous subtyping,
which is more elegant/concise: it obtains the orphan message freedom property by requiring 
{\it closure under duality} of the type relation being defined instead of making use of an explicit orphan 
message free constraint as in Definition \ref{def:subtyping}.

% The new definition is based on just adding a
%  constraint about closure under duality to the standard (non orphan message free) coinductive
%  definition of asynchronous subtyping~\cite{ESOP09}. 

For session types, we define the usual notion of duality:
%\begin{definition}[Duality]\label{def:sessiontypes}
given a session type $T$, its dual $\dual{T}$ is defined as:
$\dual{\Tselect{l}{T}} = \Tbranch{l}{\dual{T}}$,
$\dual{\Tbranch{l}{T}} = \Tselect{l}{\dual{T}}$,
$\dual{\Tend} = \Tend$, $\dual{\Tvar t} = \Tvar t$, and
$\dual{\Trec t.T} = \Trec t.\dual{T}$.
  % \begin{displaymath}
  %   \begin{array}{c}
      % \dual{\Tselect{l}{T}} = \Tbranch{l}{\dual{T}} \qquad\qquad
      % \dual{\Tbranch{l}{T}} = \Tselect{l}{\dual{T}} \\[1mm]
      % \dual{\Tend} = \Tend \qquad\qquad
      % \dual{\Tvar t} = \Tvar t \qquad\qquad
      % \dual{\Trec t.T} = \Trec t.\dual{T}
  %   \end{array}
  % \end{displaymath}
%\end{definition}
In the sequel, we say that a relation $\mathcal R$ on session types is
\emph{dual closed} if $(S,T) \in \mathcal R$ implies
$(\dual T,\dual S) \in \mathcal R$.

\begin{definition}[Asynchronous Dual Closed Subtyping,
  $\subtypedc$]\label{def:dcsubtyping}
  $\,\mathcal R\!\!\;$ is an asynchronous dual closed subtyping relation whenever it is \emph{dual closed} and 
  $(T,S)\in\mathcal R$ implies $1.$, $3.$, and $4.$ of Definition
  \ref{def:subtyping}, plus a modified version of $2.$ where the last constraint (the no orphan message constraint) 
is removed.

  $T$ is an asynchronous dual closed subtype of $S$, written
  $T \, \subtypedc \, S$, if there is an asynchronous dual closed subtyping
  relation $\mathcal R$ such that $(T,S) \, \in \, \mathcal R$.
\end{definition}  

We observe that our definition is formally different from the ones
found in literature. 
In particular, 
with respect to that 
by Mostrous and Yoshida~\cite{MY15},
it additionally requires the subtyping relation
to be dual closed. 
%In the sequel, we show that, with such a
%restriction, our subtyping is equivalent to 
%dual closeness 
Below, we state that the dual closeness requirement is equivalent to imposing the orphan message free constraint, i.e. the last item of condition 2 in Definition \ref{def:subtyping} 
(both guarantee
orphan-message freedom):
\begin{theorem}\label{thm:orphanMessageEquiv}
  Given two session types $T$ and $S$, we have $T \, \subtype \, S$ if and only if $T \, \subtypedc \, S$.
\end{theorem}

\subsection{Subtyping Relation Restrictions}\label{sec:subtyperest}

As already discussed in the Introduction, we focus on two kinds of restrictions of asynchronous subtyping: limitations to the branching/selection structure and to the communication buffer, giving rise to the numerous relations shown in the lefthand part and righthand part of Figure~\ref{fig:lattice}, respectively.

We now define fragments of $\subtype$ obtained by posing limitations to the branching/selection structure.
\begin{definition}
Restrictions of the asynchronous subtyping relation %$\subset$ 
are denoted by adding subscripts to the $\subtype$ notation, with the following meaning:
\begin{itemize}
\item whenever we add subscript $\mathsf{sout}$ ($\mathsf{sin}$, resp.) 
we additionally require in Definition \ref{def:subtyping} both $T$ and $S$ to be single-out
(single-in, resp.),
\item whenever we add subscript $\mathsf{tout}$ ($\mathsf{tin}$, resp.)
we additionally require in Definition \ref{def:subtyping} $I=J_k$ in
point 2. ($I=J$ in point 3., resp.).
\end{itemize}
\end{definition}
The latter means that each output selection (input branching, resp.) performed by the subtype is matched by an  
output selection (input branching, resp.) performed by the supertype with the exactly the same {\it total} set of labels, i.e.\ output covariance (input contravariance, resp.) is not admitted.

Notice that, while it holds that $\subtype = \subtypedc$, not all fragments of $\subtype$ are
asynchronous dual closed subtyping relations. For instance this does not hold for $\subtypesin$, $\subtypesout$, $\subtypetin$ and $\subtypetout$,
which perform a limitation, but not its ``dual'' one. It holds, instead, for the following two relations that we 
will show to be in the boundary between decidability and undecidability.

\begin{proposition}\label{prop:dualclosed}
The $\subtypetintout$ relation and the $\subtypesin \! \cup \! \subtypesout$ relation are asynchronous dual closed subtyping relations.
\end{proposition}

Dual closeness of the $\subtypesin \cup \subtypesout$ relation is a direct consequence of the fact that $T \,\subtypesin\, S$ if and only if $\dual{S} \,\subtypesout\, \dual{T}$, which obviously derives from dual closeness of $\subtype$ and from the dual of a single-in type being a single-out type and vice-versa. This fact, together with the following proposition, will be used to infer decidability of $\subtypesin$ and $\subtypesintout$ relations from that of $\subtypesout$ and $\subtypetinsout$, respectively.

\begin{proposition}\label{prop:complicated}
The $\subtypesintout$ and $\subtypetinsout$ relations are such that: $T \,\subtypesintout\, S$ if and only if $\dual{S} \,\subtypetinsout\, \dual{T}$.
\end{proposition}

We now consider variants of $\subtype$ obtained by posing limitations to the communication buffer.

We can define a
variant decidable relation by putting an upper-bound to the messages
that can be buffered.
Technically speaking, when an output in the r.h.s.
is anticipated during the subtyping simulation,
we impose a bound to the number of inputs that are in front
of such output.

We say that an input
context $\mathcal A$ is {\em ${\mathsf k}$-bounded} if the maximal number of
nested inputs in $\mathcal A$ is less or equal to $k$.

\begin{definition}[${\mathsf k}$-bounded Asynchronous Subtyping, with ${\mathsf k} \geq 0$]\label{def:kbounded}
  $\,$ The ${\mathsf k}$-bounded asynchronous subtyping $\subtype_{\mathsf k}$ is defined as
  %$\subtype$ 
  in Definition \ref{def:subtyping}, with the only
  difference that the input context $\mathcal A$ in item $2.$ is
  required to be ${\mathsf k}$-bounded.
\end{definition}

Notice that the case ${\mathsf k}=0$ yields synchronous subtyping: since  $\mathcal A=[\,]^1$
is the only $0$-bounded input context, we obviously have  $\subtype_0=\subtypes$. 

Lange and Yoshida \cite{LY17} show the decidability
of asynchronous subtyping for a subclass of session
types, called \emph{alternating}, that in our setting corresponds
to impose that every output in a subtype is immediately
followed by an input, while every input in a supertype is 
followed by an output.
For instance, this property is satisfied by the following pair of types:
$$
T=\Trec t.\oplus\{l_2: \& \{l_1:\Tvar t\}\}
\qquad
S=\Trec t.\&\{l_1: \oplus \{l_2:\Tvar t\}\}
$$ 
It is not difficult to see that $T \subtype_1 S$.
The key point of the proof of decidability
of asynchronous subtyping for alternating session types 
by Lange and Yoshida is the observation that if $T$ and
$S$ are alternating, then $T \subtype S$
if and only if $T \subtype_1 S$. 

As we explained in the Introduction, we also consider the more generic notion of {\it bounded}
asynchronous subtyping. This relation is in our opinion
of interest because it reflects real cases in which
it is possible to assume bounded buffers, without 
an a priory knowledge of the actual bound.

\begin{definition}[Bounded Asynchronous Subtyping, $\subtypebound$]\label{def:boundedSub}
  $\!\!$We say that $T$ is a 
  \emph{bounded asynchronous subtype}
  of $S$, written $T \subtypebound S$, if there exists ${\mathsf k}$ such that
  $T \subtype_{\mathsf k} S$.
%  is the
%  union of all $k$-bounded asynchronous subtypings, for every
%  $k \in \mathbb{N}$.
\end{definition}

% %%% Local Variables: 
% %%% mode: latex
% %%% TeX-master: "main"
% %%% End: 

\section{Decidability Results}\label{sec:algorithm}
We now present decidability results for ${\mathsf k}$-bounded asynchronous
subtyping
%, a variant of asynchronous subtyping, 
and asynchronous
subtyping for single-out/single-in session types.
% identify classes of session types for which asynchronous subtyping
% is decidable. Such class is the class of single-out session types
% where instances of the selection construct $\Tselect{l}{T}$ are
% always singletons ($|I|=1$).

\subsection{A Subtyping Procedure}\label{subsec:subproc} 
We start by giving a procedure (an algorithm that does not necessarily
terminate) for the general subtyping relation, which is known to be
undecidable~\cite{BCZ16,LY17}.  Such a procedure is inspired by the
one proposed by Mostrous et al.~\cite{ESOP09} for asynchronous
subtyping in multiparty session types. In order to do so, we introduce
two functions on the syntax of types. The function $\depth$ calculates
how many unfolding are necessary for bringing an output outside a
recursion. If that is not possible, the function is undefined (denoted
by $\perp$).
%
%\begin{definition}[outDepth]\label{def:depth}
%  The partial function $\depth(T,\Gamma)$, with $\Gamma$ set of
%  recursion variables, is inductively defined as:
%  \begin{displaymath}
%  \begin{array}{ll}
%    \depth(\Tselect
%    lT,\Gamma) \!=\! 0
%    \qquad
%    \depth(\Tbranch lT,\Gamma) \!=\! 
%    % \perp & \text{if }\exists j.\ \depth(T_j,\Gamma)=\perp\\
%    \mathsf{max}\{\depth(T_i,\Gamma)\ |\ i\in I\} % & \text{otherwise}
%    \\[2mm]
%    \depth(\Tend,\Gamma) \!=\! \perp 
%    \quad
%    \depth(\Trec t.T,\Gamma) \!=\! \left\{\!\!\!
%    \begin{array}{ll}
%      \perp & 
%              \text{if } \Tvar t\in\Gamma 
%              % \lor \depth(T\{\Trec t.T/\Tvar t\}, \Gamma+\{\Tvar
%              % t\})=\perp
%      \\
%      1\!+\!\depth(T\{\Trec t.T/\Tvar t\}, \Gamma\!\cup\!\{\Tvar t\}) & \text{otherwise}
%    \end{array}
%                                                             \right.
%  \end{array}
%\end{displaymath}
%where $\mathsf{max}\{\depth(T_i,\Gamma)\ |\ i\in I\}=\perp$, if
%$\depth(T_i,\Gamma)=\perp$ for some $i\in I$; similarly,
%$1+\perp = \perp$. We use $\depth(T)$ as a shorthand for
%$\depth(T,\emptyset)$.
%\end{definition}
%
%VERSIONE ALTERNATIVA:

\begin{definition}[outDepth]\label{def:depth}
  The partial function $\depth(T)$ is inductively defined 
%on open
%  terms (recursion variables could be outside the scope of a
%  corresponding definition) 
as follows:
  \begin{displaymath}
  \!\!\!\begin{array}{ll}
    \depth(\Tselect
    lT) \!=\! 0
    \quad \;
    \depth(\Tbranch lT) \!=\! 
    % \perp & \text{if }\exists j.\ \depth(T_j,\Gamma)=\perp\\
    \mathsf{max}\{\depth(T_i)\, |\, i \!\in\! I\} % & \text{otherwise}
    \\[2mm]
    \depth(\Trec t.T) \!=\! 
      1\!+\!\depth(T\{\Tend/\Tvar{t}\})                                                         
    \quad
    \depth(\Tend) \!= \perp 
   \end{array}
\end{displaymath}
where $\mathsf{max}\{\depth(T_i)\ |\ i\in I\}=\perp$, if
$\depth(T_i)=\perp$ for some $i\in I$; similarly,
$1+\!\perp = \perp$. 
\end{definition}

%Despite $\depth$ is defined on open terms, it is worth to observe
%that, in the following, it will be applied to closed terms only.
%FINO QUI (SI MODIFICA QUALCOSA NELLE PROVE??) \\
As an example of application of $\depth$ consider, for any $T_1$ and $T_2$,
$\depth(\oplus\{l_1:T_1,\ l_2:T_2\})=0$. On the other hand, consider
the type
$T_{\textsf ex}=\&\Big\{l_1:\Trec t.\oplus\!\big\{l_2:T_1\big\},\
l_3:\Trec t.\&\!\big\{l_4:\Trec t'.\oplus\!\{l_5:T_2\}\big\}\Big\}$:
clearly, $\depth\big(T_{\textsf ex}\big)=2$.  We then define
$\outunfold$, a variant of the unfolding function given in
Definition~\ref{def:unfolding}, which unfolds only where it is
necessary, in order to reach an output:
\begin{definition}[outUnf]\label{def:outunfolding}
  The output unfolding $\outunfold T$ is a partial function defined
  whenever $\depth(T)$ is defined. Given $\depth(T)=n$, $\outunfold T$
  is computed using the same inductive rules of $\unfold nT$,
  excluding the rule for $\Tselect{l}{T}$ that, instead of recursively
  unfolding $T_i$, returns the same term $\Tselect{l}{T}$.
\end{definition}
The function above differs from $\unfoldNOPAR n$: for example,
$\unfoldNOPAR 2 \big(T_{\textsf ex}\big)$ would unfold twice both
subterms $\Trec t.\oplus\!\{l_2:T_1\}$ and
$\Trec t.\&\!\big\{l_4:\Trec t'.\oplus\!\{l_5:T_2\}\big\}$.  On the
other hand, applying $\mathsf{outUnf}$ to the same term would unfold once the
term reached with $l_1$ and twice the one reached with $l_3$.

In the subtyping procedure defined below we make use of $\mathsf{outUnf}$
in order to have that recursive definitions under the scope
of an output are never unfolded. This guarantees that during the
execution of the procedure, even if the set of reached terms could be
unbounded, all the subterms starting with an output are
taken from a bounded set of terms. This is important to guarantee
termination\footnote{Technically speaking, this property of
the unfolding is used in the proof of Theorem \ref{theo:termination}.} of the algorithm that we will define in Section 
\ref{subsec:singleout} as an extension of the procedure 
described below.

\myparagraph{Subtyping Procedure.}  An environment $\Sigma$ is a set
containing pairs $(T,S)$, where $T$ and $S$ are types. %We consider
%represent the states of our subtyping procedure 
Judgements are triples of the form $\Sigma\vdash T\subtypea S$ which
intuitively read as ``in order to succeed, the procedure must check
whether $T$ is a subtype of $S$, provided that pairs in $\Sigma$ have
already been visited''.  Our {\em subtyping procedure}, applied to the
types $T$ and $S$, consists of deriving the state space of our
judgments using the rules in Figure~\ref{fig:algo} bottom-up starting
from the initial judgement $\emptyset \vdash T \subtypea S$.  More
precisely, we use the transition relation
%defined as a transition relation
% between triples of the form
$\Sigma \vdash T \subtypea S \rderiv \Sigma' \vdash T' \subtypea S'$
to indicate that 
%such that 
if $\Sigma \vdash T \subtypea S$ matches the conclusions of one of the
rules in Figure~\ref{fig:algo}, then $\Sigma' \vdash T' \subtypea S'$
is produced by the corresponding premises.  The procedure explores the
reachable judgements according to this transition relation.
We give highest priority to rule $\textsf{Asmp}$, thus ensuring that
at most one rule is applicable.\footnote{The priority of
  $\textsf{Asmp}$ is sufficient because all the other rules are
  alternative, i.e., given a judgement $\Sigma \vdash T \subtypea S$
  there are no two rules different from $\textsf{Asmp}$ that can be
  both applied.} The idea behind
% environment 
$\Sigma$ is to
avoid cycles when dealing with recursive types. Rules
$\textsf{RecR}_1$ and $\textsf{RecR}_2$ deal with the case in which
we need to unfold recursion in the type on the right-hand side. 
If the type on the left-hand side is not an output then the
procedure simply adds the current pair to $\Sigma$ and continues. On
the other hand, if an output must be found, we apply $\textsf{RecR}_2$
which checks whether such output is available. Rule \textsf{Out}
allows nested outputs to be anticipated (when not under recursion) and
condition
$\big( \mathcal{A}\neq [\,]^1\big) \Rightarrow \forall i\in I.\& \in
T_i$
makes sure there are no orphan messages. The remaining rules are
self-explanatory.
$\Sigma \vdash T \subtypea S \rderiv^* \Sigma' \vdash T' \subtypea S'$
is the reflexive and transitive closure of the transition relation
among judgements.  We write $\Sigma \vdash T \subtypea S \success$ if
the judgement $\Sigma \vdash T \subtypea S$ matches the conclusion of
one of the axioms \textsf{Asmp} or \textsf{End}, and
$\Sigma \vdash T \subtypea S \notrderiv$ to mean that no rule can be
applied to $\Sigma \vdash T \subtypea S$.  Due to input branching and output
selection,
the rules 
\textsf{In} and \textsf{Out} could generate
branching also in the state space to be explored by the
procedure. Namely, given a judgement $\Sigma \vdash T \subtypea S$,
there are several subsequent judgements
$\Sigma' \vdash T' \subtypea S'$ sucht that
$\Sigma \vdash T \subtypea S \rderiv \Sigma' \vdash T' \subtypea S'$.
The procedure could (i) successfully terminate because all the
explored branches reach a successful judgement
$\Sigma' \vdash T' \subtypea S' \success$, (ii) terminate with an
error in case at least one judgement
$\Sigma' \vdash T' \subtypea S' \notrderiv$ is reached, or (iii)
diverge because no branch terminates with an error and at least one
branch never reaches a succesful judgement.
%
%Checking whether a type $T$
%is a subtype of a type $S$ may be done by applying our procedure to
%$\emptyset \vdash T \subtypea S$.
%
% \begin{definition}[Subtyping procedure]\label{def:applyAlgo}
%   To check whether $T \subtype S$ we apply the rules in Figure
%   \ref{fig:algo} starting from the judgement
%   $\emptyset \vdash T \subtypea S$.  Given a judgement
%   $\Sigma \vdash T \subtypea S$, we write
%   $\Sigma \vdash T \subtypea S \rderiv \Sigma' \vdash T' \subtypea S'$
%   if $\Sigma \vdash T \subtypea S$ matches the consequences of one of
%   the rules, and $\Sigma' \vdash T' \subtypea S'$ is produced by the
%   corresponding premises.
%   $\Sigma \vdash T \subtypea S \rderiv^* \Sigma' \vdash T' \subtypea
%   S'$
%   is the reflexive and transitive closure of such relation.  We write
%   $\Sigma \vdash T \subtypea S \notrderiv$ to mean that no rule can be
%   applied to the judgement $\Sigma \vdash T \subtypea S$.  To avoid
%   the possibility to apply two distinct rules to the same judgement,
%   we give priority to $\textsf{Asmp}$.\footnote{The priority of
%     $\textsf{Asmp}$ is sufficient because all the other rules are
%     alternative, i.e., given a judgement $\Sigma \vdash T \subtypea S$
%     there are no two rules different from $\textsf{Asmp}$ that can be
%     both applied.}
% \end{definition}
%
\begin{figure}[t]
  \begin{displaymath}\small
    \begin{array}{c}
      \infer[\textsf{Out}]
      {
      \Sigma \vdash \Tselect{l}{T} \subtypea \context A{\Tselectindex{l}{S_n}j{J_n}}{n}
      }
      {
        \begin{array}{c}
           ( \mathcal{A}\neq [\,]^1) \Rightarrow \forall i\in I.\& \in T_i\\%[1mm]
          \forall n.I\subseteq J_n \qquad \forall i\in I\,.\,\Sigma \vdash
          T_i\subtypea \context A {S_{ni}}{n} 
        \end{array}
      %{t}
      % & t\ \mathtt{fresh}
          }\ \ \ \ \ \
       \\[.5cm]
        \infer[\textsf{In}]
        {
        \Sigma \vdash \Tbranch{l}{T} \subtypea \Tbranchindex{l}{S}jJ
        }
        {
        J\subseteq I & \forall j\in J\,.\,\Sigma \vdash T_j\subtypea S_j
                       } 
      \qquad \qquad
      \infer[\textsf{End}]
      {
      \Sigma \vdash \Tend\subtypea\Tend
      }
      {
      }
      \\[.5cm]
%      \multicolumn{2}{c}{
      \infer[\textsf{Asmp}]
      {
      \Sigma, (T,S) \vdash T\subtypea S
      }
      {
      }
      \qquad \qquad
      \infer[\textsf{RecL}]
      {
      \Sigma \vdash \Trec t.T \subtypea S
      }
      {
      \Sigma,(\Trec t.T,S) \vdash T\{\Trec t.T/\Tvar t\} \subtypea S
      }
%      }
      % &
      % \infer[\textsf{RecR}]
      % {
      % \Sigma \vdash T\subtypea S
      % }
      %   {
      %   \begin{array}{l}
      %     n=\mathsf{depth}(S,\emptyset)
      %     \quad
      %     n \geq 1 \\
      %     \Sigma, (T,S) \vdash T\subtypea \unfold nS
      %   \end{array}
      % }
      \\\\
%      \multicolumn{2}{c}{
      \infer[\textsf{RecR}_1]
      {
      \Sigma \vdash T\subtypea \Trec t.S
      }
      {
      T=\Tend \vee T=\Tbranch lT & \Sigma, (T,\Trec t.S) \vdash T\subtypea S\{\Trec t.S/\Tvar t\}
                                   }
%                                   }
      \\\\
%      \multicolumn{2}{c}{
      \infer[\textsf{RecR}_2]
      {
      \Sigma \vdash \Tselect lT\subtypea S
      }
      {
      \depth(S) \geq 1  &
                    \Sigma, (\Tselect lT,S) \vdash \Tselect lT\subtypea \outunfold S
                    }
%                    }
    \end{array}
  \end{displaymath}
  \caption{A Procedure for Checking Subtyping}
  \label{fig:algo} 
\end{figure}
\begin{example}\label{exampleInfinito}
  Consider $T=\Trec t.\oplus\big\{l_1:\&\{l_2:\Tvar t\}\big\}$ and
  $S=\Trec t.\oplus\Big\{l_1:\&\big\{l_2:\&\{l_2:\Tvar
  t\}\big\}\Big\}$.  Clearly, the two types $T$ and $S$ are related by
  asynchronous subtyping, i.e. $T\subtype S$. However, the subtyping
  procedure on $\emptyset\vdash T\subtypea S$ does not terminate:

  $$
  \!\!\!\begin{array}{l}
   \emptyset\vdash T\subtypea S \rightarrow \\
   \big\{(T,S)\big\} \vdash \oplus\big\{l_1\!:\&\{l_2\!:T\}\big\}\subtypea S \rightarrow \\
   \big\{(T,S),(\oplus\big\{l_1\!:\&\{l_2\!:T\}\big\},S)\big\} 
   \\
   \qquad \vdash \oplus\big\{l_1\!:\&\{l_2\!:T\}\big\}\subtypea \oplus\Big\{l_1\!:\&\big\{l_2\!:\&\{l_2\!:S\}\big\}\Big\} \rightarrow \\
   \big\{(T,S),(\oplus\big\{l_1\!:\&\{l_2\!:T\}\big\},S)\big\} 
   \vdash \&\{l_2\!:T\}\subtypea \&\big\{l_2\!:\&\{l_2\!:S\}\big\} \rightarrow \\
   \big\{(T,S),(\oplus\big\{l_1\!:\&\{l_2\!:T\}\big\},S)\big\} 
   \vdash T\subtypea \&\{l_2\!:S\} \rightarrow \\
   \big\{(T,S),(\oplus\big\{l_1\!:\&\{l_2\!:T\}\big\},S),(T, \&\{l_2\!:S\})\big\} 
   \vdash \oplus\big\{l_1\!:\&\{l_2\!:T\}\big\}\!\subtypea \&\{l_2\!:S\} \rightarrow \\
   \big\{(T,S),(\oplus\big\{l_1\!:\&\{l_2\!:T\}\big\},S),(T, \&\{l_2\!:S\}),(\oplus\big\{l_1\!:\&\{l_2\!:T\}\big\}, \&\{l_2\!:S\})\big\} 
 \\
   \qquad    \vdash \oplus\big\{l_1\!:\&\{l_2\!:T\}\big\}\subtypea \&\Big\{l_2\!:\oplus\big\{l_1\!:\&\big\{l_2\!:\&\{l_2\!:S\}\big\}\big\}\Big\} \rightarrow \\
      \big\{(T,S),(\oplus\big\{l_1\!:\&\{l_2\!:T\}\big\},S),(T, \&\{l_2\!:S\}),(\oplus\big\{l_1\!:\&\{l_2\!:T\}\big\}, \&\{l_2\!:S\})\big\} 
    \\
   \qquad \vdash \&\{l_2\!:T\}\subtypea \&\Big\{l_2\!:\&\big\{l_2\!:\&\{l_2\!:S\}\big\}\Big\} \rightarrow\\
    \dots\\
  \end{array}
  $$
Notice that the last step above is obtained by application of the
rule \textsf{Out} by considering the input context $\mathcal{A}=\&\{l_2:[\,]\}$.
%  : the
%  right-hand side will grow indefinitely since every unfolding adds
%  two inputs.
  % [Put here one or two examples showing that the procedure can be
  % divergent, i.e. there are infinitely many distinct $\Sigma' \vdash
  % T' \subtypea S'$ such that $\emptyset \vdash T \subtypea S \rderiv^*
  % \Sigma' \vdash T' \subtypea S'$.]
\end{example}

The example above shows that the procedure could diverge; the next
result proves that this can happen only if the checked types are in
subtyping relation. More precisely, types $T$ and $S$ are {\em not} in
subtyping relation if and only if the procedure on $\emptyset\vdash T\subtypea S$ terminates with an error; formally
%This is shown by the following
%
\begin{proposition}\label{prop:semidecidable}
  Given the types $T$ and $S$, we have that there exist $\Sigma', T', S'$
  such that
 $T \not\!\!\subtype S$   if and only if
  $\emptyset \vdash T \subtypea S
  \rderiv^* \Sigma' \vdash T' \subtypea S' \notrderiv$.
\end{proposition}
This means that: if $T \not\!\!\subtype S$ then the procedure on $\emptyset\vdash T\subtypea S$ surely
terminates with an error; if, instead, $T\subtype S$ then the
procedure  terminates successfully or
diverges.

\subsection{${\mathsf k}$-bounded Asynchronous Subtyping}\label{subsec:kbounded}

In the previous subsection we have shown that the standard subtyping procedure does not
terminate in general. In order to guarantee termination,
Lange and Yoshida \cite{LY17} have considered limitations to the
communication buffer, like half-duplex (in this case asynchronous and
synchronous subtyping coincides) or alternating protocols
(in this case the buffer will store at most one message).
We now prove a more general decidability result. We
show that,
for every ${\mathsf k}$, 
we can define an algorithm for the notion of ${\mathsf k}$-bounded asynchronous subtyping introduced in Section \ref{sec:subtyperest},
building on the subtyping procedure defined previously.
%
% \myparagraph{$k$-bounded Subtyping Algorithm.} 

We consider an algorithm, that we denote with $\subtypeak$,
obtained from the above procedure for $\subtypea$ 
simply by imposing   
%with the only difference 
that the input context $\mathcal A$, used in rule $\textsf{Out}$ in
Figure~\ref{fig:algo}, is always %assumed to be 
${\mathsf k}$-bounded.
Then, the following result holds:
\begin{theorem}\label{theo:kbounded}
  % Let $\subtypeak$ be defined as $\subtypea$ in Figure
  % \ref{fig:algo} with the unique
  % difference that the input context $\mathcal A$ in rule $\textsf
  % Out$ is
  % assumed to be $k$-bounded.
  The algorithm for $\subtypeak$ always terminates and, given the types $T$
  and $S$, 
there exist $\Sigma', T', S'$
  such that
  $\emptyset \vdash T \subtypeak S
  \rderiv^* \Sigma' \vdash T' \subtypea S' \notrderiv$
  if and only if $T \not\!\!\subtype_{\mathsf k} S$.
\end{theorem}

\subsection{Asynchronous Subtyping for Single-Out or Single-In
  Types}\label{subsec:singleout}

In Example \ref{exampleInfinito} we have seen that, if we consider
the terms $T=\Trec t.\oplus\big\{l_1:\&\{l_2:\Tvar t\}\big\}$ and
  $S=\Trec t.\oplus\Big\{l_1:\&\big\{l_2:\&\{l_2:\Tvar
  t\}\big\}\Big\}$, the subtyping procedure in Figure \ref{fig:algo}
  applied to $\emptyset\vdash T\subtypea S$ does not terminate.
The problem is that the termination
rule \textsf{Asmp} cannot be applied
because the term on the r.h.s. (i.e. the supertype)
generates always new terms
in the form 
$\&\Big\{l_2:\&\big\{l_2:\dots \&\{l_2:S\}\dots \big\}\Big\}$.
%Hence it does not happen that a pair of terms
%already in the environment $\Sigma$ is reached.

Notice that, in this particular example, these infinitely many 
distinct terms are obtained by adding single inputs 
(i.e. single-choice input branchings) in front of
the term in the r.h.s.: we call this \emph{linear input accumulation}.

For simple cases like this one, solutions have been proposed
by Lange and Yoshida \cite{LY17} and Bravetti et al. \cite{BCZ16}.
The idea is to extend the subtyping procedure in Figure \ref{fig:algo}
with additional termination rules able to detect when it is no longer
necessary to continue because 
it entered a deterministic loop (where the only possible future behavior of the procedure is to repeat
indefinitely the same linear input accumulation). This approach holds 
only under two assumptions, both satisfied by the subtyping relations
considered in  Lange and Yoshida \cite{LY17} and Bravetti et al. \cite{BCZ16}:
while checking subtyping output selections
in the l.h.s.\ (i.e.\ the subtype) are always single-choice and the same holds
for input branchings in the r.h.s.\ (i.e.\ the supertype).
This implies that there is a linear input accumulation, which is the repetition
of a specific sequence of input labels.
%, and the output selections in the l.h.s.\ (i.e.\ the subtype) are singletons. % single-out. 
The combination of these two assumptions
guarantees that the subtyping procedure proceeds deterministically:
this makes it possible to detect whether it enters a loop because the unique kind of loops are the deterministic ones.

In this section we show that it is possible to relax at least one
of these two assumptions: either deal with the case in which the input
accumulation is not linear, or deal with the case in which output selections
in the l.h.s.\ are not single-choice.
More precisely, the two cases that we consider are the following ones:
subtyping between single-out session types
(where input branchings in the r.h.s. are not constrained to be single-choice as in previous approaches)
and subtyping between single-in session types (where output selections in the l.h.s. are not
constrained to be single-choice as in previous approaches), i.e. the two relations $\subtypesout$ and $\subtypesin$, respectively, that we introduced in Section \ref{sec:subtyperest}.
The idea is to find an algorithm for one of the two cases and apply it
also to the other one by exploiting type duality.
%single-in, i.e., we restrict to terms in which inputs are always
%singletons, and single-out, i.e., we assume outputs are always 
%singletons. 

In the single-in case we surely have linear input accumulation but
%we have non-determinism in 
%the execution of 
the subtyping procedure is no longer deterministic
due to non-single output selections in the l.h.s.\ that have multiple possible continuations.
This causes the approach proposed in Lange and Yoshida \cite{LY17} and Bravetti et al. \cite{BCZ16}
to fail because now the procedure can incur in nondeterministic loops (so it is not guaranteed to repeat indefinitely
the accumulation behavior detected by the additional termination rule they consider).
%We did not find easy to extend the approach in these papers
%to deal with this form of nondeterminism. 
On the other hand, in the single-out case we loose the linear input accumulation but we do not have
output selections to cause the problematic nondeterminism discussed above. 

The latter advantage led us to opt for the single-out case, which we were able to manage by adopting a totally new approach
where the input accumulation is represented {\it in the form of a tree} (thus accounting for all possible alternative accumulated input behaviors at the same time).

We start with an example of subtyping between single-out types that 
cannot be managed with the approach in 
Lange and Yoshida \cite{LY17} and Bravetti et al.~\cite{BCZ16}
because there is non-linear input accumulation.

\begin{example}\label{exampleAntAlbero}
  Consider $T=\Trec t.\oplus\big\{l_1:\&\{l_2:\Tvar t,l_3:\Tvar t\}\big\}$ and
  $S=\Trec t.\oplus\Big\{l_1:\&\big\{l_2:\&\{l_2:\Tvar
  t\},l_3:\Tvar t\big\}\Big\}$.  
  We now comment the application of the subtyping
  procedure on $\emptyset\vdash T\subtypea S$.
  $$
  \begin{array}{l}
   \emptyset\vdash T\subtypea S \rightarrow \\
   \big\{(T,S)\big\} \vdash \oplus\big\{l_1:\&\{l_2:T,l_3:T\}\big\}\subtypea S \rightarrow \\
   \big\{(T,S),(\oplus\big\{l_1:\&\{l_2:T,l_3:T\}\big\},S)\big\}\vdash \\
   \qquad
   \oplus\big\{l_1:\&\{l_2:T, l_3:T\}\big\}\subtypea \oplus\Big\{l_1:\&\big\{l_2:\&\{l_2:S\},l_3:S\big\}\Big\} \rightarrow \\
   \big\{(T,S),(\oplus\big\{l_1:\&\{l_2:T, l_3:T\}\big\},S)\big\} 
   \vdash \\
   \qquad
\&\{l_2:T,l_3:T\}\subtypea \&\big\{l_2:\&\{l_2:S\},l_3:S\big\}
  \end{array}
  $$   
  At this point, the subtyping procedure has two continuations, 
  one for the label $l_2$ and one for the label $l_3$.
  In case of label $l_3$ we reach the judgement:
  $$
  \big\{(T,S),(\oplus\big\{l_1:\&\{l_2:T, l_3:T\}\big\},S)\big\} 
   \vdash T\subtypea S
  $$ 
  on which the termination rule $\textsf{Asmp}$ can be applied.
  In case of label $l_2$ we have:
  $$
\!\!\!  \begin{array}{l}   
   \big\{(T,S),(\oplus\big\{l_1\!:\&\{l_2:T, l_3:T\}\big\},S)\big\} 
   \vdash T\subtypea \&\{l_2\!:S\} \rightarrow \\
   \big\{(T,S),(\oplus\big\{l_1\!:\&\{l_2\!:T, l_3\!:T\}\big\},S),(T, \&\{l_2\!:S\})\big\} 
\\
 \qquad      \vdash \oplus\big\{l_1\!:\&\{l_2\!:T,l_3\!:T\}\big\}\subtypea \&\{l_2\!:S\} \rightarrow \\
   \big\{(T,S),(\oplus\big\{l_1\!:\&\{l_2\!:T, l_3\!:T\}\big\},S),(T, \&\{l_2\!:S\}), \\
\,\,\;(\oplus\big\{l_1\!:\&\{l_2\!:T,l_3\!:T\}\big\}, \&\{l_2\!:S\})\big\} 
\\
 \qquad   \vdash    \oplus\big\{l_1\!:\&\{l_2\!:T,l_3\!:T\}\!\big\}\subtypea \&\Big\{l_2\!:\oplus\big\{l_1\!:\&\big\{l_2\!:\&\{l_2\!:S\},l_3\!:S\big\}\big\}\Big\} \rightarrow \\
      \big\{(T,S),(\oplus\big\{l_1\!:\&\{l_2\!:T, l_3\!:T\}\big\},S),(T, \&\{l_2\!:S\}), \\
\,\,\;(\oplus\big\{l_1\!:\&\{l_2\!:T,l_3\!:T\}\big\}, \&\{l_2\!:S\})\big\} 
  \\
 \qquad   \vdash \&\{l_2\!:T,l_3\!:T\}\subtypea \&\Big\{l_2\!:\&\big\{l_2\!:\&\{l_2\!:S\},l_3\!:S\big\}\Big\} \rightarrow\\
      \big\{(T,S),(\oplus\big\{l_1\!:\&\{l_2\!:T, l_3\!:T\}\big\},S),(T, \&\{l_2\!:S\}), \\
\,\,\;(\oplus\big\{l_1\!:\&\{l_2\!:T,l_3\!:T\}\big\}, \&\{l_2\!:S\})\big\} 
\\
 \qquad   \vdash    T \subtypea \&\big\{l_2\!:\&\{l_2\!:S\},l_3\!:S\big\} \rightarrow\\
   \dots
  \end{array}
$$
Notice that in the last judgement, the r.h.s. has a non-linear input accumulation
starting with an input choice on two labels $l_2$ and $l_3$.
%We will show that our novel approach to asynchronous subtyping
%will be able to detect that it is not necessary to continue
%from this last reached judgement because even if the subtyping 
%algorithm is not deterministic, previous explorations of the impact of
%such nondeterminism were sufficient.
\end{example}

\subsubsection{Asynchronous Subtyping for Single-Out Types}

We now present our novel approach to asynchronous subtyping
that can be applied to single-out types, hence also to the
types in the above Example \ref{exampleAntAlbero},
that will be used as a running example in this section.
As anticipated, the main novelty is the ability to deal with non-linear
input accumulation by representing it as a tree. We need to 
be able to extract the leafs from these trees: this is done
by the \emph{leaf set} function defined as follows.

%We now move to another decidability result where, instead of
%restricting the definition of asynchronous subtyping, we restrict the
%set of types over which the relation is used. In our development, we
%focus on single-out types, session types where outputs are always
%singletons. In order to present our algorithm that determines the
%decidability of asynchronous subtyping for single-out types, we define
%the notions of leaf set, output anticipation and reachable types.
\begin{definition}[Leaf Set]\label{def:leafset}
  Given a session type $S$, we write $\noIn S$ if $S$ is not of the form $\Tbranch{l}{S}$.
    Given a session type $T$, we define 
  $$\leafset T \!=\!\{T_1,\dots,T_n\ |\ \mbox{$\noIn {T_i}\!$ %for every $i \in \{1\dots n\}$ and  
                                    and $\exists$ input context $\mathcal A$ s.t. 
                                    $T\!=\!\context A{T_k}{k\in\{1\dots n\}}$} \}$$
\end{definition}
The leaf set of a session type $T$ is the set of subterms different from inputs that are reachable
from its root through a path of inputs. For example, the leaf set of
the term
$\&\{ l_1: \Trec t.\oplus\{l_2:\Tvar t\}, l_3:
\&\{l_4:\oplus\{l_2:\Trec t.\oplus\{l_2:\Tvar t\}\}\}\}$
is
$\{\Trec t.\oplus\{l_2:\Tvar t\}, \oplus\{l_2:\Trec
t.\oplus\{l_2:\Tvar t\}\}$.
If we consider the r.h.s. term in the last judgement in Example \ref{exampleAntAlbero},
we have that $\leafset{\&\big\{l_2:\&\{l_2:S\},l_3:S\big\}} = \{S\}$.

During the check of subtyping,
according to Figure \ref{fig:algo}
(rule \textsf{Out}),
when a term in the r.h.s. having input accumulation has to mimic an output
in front of the l.h.s., such output must be present in front of all the 
leafs of the tree. In this case, the checking continues by  
anticipating the output from all the leafs. The following
auxiliary function \emph{output anticipation} indicates
the way a term changes after having anticipated a sequence of outputs.
Notice that in the definition we make use of the assumption
on single-out session types, by considering single-choice output selections.

\begin{definition}[Output Anticipation]\label{def:outAnt}
  Partial function $\antOut{T}{l_{i_1}\!\cdots l_{i_n}}$, with $T$ single-out session type
  and  $l_{i_1}\cdots l_{i_n}$ sequence of labels, is inductively defined as follows:
  $$
\!\!\!  \begin{array}{l}
  \antOut{T}{l_{i_1}\!\cdots l_{i_n}} \!=\! 
  \left \{\!\!\!
  \begin{array}{ll}
  T & \mbox{if $n=0$} \\
  \context A{T_k}{k}  \!\!\! & 
%     \begin{array}{l}
         \mbox{if $\outunfold {\antOut{T}{l_{i_1}\!\cdots l_{i_{n\!\!\;-\!\!\;1}}}} \!=\!
               \context A{\Tselectsingle{l_{i_n}\!}{T_k}}{k}$} 
 %    \end{array}                
  \end {array}
  \right .
  \end {array}
  $$
  We say that $T$ can infinitely anticipate outputs, written $\antOutInf{T}$, if there
  exists an infinite sequence of labels $l_{i_1}\cdots l_{i_j}\cdots$ such that
  $\antOut{T}{l_{i_1}\cdots l_{i_{n}}}$ is defined for every $n$.  
\end{definition}
The function $\antOut{T}{\tilde l}$ anticipates all outputs in the
sequence $\tilde l$.  For example, the function applied to
$\&\{ l: \Trec t.\oplus\{l_1:\oplus\{l_2:\Tvar t\}\},\ l':
\oplus\{l_1:\Trec t.\oplus\{l_2:\oplus\{l_1:\Tvar t\}\}\}\}$
and the sequence $(l_1,l_2)$ would return the same term, while it
would be undefined with the sequence $(l_1,l_1)$.
If we go back to our running Example \ref{exampleAntAlbero},
we have that $\antOut{S}{l_{1}}=\&\big\{l_2:\&\{l_2:S\},l_3:S\big\}$. Moreover,
we have that $\antOutInf{S}$ holds because the label $l_1$ can be infinitely
anticipated.

The definition of $\antOutInf{T}$ is not algorithmic in that it
quantifies on every possible natural number $n$. Nevertheless, as we show below, it can
be decided by checking whether for every session type obtained from
$T$ by means of output anticipations, all the terms populating its
leaf set can anticipate the same output label. Although such process
may generate infinitely many session types, the terms populating the
leaf sets are finite and are over-approximated by the function
$\reach T$, which always returns a finite set and is defined as:
\begin{definition}[Reachable Types]\label{def:reach}
  $\!\!\!\!\!\;$ Given a single-out session type $T\!\!\;$, $\!\!\;\reach{T}$ is the minimal set of session types such that:
  \begin{enumerate}
  \item[{\bf 1.}] $T \in \reach{T}$;\qquad\qquad
    \item[{\bf 2.}] $\Tbranch{l}{T} \in \reach{T}$ implies $T_i \in \reach{T}$ for every $i \in I$;
  \item[{\bf 3.}] $\Trec t.T' \in \reach{T}$ implies $T'\{\Trec t.T'/\Tvar t\} \in \reach{T}$;
  \item[{\bf 4.}] $\Tselectsingle{l}{T'} \in \reach{T}$ implies $T' \in \reach{T}$.
  \end{enumerate}
\end{definition}

Notice that $\reach T$ is populated by those session types
%that are generated starting
%from $T$ and by applying the term transformations in the rules 
%of Figure \ref{fig:algo}, read from bottom to top, by excluding the unfolding of 
%recursion not at the top level (possible in some cases to $\textsf{RecR2}$) and 
%the output anticipation (done by rule $\textsf{Out}$ when $\mathcal A \neq [\,]^1$). 
%The terms generated by these transformations are finite: they can be 
obtained by consuming in sequence the initial inputs and outputs, and
by unfolding recursion only when it is at the top level. 
As an example, consider the session type $S$ of the Example \ref{exampleAntAlbero}.
We have\\[.2cm]
\hspace*{-.15cm}$
\begin{array}{l}
\reach{S} \!=\!
\Big\{
S,
\oplus\big\{l_1\!:\&\big\{l_2\!:\&\{l_2\!:S\},l_3\!:S\big\}\big\},
\&\big\{l_2\!:\&\{l_2\!:S\},l_3\!:S\big\},
\&\{l_2\!:S\}
\Big\}\\[.3cm]
\end{array}
$

For every type $T$,
we have that the terms in $\reach{T}$
are finite; in fact, during the generation
of such terms, eventually the term $\Tend$ or a term already
considered is reached. The latter occurs after consumption of all the
inputs and outputs in front of a recursion variable already unfolded.

%The function $\reach T$ returns the set of types that are reachable
%from the session type $T$, which can obviously be infinite. However,
%the following Lemma says that we can decide whether $\antOutInf{T}$
%and $\reach{T}$ are finite:
%
\begin{proposition}\label{prop:outAntInfDecidable}
Given a single-out session type $T$, $\reach{T}$ is finite and it is 
decidable whether $\antOutInf{T}$.
\end{proposition}

% \begin{definition}
\myparagraph{Subtyping algorithm for single-out
  types.}\label{def:singleOutAlgo} 
We are now ready to present the new termination condition
that once added to the subtyping procedure in Figure \ref{fig:algo}
makes it a valid algorithm for checking subtyping for single-out
types. The termination condition is defined as an additional
rule, named \textsf{Asmp2}, that
complements the already defined \textsf{Asmp} rule
by detecting those cases in which the subtyping procedure
in Figure \ref{fig:algo} does not terminate.

The new rule is defined parametrically on the
session type $Z$, which is the type on the right-hand side of the
initial pair of types to be checked (i.e. the algorithm is intended to
check $V \subtype Z$, for some type $Z$).  We start from the initial
judgement $\emptyset \vdash V \subtypet Z$ and then apply from bottom
to top the rules in Figure \ref{fig:algo}, where
% $\vdash$ is replaced by $\vdash$ and
$\subtypea$ is replaced by $\subtypet$, plus the following additional
rule:
$$
\infer[\textsf{Asmp2}]
      {
      \Sigma,(T, \antOut{S}{\gamma}) \vdash T\subtypet \antOut{S}{\beta}
      }
      {
\begin{array}{c}
S \in \reach{Z} \qquad %&\!\!\!
      \antOutInf{S} \qquad %&\!\!\!
      |\gamma| < |\beta| %&\!\!\! 
\\[-.15cm]
      \leafset{\antOut{S}{\gamma}}=\leafset{\antOut{S}{\beta}}\\[-.1cm]
\end{array}      
      }
$$
We first observe that this termination rule can be applied
to the last judgement of our running Example \ref{exampleAntAlbero}.
We have already seen that $S \in \reach S$, $\antOutInf{S}$ holds,
$\antOut{S}{l_{1}}=\&\big\{l_2:\&\{l_2:S\},l_3:S\big\}$ and that
$\leafset{\&\big\{l_2:\&\{l_2:S\},l_3:S\big\}}=\{S\}$.
We now observe that $\antOut{S}{\varepsilon}=S$
and $\leafset{S}=\{S\}$,
hence we can conclude that we can apply the 
above termination rule \textsf{Asmp2}
to the last judgement in Example \ref{exampleAntAlbero}
by instantiating $\gamma =  \varepsilon$ and $\beta =  l_1$.

%[INTUITION BEHIND THE NEW RULE] 
The first property of the new algorithm that we prove
is termination.
Intuitively, we have that 
this new termination rule guarantees
to catch all those cases where the term on the
right grows indefinitely, by anticipating outputs and accumulating
inputs.  These infinitely many distinct types are anyway obtainable
starting from the finite set $\reach{Z}$, by means of output
anticipations. Hence there exists $S \in \reach{Z}$ that can generate
infinitely many of these types: this guarantees $\antOutInf{S}$ to be
true.  As observed above, the leaves of such infinitely many terms are
themselves taken from the finite set $\reach Z$.  This guarantees that
the algorithm, among the types that can be obtained from $S$, visits
two terms having the same leaf set. These, even if syntactically
different, are equivalent as far as the subtyping game is regarded.

Concerning the precise definition of the algorithm,
in order to avoid the possibility of applying two distinct rules to
the same judgement, we give rule $\textsf{Asmp2}$ the same priority as
rule $\textsf{Asmp}$ (both rules have highest priority).  Also in this
case, we use
$\Sigma \vdash T \subtypet S \rderiv \Sigma' \vdash T' \subtypet S'$
to denote that the latter can be obtained from the former by one rule
application, and $\Sigma \vdash T \subtypet S \notrderiv$, to denote
that there is no rule that can be applied to the judgement
$\Sigma \vdash T \subtypet S$.
%\end{definition}

We can now state the termination and soundness of the algorithm:
\begin{theorem}\label{theo:termination}
Given two single-out session types $T$ and $S$, the algorithm applied to the 
initial judgement $\emptyset \vdash T \subtypet S$ terminates.
\end{theorem}

\begin{theorem}\label{theo:soundness}
Given two single-out session types $T$ and $S$, we have that 
%for every $\Sigma', T', S'$,
there exist $\Sigma', T', S'$
such that
$\emptyset \vdash T \subtypea S
  \rderiv^* \Sigma' \vdash T' \subtypea S' \notrderiv$
if and only if
there exist $\Sigma'', T'', S''$
such that
$\emptyset \vdash T \subtypet S
  \rderiv^* \Sigma'' \vdash T'' \subtypet S'' \notrderiv$.
\end{theorem}

Finally, we can conclude the decidability of asynchronous subtyping
for single-out session types. 
\begin{corollary}[Decidability for Single-out Types]\label{singleoutdec}
  Asynchronous subtyping for single-out session types $\subtypesout$ is
  decidable.
\end{corollary}

We now show that the above decidability results hold also for the $\subtypetinsout$ relation (where
we further restrict the asynchronous subtyping relation not to admit contravariance on input branchings).
In the algorithm we just modify the rule $\mathsf{In}$ of Figure~\ref{fig:algo}
%algorithm considered above 
by changing the constraint $J \subseteq I$ in the premise %of rule $\mathsf{In}$ of Figure \ref{fig:algo}
into $J = I$,
thus obtaining modified versions of 
%the procedure
$\Sigma \vdash T \subtypea S
  \rderiv \Sigma' \vdash T' \subtypea S'$
(and $\Sigma \vdash T \subtypea S
\notrderiv$)
and 
%algorithm 
$\Sigma \vdash T \subtypet S
  \rderiv \Sigma' \vdash T' \subtypet S'$
(and $\Sigma \vdash T \subtypet S
\notrderiv$). We have that Proposition~\ref{prop:semidecidable}, where relation $\subtypetin$ is considered instead of $\subtype$, termination Theorem \ref{theo:termination} and soundness Theorem \ref{theo:soundness}, where the modified judgments $\subtypea$ and $\subtypet$ are considered,
still hold (they are proved with exactly the same proofs as those reported in \ref{AppendixDecidible}
for the original statements).
  
%CTRL

\begin{corollary}
  Asynchronous subtyping for single-out session types without input contravariance $\subtypetinsout$ is
  decidable.
\end{corollary}

\subsubsection{Asynchronous Subtyping for Single-in Types}

First of all we notice that an obvious consequence of Corollary \ref{singleoutdec} is that also $\subtypesinsout$ is decidable
(we just have to add a preliminary check verifying that both types are single-in).
Moreover, exploiting dual closeness, i.e.\ the fact that $T \,\subtypesin\, S$ if and only if $\dual{S} \,\subtypesout\, \dual{T}$ (see Section \ref{sec:subtyperest}), we can use the algorithm presented for single-out
types also for the case of single-in types.

\begin{corollary}[Decidability for Single-in Types]
$\!\!$  Asynchronous subtyping for single-in session types $\subtypesin$ is
  decidable.
\end{corollary}

We can therefore identify an asynchronous dual closed subtyping relation that stands in the boundary of decidability.

\begin{corollary}[Decidability for Single-in or Single-out Types]
$\;\;\;$  The asynchronous dual closed subtyping relation $\subtypesin \! \cup \! \subtypesout$ is
  decidable.
\end{corollary}

Finally, similarly as we did for $T \,\subtypesin\, S$, by exploiting Proposition \ref{prop:complicated} we can use the modified algorithm employed for $\subtypetinsout$ 
subtyping for deciding the remaining relation $\subtypesintout$.

\begin{corollary}
  Asynchronous subtyping for single-in session types without output covariance $\subtypesintout$ is
  decidable.
\end{corollary}

% %%% Local Variables: 
% %%% mode: latex
% %%% TeX-master: "main"
% %%% End: 

\section{Undecidability Results}\label{sec:undecidability}
We now move to undecidability results.
We first consider bounded asynchronous subtyping $\subtypebound$.
%defined as the union of all $k$-bounded subtypings,
%for every possible $k$. This relation is in our opinion
%of interest because it reflects real cases in which
%it is possible to assume bounded buffers, without 
%a priory knowledge of the actual bound.
The proof in this case is a variant of the
proof we already presented in our previous work \cite{BCZ16},
where we encoded the problem of checking
(non) termination in queue machines (a well-known Turing 
powerful formalism) into checking session subtyping.
Technically speaking, 
we resort to a different property, namely {\it bounded non termination}, that we here show to be 
undecidable for queue machines.

The second, and main, undecidability result concerns subtyping without output covariance and input contravariance
$\subtypetintout$. The proof in this case requires deep modifications
to our proof technique, due to the impossibility to
exploit covariance/contravariance in the queue machine encoding.
We deal with the absence of covariance/contravariance 
by saturating each point of choice on the entire
considered alphabet. This has a strong impact on the encoding
because it introduces additional choices, in the session types, 
whose continuations do not correspond to the behaviour of the considered queue machine.
This problem is solved by ensuring that these additional choices
and the corresponding continuations 
are irrelevant as far as subtyping checking is concerned.
Such solution, however, works only for a fragment of queue machines
(that we call single-consuming queue machines) that we prove to 
be Turing complete as well.

We consider this second result interesting for the following reason:
the previous 
undecidability proofs~\cite{BCZ16,LY17} made use of both 
output covariance/input contravariance (already present
in synchronous session subtyping) and output anticipation
(specific for asynchronous subtyping),
hence our new proof shows that (provided that the syntax of types is not constrained, 
e.g, imposing single choices for output selections or input branchings) 
the source of undecidability 
is to be precisely localized into the latter,
as the former is not necessary to prove undecidability.

We first report the definition of queue machines.

%% We now focus on the undecidability of asynchronous subtyping. 
%Existing results~\cite{BCZ16,LY17} show that asynchronous subtyping is
%undecidable by encoding termination of a Turing equivalent formalism
%into the subtyping game. However, covariance of outputs and
%contravariance of inputs in the subtyping relation are determinant for
%the construction of their encoding. Hereby, we show that subtyping
%remains undecidable, even if we consider a
%%
%%coarser
%%
%variant % of subtyping
%% that does not allow
%without covariance and contravariance.
%% (thus a coarser relation),it is still undecidable.
%Moreover, we show that 
%%the union, for all $k$, of all $k$-bounded
%%subtyping relations 
%bounded asynchronous subtyping
%(relating types that do not unboundedly put
%messages in a queue) is undecidable.

%This is the section where we provide further undecidability results on
%subtyping.
%
%
%
\subsection{Queue Machines}\label{subsec:queuemachines}
%We prove undecidability of asynchronous subtyping without using output
%covariance and input contravariance, by reduction from the acceptance
%problem for queue machines. 
Queue machines are a formalism similar to
pushdown automata, but with a queue instead of a stack. Queue machines
are Turing-equivalent~\cite{KozenBook}.
\begin{definition}[Queue Machine]
  A queue machine $M$ is defined by a six-tuple
  $(Q , \Sigma , \Gamma , \$ , s , \delta )$ where:
 \begin{itemize}
  \item $Q$ is a finite set of states;
  \item $\Sigma \subset \Gamma$ is a finite set denoting the input
    alphabet;
  \item $\Gamma$ is a finite set denoting the queue alphabet;
  \item $\$ \in \Gamma -\Sigma$ is the initial queue symbol;
  \item $s \in Q$ is the start state;
  \item $\delta : Q \times \Gamma \rightarrow Q\times \Gamma ^{*}$ is
    the transition function.
  \end{itemize}
\end{definition}

A {\em configuration} of a queue machine is an ordered pair
$(q,\gamma)$ where $q\in Q$ is its current state and
$\gamma\in\Gamma ^{*}$ is the content of the queue ($\Gamma ^{*}$ is
the Kleene closure of $\Gamma$).  The starting configuration on an
input string $x\in \Sigma^*$ is $(s , x \$)$.  The transition relation
$\rightarrow _{M}$ from one configuration to the next one is defined
as $(p,A\alpha )\rightarrow _{M}(q,\alpha \gamma)$, when
$\delta (p,A)=(q,\gamma)$.  A machine $M$ accepts an input $x$ if it
blocks by emptying the queue. Formally, $x$ is accepted by $M$ if
$(s,x\$)\rightarrow _{M}^{*}(q,\epsilon)$ where $\epsilon$ is the
empty string and $\rightarrow _{M}^{*}$ is the reflexive and
transitive closure of $\rightarrow _{M}$.  Intuitively, a queue
machines is a Turing machine with a special tape that works as a FIFO
queue.

%[CUT AND PASTE DA IandC]
 The Turing completeness of queue machines is
discussed by Kozen~\cite{KozenBook} (page 354, solution to exercise
99). A configuration of a Turing machine (tape, current head position
and internal state) can be encoded in a queue, and a queue machine can
simulate each move of the Turing machine by repeatedly consuming and
reproducing the queue contents, only changing the part affected by the
move itself.  
The undecidability of termination for queue machines 
follows directly from such encoding.
%
%
%
%\subsection{Undecidability Result}

%\begin{proof}
%  Let $M=(\{q_1,..,q_n\} , \Sigma , \Gamma , \$ , s , \delta )$ and
%  let $\#$ be a special character not in $\Sigma$. We can then build a
%  new machine
%  $M'=(\{q_1',..,q_n'\} , \Sigma\cup\{\#\} , \Gamma , \$ , s' ,
%  \delta' )$ such that $\delta'$ is defined as
%  \begin{itemize}
%  \item  $\delta’(q_i',a)=q_j’,\epsilon$ if
%    $\delta(q_i,a)=q_j,\epsilon$
%
%  \item $\delta’(q_i',a)=q_j',\gamma$ if
%    $\delta(q_i,a)=q_j,\gamma \quad (\gamma\neq \epsilon)$
%
%  \item $\delta’(q_i’,a)=q_j',\#$ if $\delta(q_i,a)=q_j,\epsilon$
%
%  \item $\delta’(q_i',\#)=q_i’,\epsilon$
%    
%  \item $\delta’(q_i’,\#)=q_i',\#$
%
%  \end{itemize}
%
%  Clearly, the new machine $M'$ is such that every read-only state can
%  only move to a write-only state.
%\end{proof}

%\begin{proposition}
%  If $(q_0,x\$) \rightarrow_M^* (q,\gamma)$ then
%  $\exists (q’,\gamma’).(q_0,x\$) \rightarrow_{M’}^* (q’,\gamma’)$ and
%  $f(q’,\gamma’)=(q,\gamma)$. 
%% [con f che toglie # e primati]  
%\end{proposition}
%
%\begin{proposition}
%  If $(q_0,x\$) \rightarrow_{M’}^* (q,\gamma)$ then $(q_0,x\$) \rightarrow_M^* f(q,\gamma)$
%\end{proposition}
%
%\begin{theorem}
%  Let $M$ be a queue machine and $M'$ the corresponding machines
%  obtained from the construction from
%  Theorem~\ref{thm:construction}. Then, $M$ terminates iff $M’$
%  terminates.
%\end{theorem}

%\begin{proof}
%Our proof proceeds...
%
%\bigskip

\subsection{Bounded Asynchronous Subtyping}\label{subsec:undec_bounded}
We now consider the notion of bounded asynchronous subtyping $\subtypebound$ we introduced in Section \ref{sec:subtyperest},

The proof of undecidability of $\subtypebound$
follows the approach we already used to prove the undecidability
of single-choice asynchronous subtyping $\selsubtype$ \cite{BCZ16}
(that we have commented in the Introduction).
The idea is to define, given a queue machine $M$ and its input $x$,
two session types $S$ and $T$, such that $S$ is a subtype of $T$
if and only if $M$ does not accept $x$.
More precisely, the type $S$ models the finite control of the 
queue machine $M$ while the type $T$ models the queue that initially
contains the sequence $x\$$.

\begin{figure}[t]
  \begin{displaymath}
\begin{array}{l}
    \begin{array}{l}
          \textit{a) Finite Control}\\[1mm]
      \semT{q}{\mathcal S} = 
      \left \{
      \begin{array}{l}
          \Trec{q}. 
          \Tbranchset{A}{
          \Tselectsingle{B^A_1}{\cdots
%          \Tselectsingle{B^A_2}{\cdots
          \Tselectsingle{B^A_{n_A}}{\semT{q'}{\mathcal S \cup q}
          }
%          }
          }
          }                  
          {\Gamma} 
        \\[1mm]
        \hspace{0.9cm}\text{if }q\not\in {\mathcal S} \text{ and } \delta(q,A)=(q',B^A_1\cdots B^A_{n_A})
        \\
        \\
        \Tvar{q}\qquad \mbox{if $q \in {\mathcal S}$}
      \end{array}
      \right.
    \end{array}
\\
    \begin{array}{l}
          \textit{b) Queue}\\[1mm]
    \semS{C_1\!\cdots C_m} =
    \Tbranchsingle{C_1}{
    \!\ldots
%    \Tbranchsingle{C_2}{\cdots
    \Tbranchsingle{C_m}
    {\Trec{t}.\Tselectset{A}{\Tbranchsingle{A}{\Tvar{t}}}{\Gamma}}
    }
%    }
    \end{array}
\end{array}
  \end{displaymath}
  \caption{Encoding of the Finite Control and the Queue of a Queue
    Machine}
  \label{fig:encoding}
\end{figure}

More precisely, the encoding of queue
machines is as follows \cite{BCZ16}.

\begin{definition}[Queue Machine Encoding]
  Let $M= (Q , \Sigma , \Gamma , \$ , s , \delta )$ be a queue machine,
  and let $C_1, \cdots, C_m \in \Gamma$, with $m \geq 0$,
  $q\in Q$ and $\mathcal S\subseteq Q$.
The {\em finite
    control encoding function} $\semT{q}{\mathcal S}$ and the
  \emph{queue encoding function} $\semS{C_1 \cdots C_m}$ are defined
  as in Figure~\ref{fig:encoding}(a) and Figure~\ref{fig:encoding}(b)
  respectively. The initial encoding of $M$ with input $x$
  is given by the pair of types $\semT{s}{\emptyset}$
  and $\semS{x\$}$.
\end{definition}

The basic idea behind the encoding of the finite control
is to use a recursively defined
type with a recursion variable $\Tvar q$ for each state $q$ of the
encoded queue machine $M$.
The type corresponding to the recursion variable $\Tvar q$
starts with an input with multiple choices,
one for each possible symbol that can be consumed from the queue.
The continuation is composed of a sequence
of single-choice inputs labeled with the symbols $B_1^A \dots B_{n_A}^A$, where
$B_1^A \dots B_{n_A}^A$ are the symbols enqueued by the queue machine
when, in state $q$, consumes $A$ from the queue. 
Assuming that $q'$ is the new state of $M$ after execution
of this step (i.e. $\delta(q,A)=(q',B_1^A \dots B_{n_A}^A)$), 
the type becomes the one corresponding to the recursion variable $\Tvar q'$.

On the other hand, the type modeling the queue with contents
$C_1\dots C_m$
is denoted with $\semS{C_1\dots C_m}$:
this type starts with a sequence of single-choice inputs labeled
with the symbols $C_1\dots C_m$, followed by a recursive type.
Such type
starts with an output with multiple-choices, one for each symbols
that can be enqueued, followed by a single-choice input having the same
label. This particular type has the following property: if one
label $A$ of the multiple-choice output is selected for anticipation
during the subtyping simulation game, the corresponding single-choice input 
labeled with $A$ is enqueued at the end of the sequence
of inputs preceding the recursive definition.
This perfectly corresponds to the behaviour of the queue in the modeled
queue machine.

As mentioned above, this encoding has been already used to prove the undecidability
of $\selsubtype$ \cite{BCZ16}. More precisely, we proved that given a queue 
machine $M= (Q , \Sigma , \Gamma , \$ , s , \delta )$ and an initial
input $x$, we have that $\semT{s}{\emptyset} \selsubtype \semS{x\$}$
if and only if $x$ is not accepted by $M$ (i.e. $M$ does not terminate
on input $x$).
The same result does not hold for the bounded asynchronous subtyping
because there are cases in which $M$ does not accept $x$ but
$\semT{s}{\emptyset} \not\!\!\subtypebound \semS{x\$}$,
in particular, those cases in which
the subtyping simulation game generates unbounded accumulation
of inputs. For this reason we have to consider a more complex 
undecidable property for queue machines: 
\emph{bounded non termination}, i.e., the ability of a queue machine
to have an infinite computation while keeping the length of the queue bounded.
We now define the notion of boundedness for queue machines
and then prove that bounded non termination is undecidable.

%That result is not sufficient to prove the undecidability of our
%new relation $\subtypebound$. To prove this new result, we need
%a preliminary result about queue machines: we define the boundedness
%property on queue machines and then show that this property
%is undecidable. 

\begin{definition}[Queue Machine Boundedness]\label{def:boundedMachine}
Let $M$ be a queue machine and $x$ a possible input. We say that $M$ is bound on input $x$
if there exists $k$ such that, for every $q$ and $\gamma$ such that
$(s,x\$)\rightarrow _{M}^{*}(q,\gamma)$, we have that $|\gamma| \leq k$. 
\end{definition}

\begin{lemma}\label{lem:boundedUnd}
  Given a queue machine $M$ and an input $x$, it is undecidable whether $M$ 
  does not terminate and is bound
  on $x$.
\end{lemma}

Following the proof technique we already used to prove
undecidability of $\selsubtype$, i.e. by reducing the
termination problem for queue machines into subtyping 
checking \cite{BCZ16}, 
we can prove also the undecidability of $\subtypebound$ 
by reduction from the bounded non 
termination problem.

%We can finally prove that $\subtypebound$ is undecidable
%as a consequence of the undecidability of termination and 
%boundedness on queue machines.

%We can then give a different encoding of queue machines into bounded
%subtyping:
%\begin{definition}[Queue Machine Encoding for Bounded Subtyping]\mbox{}\\
%  Let $M= (Q , \Sigma , \Gamma , \$ , s , \delta )$ be a queue machine,
%  and let $C_1, \cdots, C_m \in \Gamma$, with $m \geq 0$,
%  $q\in Q$ and $\mathcal S\subseteq Q$.
%  We define the $\BsemS{C_1\ldots C_m}$, resp. $\BsemT{q}{\mathcal S}$,
%  in the same way as $\semS{C_1\ldots C_m}$, resp. $\semT{q}{\mathcal S}$,
%  with the difference that the $\labelunion$ operator is interpreted in 
%  the following way: $\{l_i:T_i\}_{i\in I} \labelunion \{l_j:T_j\}_{j\in J} = \{l_i:T_i\}_{i\in I}$.
%\end{definition}
%Finally, we can state the following impossibility result:
\begin{theorem}\label{thm:undecidableBounded}
  Given a % single consuming
  queue machine $M= (Q , \Sigma , \Gamma , \$ , s , \delta )$ and
  an input string $x$, 
  we have that $\semT{s}{\emptyset} \subtypebound \semS{x\$}$ if and only if
  $M$ does not terminate and is bound on $x$.
\end{theorem}

\begin{corollary}
Bounded asynchronous subtyping $\subtypebound$
is undecidable.
\end{corollary}

%Bounded asynchronous subtyping can be related to the
%notion of boundedness for
%%also be seen as bound
%%to the size of the queue of a 
%queue machines:
%We can then give a different encoding of queue machines into bounded
%subtyping:
%\begin{definition}[Queue Machine Encoding for Bounded Subtyping]
%  Let $M= (Q , \Sigma , \Gamma , \$ , s , \delta )$ be a queue machine,
%  and let $C_1, \cdots, C_m \in \Gamma$, with $m \geq 0$,
%  $q\in Q$ and $\mathcal S\subseteq Q$.
%  We define the $\BsemS{C_1\ldots C_m}$, resp. $\BsemT{q}{\mathcal S}$,
%  in the same way as $\semS{C_1\ldots C_m}$, resp. $\semT{q}{\mathcal S}$,
%  with the difference that the $\labelunion$ operator is interpreted in 
%  the following way: $\{l_i:T_i\}_{i\in I} \labelunion \{l_j:T_j\}_{j\in J} = \{l_i:T_i\}_{i\in I}$.
%\end{definition}
%Finally, we can state the following impossibility result:
%\begin{theorem}\label{thm:undecidableBounded}
%  Given a % single consuming
%  queue machine $M= (Q , \Sigma , \Gamma , \$ , s , \delta )$ and
%  an input string $x$, 
%  we have that $\BsemT{s}{\emptyset} \subtypebound \BsemS{x\$}$ if and only if
%  $M$ does not terminate and is bound on $x$.
%\end{theorem}

\subsection{Undecidability of Asynchronous Subtyping without Output
  Covariance and Input Contravariance}\label{subsec:undec_first}

We now move to the proof of undecidability of 
$\subtypetintout$, the asynchronous subyping relation, we introduced in Section \ref{sec:subtyperest},
that does not admit output covariance and 
input contravariance by imposing matching 
choices to have the same set of labels.

The proof technique is still based on an encoding
of queue machines, but we have to significantly
improve the encoding discussed in the previous
subsection. In fact, the encoding of Figure \ref{fig:encoding}
exploits both input contravariance (in the matching
between the multiple-choice input at the beginning
of the encoding of the finite control and the initial
single-choice inputs of the queue encoding)
and output covariance (in the matching
between the multiple-choice output at the beginning
of the recursive part of the queue encoding
and the single-choice outputs in 
the encoding of the finite control).

The new encoding that we propose saturates all choices,
both inputs and outputs,
with labels corresponding to the entire queue alphabet. The addition of these labels and
of the corresponding continuations, introduces new possible
paths in the subtyping simulation game. We are
able to make these additional behaviour irrelevant, but
at the price of restricting the class of encoded queue
machines. These queue machines are named \emph{single
consuming queue machines}; their characteristic
is to guarantee that in two subsequence actions,
at least one of the two will enqueue symbols.

\begin{definition}[Single Consuming Queue machine]
  We say that a queue machine
  $M=(Q , \Sigma , \Gamma , \$ , s , \delta )$ is \emph{single
    consuming} if $\delta(q,a)=(q',\epsilon)$, for some $q$, $a$ and
  $q'$, implies that there exist no $b$ and $q''$ such that
  $\delta(q',b)=(q'',\epsilon)$.
\end{definition}

We have that 
single consuming queue machines are still Turing-complete
(see \ref{app:scqm} for the detailed proof based on an encoding of queue machines into single consuming queue machines):
\begin{theorem}\label{thm:construction}
  Given a single consuming queue machine $M$ and an input $x$, the
  termination of $M$ on $x$ is undecidable.
%  Given a queue machine $M$, there exists a single-machine 
%  
%   we can always build an equivalent machine
%  $M'$ such that after consuming an element from the queue, it
%  immediately reaches a state that can only insert symbols to the
%  queue.
\end{theorem}

We prove the undecidability of 
$\subtypetintout$
by encoding single consuming queue
machines into the subtyping simulation game. 
Following the approach already discussed in the previous subsection,
given a queue machine, our encoding generates a
pair of types, say $T$ and $S$, such that $T$ encodes the finite
control and $S$ encodes the queue. Then, the subtyping $T\,\subtypetintout\, S$
simulates the execution of the machine.
%We do not make use of output
%covariance and input contravariance in that all the choices in 
%both $T$ and $S$ are always defined on the same set of labels.
% We now give the formal
% definition of our encoding and then give a detailed explanation.

% [ REMOVE. We recall that a subtyping game is the process of applying
% the coinductive definition found in
% Definition~\ref{def:subtyping}. Such game is based on the idea that,
% whenever we wish to show that $T\subtype S$, we ask type $S$ to match
% all actions that $T$ may perform. However, if $T$ performs an output,
% type $S$ is allowed to match such output with one that is nested under
% some inputs. Moreover, outputs can use covariance and inputs can use
% contravariance. The goal of this section is to show that it is
% possible to encode a queue machine in a subtyping game without the
% game using covariance and contravariance. ]

%As in the previous subsection, our encoding associates each state $q$ of a queue
%machine with a unique recursion variable $\Tvar q$.  Symbols of the
%alphabet $\Gamma$ are used as labels by the session types generated by
%the encoding. We need a new notation,
We are now ready to present the definition of the new encoding where 
we make use of the following new notation:
$\{l_i:T_i\}_{i\in I} \labelunion \{l_j:T_j\}_{j\in J} =
\{l_k:T_k\}_{k\in I\cup J}$.

\begin{definition}[Encoding Single Consuming Queue machines]\label{def:encoding}
  Let $M= (Q , \Sigma , \Gamma , \$ , s , \delta )$ be a queue machine
  such that $q\in Q$, $\mathcal S\subseteq Q$ and
  $C_1, \cdots, C_m \in \Gamma$, with $m \geq 0$. The {\em finite
    control encoding function} $\BsemT{q}{\mathcal S}$ and the
  \emph{queue encoding function} $\BsemS{C_1 \cdots C_m}$ are defined
  as in Figure~\ref{fig:encoding2}(a) and Figure~\ref{fig:encoding2}(b)
  respectively.
\end{definition}

\begin{figure}[t]
  \begin{displaymath}
    \begin{array}{llllll}
      %%%%% FINITE CONTROL ENCODING
      \textit{a) Finite Control}\\[1mm]
      \BsemT{q}{\mathcal S} = 
      \left \{
      \begin{array}{ll}
        \Trec{q}.\Tbranchset{A}{\semTcont{B^A_1\cdots B^A_{n_A}}_{q'}^{\mathcal S \cup \{q\}}}{\Gamma} 
        &
\begin{array}{l}
\!\!\!\!\, \text{if }q\not\in {\mathcal S} \text{ and } \\[-.1cm]
 \; \,        \delta(q,A)=(q',B^A_1\cdots B^A_{n_A})\\[.2cm]
\end{array}
        \\
        \Tvar{q} & \mbox{if $q \in {\mathcal S}$}
      \end{array}
                   \right.
      \\[9mm]
      %%%%% QUEUE ENCODING
      \textit{b) Queue}\\[1mm]
      \BsemS{C_1\ldots C_m} = \left\{
      \begin{array}{ll}
        \Trec t \oplus\big\{A : \& \big( \big\{A:\Tvar
                                   t\big\} \labelunion
                                   \big\{A':T''\big\}_{A'\in
                                   \Gamma\setminus\{A\}}\big)   \big\}_{A \in \Gamma} \ 
        &\ \text{if }m=0\\[1mm]
        \&  \big( \big\{C_1:  \BsemS{C_2\ldots C_m}\big\} \labelunion
           \big\{{A':T''}\big\}_{A'\in \Gamma\setminus\{C_1\}} \big) \
        &\ \text{otherwise}
      \end{array}
          \right.\\\\
      \quad\qquad\rule{2cm}{0.4pt}
      \\
       \text{where:} \\
      \begin{array}{l}
           \semTcont{B_1\cdots B_{m}}_{r}^{\mathcal T} \!=\! \left\{\!\!
          \begin{array}{ll}
            \!\BsemT{r}{\mathcal T} 
            & \text{if }m=0\\
            \!\oplus  \big( \big\{B_1:  \semTcont{B_2\ldots B_m}_{r}^{\mathcal
            T}\big\} \labelunion
            \big\{{A':T'}\big\}_{A'\in\Gamma\setminus\{B_1\}} \big) \!
            & \text{otherwise}
          \end{array}
              \right.
        \\[5mm]
         T'=\Trec t. \&\big\{A_1: \oplus\{A_2:\Tvar t\}_{A_2 \in
                         \Gamma} \big\}_{A_1 \in \Gamma}
        \\[2mm]
         T''=\Trec t. \&\big\{A_1: \&\big\{A_2: \oplus\{A_3:\Tvar
                                       t\}_{A_3\in \Gamma} \big\}_{A_2
                                       \in\Gamma} \big\}_{A_1 \in
                                       \Gamma}
      \end{array}
    \end{array}
  \end{displaymath}
  \caption{Encoding of the Finite Control and the Queue of a Single Consuming Queue
    Machine}
  \label{fig:encoding2}
\end{figure}
%

% The encoding generates two types $T$ and $S$ such that $T$ encodes the
% transition function of the encoded queue machine (finite control),
% while $S$ encodes the queue. Then, playing the subtyping game between
% type $T$ and type $S$ will simulate executing $M$.

% The basic idea is that $T$ performing an input on label $A$ simulates
% the control willing to read symbol $A$ from the queue. That has to be
% matched by the $S$, the encoding of the queue. Then, $T$ performs an
% output with a label $B$ to simulate that the control is willing to put
% symbol $B$ into the queue. When such operation is matched by the
% queue, another input is released by $S$. 

As discussed in the previous subsection, the idea is that the 
type encoding the finite control is
%The encoding generates two types $T$ and $S$ such that $T$ encodes the
%transition function of the encoded queue machine (finite control),
%while $S$ encodes the queue.  The main idea is for type $T$ to be 
able
to perform an input on each of the symbols in $\Gamma$, and continue
according to the definition of the transition function $\delta$. 
The type representing the queue
then matches such input with the correct symbol depending on the
state of the queue. For instance, in the encoding described 
in the previous subsection,
if we denote with $T$ and $S$ the types representing the
finite control and the queue respectively, and
if $\Gamma=\{A,B\}$ and symbol $A$ is
on the head of the queue, we have $T=\&\{A: \ldots, B: \ldots\}$ and
$S=\&\{A:\ldots\}$: type $T$ is able to react to any symbol that may
be present on the queue (like the transition function $\delta$), while
type $S$ reacts with the actual value on the queue, symbol
$A$. Unfortunately, such idea exploits contravariance for
inputs. Therefore, it must be the case, in the new encoding, 
that the input in $S$ is of the
form $\&\{A: \ldots, B: \ldots\}$.  We make sure that if
label $A$ is selected then the simulation of the queue machine
continues. Otherwise, an infinite subtyping simulation game 
is started (starting from $B$ in the example).  

Also the insertion of symbols in the queue was simulated 
in the encoding of the previous subsection by 
exploiting output covariance. The type representing the 
finite control
performs a single-choice output that is matched by a 
multiple-choice output having the effect of adding a
corresponding symbol
at the end of the input accumulated in the type
modeling the queue. 
Also in this case, we have to add choices to the type
modeling the finite control: also in this case we ensure that
these extra paths start an infinite subtyping simulation game.

These additional paths make the subtyping simulation 
game highly non-deterministic and such
that several paths that the game can take differ from what the encoded
machine does. We discuss in detail the various cases which our
encoding in Figure~\ref{fig:encoding2} can be in:
\begin{enumerate}
\item {\em The encoding of the finite control reads the correct
    symbol.}  We represent the machine reading a symbol $A$ from the
  queue while being in state $q$, with an input type of the form
  $\Tbranchset{A}{\semTcont{B^A_1\cdots B^A_{n_A}}_{q'}^{\mathcal S
      \cup \{q\}}}{\Gamma} $,
  where each branch corresponds to a possible symbol that can be
  read. On the other hand, a queue $C_1\cdots C_m$ is encoded as an
  input type of the form
  $\& \big( \big\{C_1: \BsemS{C_2\ldots C_m}\big\} \labelunion
  \big\{{A':T''}\big\}_{A'\in \Gamma\setminus\{C_1\}} \big)$
  where the branch with label $C_1$ represents the actual content of
  the queue. Hence, in the simulation game, if the finite control
  reads symbol $A$ and this is matched by the correct symbol in the
  queue, then the type
  $\semTcont{B^A_1\cdots B^A_{n_A}}_{q'}^{\mathcal S \cup \{q\}}$
  deals with inserting symbols $B^A_1\cdots B^A_{n_A}$ into the queue.
  
\item {\em The encoding of the finite control reads the wrong symbol.}
  In this case, the encoding of the finite control picks a symbol that
  is not that in the queue head. In order to match it, the encoding of
  the queue will take $\big\{{A':T''}\big\}_{A'\in
    \Gamma\setminus\{C_1\}} $.  Type $T''$ is designed in a way that
  it can match every move of the finite control, by repeatedly
  alternating two inputs with a subsequent output on every queue
  symbol.  Note that, since inputs cannot be anticipated, matching
  every move is feasible only if the encoded machine is single
  consuming.

\item {\em The encoding of the finite control writes the correct
    symbol.} Once the finite control has read a symbol, it performs
  $\semTcont{B_1\cdots B_{m}}_{r}^{\mathcal T}$, % whose objective is to
  % simulate
  which simulates the writing of $B_1\cdots B_m$ into the queue. If $m=0$
  % such string is empty
  then it moves to the encoding of the next state
  according to function $\delta$. Otherwise, it translates to the type
  $ \oplus \big( \big\{B_1: \semTcont{B_2\ldots B_m}_{r}^{\mathcal
    T}\big\} \labelunion\big\{{A':T'}\big\}_{A'\in\Gamma\setminus\{B_1\}}
  \big)$.  The queue, in order to match $B_1$ (and $B_2$, \ldots,
  $B_m$) can always anticipate outputs with the term $\Trec t
  \oplus\big\{A : \& \big( \big\{A:\Tvar t\big\} \labelunion
  \big\{A':T''\big\}_{A'\in \Gamma\setminus\{A\}}\big) \big\}_{A \in
    \Gamma}$ which, after consuming a label $A$ will add an input with
  label $A$, simulating the adding of $A$ to the queue.

\item {\em The encoding of the finite control writes the wrong
    symbol.} In this case, the finite control writes a symbol to the
  queue with $ \oplus \big( \big\{B_1: \semTcont{B_2\ldots
    B_m}_{r}^{\mathcal
    T}\big\} \labelunion\big\{{A':T'}\big\}_{A'\in\Gamma\setminus\{B_1\}}
  \big)$.  However, the simulation executes the wrong output (with any
  $A'\neq B_1$) and continues as $T'$. In this case, $T'$ continues
  removing and adding any value from the queue, indefinitely. Note
  that it may remove the wrong value from the queue overlapping with
  case 2. In this case, the requirement that the queue machine is
  single consuming is not necessary.
\end{enumerate}

\begin{example} In order to further clarify our encoding, consider a
  queue machine with states $\{s,q\}$ (where $s$ is the starting
  state), queue alphabet $\Gamma=\{X,Y\}$ and transition relation
  $\delta$ such that $\delta(s,A)=(q,A)$ and
  $\delta(q,A)=(s,\epsilon)$, for every $A\in\Sigma$.
  Clearly, the machine terminates on any input. The encoding of the
  finite control is the following session type:
  \begin{displaymath}
%    \begin{array}{llll}
      \BsemT{s}{\emptyset} = 
      \Trec{s}.\&
                  \left\{
                  \begin{array}{lll}
                    X: \oplus\{X: \BsemT{q}{s}, \quad Y:T'\}
                    \\
                    Y: \oplus\{Y: \BsemT{q}{s}, \quad X:T'\}
                  \end{array}
      \right\}
  \end{displaymath}
  \begin{displaymath}
%      \quad&\quad
      \BsemT{q}{\{s\}} = 
      \Trec{q}.\&
                  \left\{
                  \begin{array}{lll}
                    X: \oplus\{X: \Tvar s, \quad Y:T'\}
                    \\
                    Y: \oplus\{Y: \Tvar s, \quad X:T'\}
                  \end{array}
      \right\}
%    \end{array}
  \end{displaymath}
  %Let us now assume
  Assume, e.g., that the queue initially contains
  the string $XY$. The machine will empty the queue by visiting state
  $q$ twice and terminate in state $s$ with the empty queue. 
  If we now run the subtyping
  simulation game between
  the encoding of finite control above and the encoding of the queue
  we will end up with two types that are not in subtyping:
  %, since
  the encoding of the state $s$ starting with an input %(on $X$ or $Y$)
  and
  % which
  the encoding of the empty queue that does not match it.
  %
  % $\semT{s}{\emptyset} = \Trec{s}.  \Tbranchset{A} {\Tselectsingle{A}
  %   {\semT{q}{\{s\}}} } {\Gamma} = \Trec{s}.  \Tbranchset{A}
  % {\Tselectsingle{A} {\Trec{q}.\Tbranchset{A} {\semT{s}{\{s,q\}}}
  %     {\Gamma}} } {\Gamma} = \Trec{s}.  \Tbranchset{A}
  % {\Tselectsingle{A} {\Trec{q}.\Tbranchset{A} {\Tvar s} {\Gamma}} }
  % {\Gamma} $.
\end{example}
%%%%%%%%%%%%%% 
%% END EXAMPLE
%%%%%%%%%%%%%% 

\bigskip
% In the light of the properties listed above, t
The encoding of the finite-control and of the queue are such that the
following properties hold: given a queue machine $M$ with initial
state $s$ and initial queue symbol $\$$, if $M$ does not accept
$x$
then it is possible to define an asynchronous subtyping relation that
includes the pair $(
\BsemT{s}{\emptyset} , \BsemS{x\$})$; moreover, if $\BsemT{s}{\emptyset}
\subtypetintout \BsemS{x\$}$ then it is possible to conclude that
$M$
does not terminate (i.e. does not accept) on input $x$.
We thus have the following:

\begin{theorem}\label{thm:undecidableNoBranching}
  Given a single consuming queue machine $M= (Q , \Sigma , \Gamma , \$ , s , \delta )$ and
  an input string $x\in \Sigma^*$, we have $\BsemT{s}{\emptyset} \subtypetintout \BsemS{x\$}$ 
  if and only if $M$ does not terminate on $x$.
\end{theorem}
%\begin{proof}[Proof (sketch)]
%  The proof of the theorem is based on two Lemmas that can be found in
%  the Appendix. Intuitively, ...
%\end{proof}

We can therefore conclude that subtyping without output covariance and input contravariance is undecidable.

%HO TOLTO QUESTO SOTTO!!!!!!!!!!!!:
%Below, subtyping without output covariance and input contravariance is
%defined as in Definition~\ref{def:subtyping} except that $I=J_k$ in
%point 2. and $I=J$ in point 3.
\begin{corollary}[Undecidability of  Subtying without Co/contravariance]%\label{}
  The asynchronous dual closed subtyping relation $\subtypetintout$ is
  undecidable.
\end{corollary}

In the same way we can also show that $\subtypetin$ and $\subtypetout$
are undecidable and
provide an alternative proof of undecidability of $\subtype$. This because,
since for the types obtained 
with the encoding (for which the ability to match via covariance/contravariance is irrelevant) obviously such relations coincide, i.e. $\BsemT{s}{\emptyset} \subtypetintout \BsemS{x\$}$ 
if and only if $\BsemT{s}{\emptyset} \subtypetin \BsemS{x\$}$ if and only if $\BsemT{s}{\emptyset} \subtypetout \BsemS{x\$}$ if and only  if $\BsemT{s}{\emptyset} \subtype \BsemS{x\$}$,
Theorem \ref{thm:undecidableNoBranching}  
holds also if we replace the $\subtypetintout$ relation with one of such relations.
%subtyping from Definition~\ref{def:subtyping}. % is undecidable.

\begin{corollary}%\label{}
  Asynchronous subtyping relations $\subtypetin$, $\subtypetout$ and $\subtype$ are undecidable.
\end{corollary}

\section{Related Work}\label{sec:relatedwork}
%%% DECIDABILITY
% The articles closest to ours are those by Bravetti et al.~\cite{BCZ16}
% and by Lange and Yoshida~\cite{LY17}. 
\noindent{\bf Subtyping for Session Types.} 
Subtyping for session types was first introduced by Gay and
Hole~\cite{GH05}\footnote{The Gay and Hole 
%In~\cite{GH05} 
%, differently from the present paper, 
subtyping is contravariant on outputs and covariant on inputs.
This is because a channel-based subtyping~\cite{G16} is
considered instead of our process-oriented subtyping.} 
for a session-based $\pi$-calculus where
communication is synchronous, i.e., an output directly synchronises
with an input. In such case, the relation allows no output anticipation. However, as in our case, outputs are
covariant and inputs are contravariant.

To the best of our knowledge, Mostrous et al.~\cite{ESOP09} were the
first to adapt the notion of session subtyping to an asynchronous
setting. Their computation model is a session $\pi$-calculus with
asynchronous communication that makes use of session queues for
maintaining the order in which messages are sent. They introduce the
idea of output anticipation, which is also a main feature of our theory.
% adapted the Gay and Hole subtyping algorithm to multiparty
% asynchronous session types.  Differently from what stated therein,
% the algorithm does not terminate due to unbounded message
% accumulation in the queues, e.g. for the pair of terms in Example
% \ref{exampleInfinito}.  Any way, such algorithm inspired our
% procedure in \S~\ref{subsec:subproc}. The problem of unbounded
% accumulation has been observed by
Mostrous and Yoshida~\cite{MY15}
%who 
extended the notion of asynchronous subtyping to session types for the
higher-order $\pi$-calculus.  In the same article, Mostrous and
Yoshida observe that their definition of asynchronous subtyping allows
orphan messages. Orphan message are prohibited with the definition of
subtyping given by Chen et al.~\cite{MariangiolaPreciness}. In their
article, they show that such a definition is both sound and complete
w.r.t.\ type safety and orphan message freedom.

%%% UNDECIDABILITY
\noindent{\bf Undecidability Results.} 
Mostrous et al.~\cite{ESOP09} proposed a procedure to check
asynchronous subtyping for multiparty session types.  Differently from
what stated therein, the procedure does not terminate due to unbounded
message accumulation in the queues, e.g. for the terms in Example
\ref{exampleInfinito}.  Such a procedure inspired the one we presented in
Section~\ref{subsec:subproc}.  The problem of unbounded accumulation was
observed by Mostrous and Yoshida~\cite{MY15}.  The impossibility to
define a correct algorithm has been independently proved by Lange and
Yoshida~\cite{LY17} and Bravetti et. al~\cite{BCZ16}.  Lange and
Yoshida~\cite{LY17} reduce Turing machine termination into a notion of
compatibility for communicating automata and, then, transfer such a
result to session types.  This proof technique applies only to dual
closed subtyping relations, like the one by Chen et
al.~\cite{MariangiolaPreciness}.  The proof by Bravetti
et.\ al~\cite{BCZ16}, on the other hand, exploits a direct encoding of
queue machines into session subtyping. This made it possible to prove
undecidability of all the other notions of asynchronous subtyping in
the literature.
% Our results use a notion of subtyping where orphan messages are not
% allowed, while the subtyping relation by Bravetti et al.  allows for
% orphan messages.
Unlike the encoding in this paper (Figure \ref{fig:encoding2}), both encodings take advantage of the % do need to
use of output covariance and input contravariance. 
% (when the finite control reads from the queue)
For example, by exploiting this feature, the
queue machine encoding by Bravetti et al.~\cite{BCZ16}  (Figure \ref{fig:encoding}) is much simpler than the encoding we need to use here.
% (when the finite control writes on the queue).
We notice that our results on undecidability focus on binary session
types. However, it is immediate to generalise this kind of
undecidability results from binary to multiparty sessions
(binary session types are just multiparty session types
with only two roles~\cite{BCZ16}).

\noindent{\bf Decidability Results.} Synchronous subtyping for binary
session types is decidable~\cite{GH05}.  Both Bravetti et
al.~\cite{BCZ16} and Lange and Yoshida~\cite{LY17} investigate
fragments of session types for which asynchronous subtyping becomes
decidable. However, such fragments are  much
more limited, and far from having practical applications, with respect
to those considered here. Both address cases where 
one of the compared
types is a {\it single-choice session type}, i.e.\ all its branchings and selections are single-choice. Thus they are both, basically, 
special cases of our subtyping for single-in or single-out types ($\subtypesin \cup \subtypesout$).  In particular, Lange and Yoshida
give an algorithm for deciding subtyping between a general session
type and a single-choice session type. Although it may seem that such
case is not properly included in our decidable subtyping relation for single-out/single-in types,
covariance and contravariance ensure that all types containing at
least one multiple input branch and one multiple output selection (both
reachable in the subtyping simulation game) cannot be related with a
single-choice type.
%
% Bravetti et al.  consider fragments of our single-out and single-in
% session types. In particular,
Bravetti et al.~\cite{BCZ16} prove decidability for  
relations $\selsubtypesin$ and $\selsubtypesout$ that pose an analogous restriction to the branching/selection structure, but that allow for orphan messages.  $\selsubtypesin$ and $\selsubtypesout$ are fragments
where related types $(T,S)$ are such that, either $T$ is single-choice and $S$ is
single-in ($\selsubtypesin$), or $T$ is single-out and $S$ is single-choice  ($\selsubtypesout$).  
For types that do not produce orphan messages, the
sutyping of Bravetti et al.~\cite{BCZ16} is just a special case of our single-in ($\subtypesin$)
and single-out ($\subtypesout$) session subtyping.
% considered in the present paper.

Additionally, Lange and Yoshida state the decidability of subtyping
for half-duplex communication~\cite{CF05} and alternating machines:
the former coincides with synchronous subtyping %~\cite{GH05}
while the latter can be reduced to $1$-bounded asynchronous subtyping
as discussed in Section~\ref{subsec:kbounded}.

%
%Lange and Yoshida consider, instead, a subtyping relation without
%orphan messages like ours, based on the original definition given by
%Chen et al.~\cite{MariangiolaPreciness}.  They give an algorithm for
%the case where one of the two types for which we wish to check
%subtyping has no branching. % This setting, although quite limited, is
%% not fully comparable to ours: one type has no branching at all, but
%% the other type is a general type.
%%
%Note that our subtyping procedure is inspired by the work of Mostrous
%et al.~\cite{ESOP09} on asynchronous subtyping for multiparty session
%types.

\noindent{\bf Comparison with our Previous Work \cite{BCZ16}.} 
Concerning decidability results, the approaches taken in this and our previous paper
are both based on initially considering a (non always terminating) subtyping procedure $\!\subtypea\!$, inspired from Mostrous et al.~\cite{ESOP09}, and then presenting a subtyping algorithm $\!\subtypet\!$ obtained by adding a new termination $\textsf{Asmp2}$  rule to the definition of the $\!\subtypea\!$ procedure (the further additional $\textsf{Asmp3}$ rule in our previous work is not needed when
dealing with orphan message-free subtyping, as we do here). 
Such an $\textsf{Asmp2}$  rule is crucial to guarantee the termination of the $\!\subtypet\!$ algorithm: the structure of the $\textsf{Asmp2}$  rule and the related proof of algorithm correctness and termination constitute the main contribution  of this paper %with respect to \cite{BCZ16} 
(as far as decidability results are concerned). 
In particular, since here multiple choices are admitted for input branchings,
the termination condition $\textsf{Asmp2}$  needs to deal with input accumulation in the form of a tree, instead of simple linear accumulation as in our previous work 
(see the discussion at the beginning Section~\ref{subsec:singleout} for details about this comparison).
As a consequence $\textsf{Asmp2}$  is completely modified:
it has to deal with complex recurrent patterns to be checked on the leaves of trees representing input branchings (with multiple choices), 
instead of detecting simple repetitions on strings representing sequences of single-choice inputs.
% as in \cite{BCZ16}. 
A detailed comparison follows.
Concerning the subtyping procedure $\!\subtypea\!$, the one considered in this 
paper (Figure \ref{fig:algo}) is novel in just two details:
an additional orphan message-free condition is considered in the
$\textsf{Out}$ rule (because here, %differently from \cite{BCZ16} 
we consider orphan message-free subtyping) and the usage of the $\mathsf{outUnf}$ unfolding instead of 
$\unfoldNOPAR n$ (so to perform the minimal needed per-branch unfolding).
The latter is needed in this paper because, 
when the procedure is turned into a subtyping algorithm $\!\subtypet\!$, we have to deal, in the 
new $\textsf{Asmp2}$ termination condition, with input trees instead of strings.
Correctness of $\!\subtypea\!$ w.r.t. asynchronous subtyping is stated by Proposition \ref{prop:semidecidable} in this paper, which corresponds to
Lemma 5.1 of our previous work \cite{BCZ16}. %However here,  differently from \cite{BCZ16}, 
Differently from the proof of that Lemma, here we need to cope with the different way of performing unfolding of right-hand terms in the asynchronous subtyping definition (via $\unfoldNOPAR n$) and in $\!\subtypea\!$ (via $\mathsf{outUnf}$).
Termination of the $\!\subtypet\!$ algorithm is stated by Theorem \ref{theo:termination} in this paper, which corresponds to
Lemma 5.2 of our previous work \cite{BCZ16}. 
Here due to the new structure of the $\textsf{Asmp2}$  rule, the proof is based on a much more complex characterization
of right-hand types produced, starting from the initial $Z$ one, by
the subtyping algorithm. Such types are characterized as $\antOut{S}{\gamma}$ with $\gamma$ being 
a sequence of output labels and $S$ being a type such that 
$S \in \reach{Z}$ and $\antOutInf{S}$.  
For the proof in this paper it is thus crucial to show, that
$\reach{Z}$ is finite and $\antOutInf{S}$ is decidable, as stated
by Proposition \ref{prop:outAntInfDecidable}, which has no counterpart in our previous work.
Correctness of the $\!\subtypet\!$ algorithm w.r.t. $\!\subtypea\!$ procedure is stated by Theorem \ref{theo:soundness} in this paper, which corresponds to
Lemma 5.3 of our previous work.
% \cite{BCZ16}. 
%As in \cite{BCZ16} we need to show that, 
In both papers correctness follows from the following property:
whenever the termination rule $\textsf{Asmp2}$  is applied by the $\!\subtypet\!$ algorithm, the
subtyping procedure $\!\subtypea\!$ is guaranteed to proceed forever.
However, being $\textsf{Asmp2}$  completely different, here we need to perform a totally new proof. In particular, %while in \cite{BCZ16} 
in our previous work we simply had to prove the existence of a cycle causing the initial input string in the right-hand type to get longer in a repetitive way, here, instead, we have to manage the more complex case of input trees: this entails
considering all possible branchings in order to prove
that the subtyping procedure proceeds forever.

Concerning undecidability results, both the approaches taken in the two papers
%in this paper 
%and that in \cite{BCZ16} 
are based on resorting to undecidability of
queue machines via a suitable encoding of them into subtyping. In particular, for 
showing undecidability of bounded asynchronous subtyping
($\subtypebound$) we resort to the same encoding as in our previous work,
% \cite{BCZ16} 
but we consider a more complex property of queue machines that we show to be undecidable:
bounded non termination. On the other hand, for showing undecidability of 
asynchronous subtyping without output covariance and input contravariance ($\subtypetintout$),
we need to introduce the novel class of single consuming queue machines
and to perform a much more complex encoding which, differently from the 
previous one,
uses nondeterminism (in order to avoid usage of covariance and contravariance, spurious 
nondeterministic paths that proceed forever have to be added by the encoding). 
The undecidability proof thus becomes significantly more complex %with respect to \cite{BCZ16} 
in that the above encoding makes it much more intricated to relate the
queue machine behaviour with the subtyping game and 
the novel class of queue machines. Moreover, introducing a new restricted class
of queue machines requires showing them
to be Turing-equivalent (by providing an encoding of standard queue machines into them).

% %%% Local Variables: 
% %%% mode: latex
% %%% TeX-master: "main"
% %%% End: 

\section{Conclusion}\label{sec:conclusions}
In this article, we have shed light on the boundaries between
decidability and undecidability of asynchronous session subtyping by
analyzing % the impact of
two kinds of restrictions: to the branching/selection
structure of inputs/outputs and to the capabilities of the
communication buffer.  In particular, considering all the relations in Figure \ref{fig:lattice}, we have shown: decidability for those in the lower part, notably
of ${\mathsf k}$-bounded subtyping and of subtyping over single-out
or single-in session types; and the undecidability for those in the upper part, notably
of bounded subtyping and of subtyping without
output covariance and input contravariance.

As future work, we plan to develop typing systems for server/client
code in the context of web services, exploiting our subtyping
algorithms for single-out/single-in session types.  Note that, in
practice, server code typically connects, as a client, to other
services (e.g.\ a database server) using another binary session,
according to the commonly used multitier architecture. Thus, in general, when
typing code, we would use for a specific session one of the two
algorithms above depending if the code is playing the role of the
client or of the server in that session.

Moreover, we plan to investigate whether other kinds of restriction
w.r.t. the two above allow us to obtain a decidable relation (thus
retaining general branching/selection structure for both inputs and outputs and not
limiting communication buffers).

% Our goal for the future is to identify a decidable subtyping relation
% that can partly deal with asynchrony while still allowing general
% session types. 
% loosens things in a way that everything can
% easily become undecidable. However, driven by practice, as we did for
% this paper, we hope we can find the right decidable subtyping
% relation.

% %%% Local Variables: 
% %%% mode: latex
% %%% TeX-master: "main"
% %%% End: 

\section*{References}

\bibliography{biblio}

 \newpage
 \appendix
 \section{Proofs of Section \ref{sec:preliminaries}}
 \subsection{Proof of Theorem \ref{thm:orphanMessageEquiv} and Propositions \ref{prop:dualclosed} and \ref{prop:complicated}}

We start by proving Theorem \ref{thm:orphanMessageEquiv}. This is done by separately showing, as preliminary lemmas, 
both implications (one in each direction) to hold. The proof of such lemmas will be then also exploited to prove Propositions \ref{prop:dualclosed} and \ref{prop:complicated}.

\begin{lemma}\label{lemma:easyOrphEquiv}
Given two session types $T$ and $S$, we have that $T \subtypedc S$ implies $T \subtype S$. 
\end{lemma}
\proof
Given an asynchronous  dual closed subtyping relation $\mathcal R$ we show that $\mathcal R$ is also a
(orphan-message-free) subtyping relation. To this aim we need to prove that if $(T,S) \in \mathcal R$
and $T=\Tselect{l}{T}$ then the additional item in $2.$ of Definition \ref{def:subtyping} holds, i.e.
  \begin{itemize}
  \item if $\mathcal A\neq[\,]^1$ then $\forall i\in I.\& \in T_i$
  \end{itemize}
%We proceed by contraposition showing that $\mathcal A \neq [\,]^1$ implies that 
%all $T_i$ contain at least one external choice.
From $\mathcal A \neq [\,]^1$ it follows that $S$, after some possible unfoldings, starts with an input (it must be in the form $\context {A} {{S_{k}}} {k\in \{1,\ldots, m\}}$).
As $\mathcal R$ is an asynchronous dual closed subtyping relation we have 
$(\dual S, \dual T) \in \mathcal R$.
We observe that $\dual S$, after some possible unfoldings, starts with an output and $\dual T=\Tbranch{l}{\dual T}$.
For item $2.$ of  Definition \ref{def:dcsubtyping},
we have that $\dual T=\context {A'} {\Tselectindex l{V_k}{j}{J_k}} {k\in \{1,\ldots, m\}}$,
for some input context $\mathcal{A'}$.
This means that all $\dual T_i$ contain at least an output selection,
which implies that all $T_i$ contain at least one input branching.
\qed

The following lemma states that $T \subtype S$ implies $T \subtypedc S$.
This is proved by showing that given 
$T \subtypeo S$ we have also $T \subtypedc S$ because
there exists an asynchronous dual closed
subtyping relation $\mathcal R$ s.t. $(T,S) \in \mathcal R$.
Such relation is defined as follows: $\mathcal R = \{(T,S),(\dual S,\dual T)\ |\ T \subtypeo S\}$.
The proof that each pair $(T,S) \in \mathcal R$ satisfies the items in
Definition \ref{def:dcsubtyping} has only one complex case, namely, the one
in which we assume $\dual S \subtypeo \dual T$, $T=\Tselect{l}{T}$ 
and $S = \Trec t_1. \dots \Trec t_n.\Tbranchindex lSjJ$ (case numbered
\ref{caso:difficile} in the proof).
In this case we have to reason on all initial output paths of $\dual S$
and use the \emph{no orphan message constraint} of Definition \ref{def:subtyping}
to be sure that such paths cannot be infinite, i.e. they eventually end in a state performing
an input choice, and moreover each of these reachable input
choices must match with the initial input choice of $\dual T$.

\begin{lemma}\label{lemma:hardOrphEquiv}
Given two session types $T$ and $S$, we have that $T \subtype S$ implies $T \subtypedc S$. 
\end{lemma}
\proof
We show that, given $T \subtypeo S$, it is possible to define an asynchronous dual closed
subtyping relation $\mathcal R$ s.t. $(T,S) \in \mathcal R$.
Consider
$$
\mathcal R = \{(T,S),(\dual S,\dual T)\ |\ T \subtypeo S\}
$$
The relation $\mathcal R$ is dual closed by definition. It remains to show that
it satisfies the four items in Definition \ref{def:dcsubtyping}.
Let $(T,S) \in \mathcal R$.
There are two cases: $T \subtypeo S$ or $\dual S \subtypeo \dual T$.
In the first case all the item holds by definition of orphan-message-free subtyping relation.
We consider now the second case, i.e. $\dual S \subtypeo \dual T$, and proceeds 
with a case analysis.
\begin{enumerate}
\item
$T = \Tend$.\\
We have $\dual T= \Tend$. Having $\dual S \subtypeo \Tend$, by definition of $\subtypeo$,
in particular by $n$ applications of item $4.$ (with $n \geq 0$) and one application of item $1.$,
it follows that $\dual S = \Trec t_1. \dots \Trec t_n.\Tend$.
Hence $S = \Trec t_1. \dots \Trec t_n.\Tend$, then we can conclude
what requested, i.e., $\exists n\geq 0$ such that $\unfold nS = \Tend$.    

\item
$T=\Tselect{l}{T}$.\\
We have $\dual T= \Tbranch{l}{\dual T}$.
Having $\dual S \subtypeo \Tbranch{l}{\dual T}$, by definition of $\subtypeo$,
we have two possible cases.
\begin{enumerate}
\item
By $n$ applications of item $4.$ (with $n \geq 0$) and one application of item $3.$,
it follows that $\dual S = \Trec t_1. \dots \Trec t_n.\Tbranchindex l{\dual S}jJ$, with $I\subseteq J$
and $\unfold n{\dual S} = \Tbranchindex l{\dual{S'}}jJ$ with 
$\dual{S'}_i \subtypeo \dual T_i$ for every $i \in I$.
Hence $S = \Trec t_1. \dots \Trec t_n.\Tselectindex lSjJ$, then we can conclude
what requested, i.e., 
    $\unfold nS = [\Tselectindex l{S'}jJ]^1$, $I\subseteq J$ and
    $\forall i\in I. (T_i,S'_i)\in\mathcal R$.
Notice that we have used the fact that $\unfold n{\dual S} = \dual{\unfold n{S}}$
and we have considered an input context $\mathcal A=[]^1$.     
\item \label{caso:difficile}
By $n$ applications of item $4.$ (with $n \geq 0$) and one application of item $2.$,
%with an input context different from $[]^1$,
it follows that 
$\dual T= \Tbranch{l}{\dual T}=\context {A} {\Tselectindex l{\dual{T_k}}{p}{J_k}} {k\in
        \{1,\ldots, m\}}$ (hence with $\mathcal A \neq []^1$),
and    $\dual S = \Trec t_1. \dots \Trec t_n.\Tselectindex l{\dual S}jJ$, with    
        $\forall k\in\{1,\ldots, m\}.  J\subseteq J_k$ and
 $\unfold n{\dual S} = \Tselectindex l{\dual{S'}}jJ$       
        with
      $\forall j\in J.\dual{S'}_j \subtypeo \context {A} {{\dual {T_{k}}_j}} {k\in \{1,\ldots, m\}}$.
%      ,
%and moreover every 
%$\dual{S'}_j$ contains at least one external choice.
Hence $S = \Trec t_1. \dots \Trec t_n.\Tbranchindex lSjJ$.
% with every $S_j$ that contains
%at least one internal choice.
We now observe that there exists an input context $\mathcal {A}'$
and $n',m'$ such that $\unfold{n'}S = \context {A'}{\Tselectindex l{S_k}{h}{L_k}}{k\in \{1,\ldots, m'\}}$
with $\forall k \!\in\! \{1,\ldots, m'\}. I$ $\subseteq\! L_k$.
This follows from the fact that $\dual S \subtypeo \Tbranch{l}{\dual T}$: by repeated application
of the rule 2. of Definition \ref{def:subtyping} (that includes the no orphan message constraint),
we have the guarantee that along all branches of $\dual S$ (and its unfoldings) it is guaranteed to
reach an input branching, and by application of rule 3. (in particular the contra-variance on input branchings), the labels of such choices include the set of labels of the initial input branching of $\Tbranch{l}{\dual T}$.
We conclude by showing that what is requested, i.e., $\forall i \in I.(T_i,\context {A'}{S_{ki}}{k\in \{1,\ldots, m'\}}) \in \mathcal R$, actually holds.
This follows from the fact that $\dual{\context {A'}{S_{ki}}{k\in \{\!1,\ldots, m'\}}} \subtypeo \dual{T_i}$,
which is a consequence of $\dual S \subtypeo \dual T$. In fact,
this implies that also $\unfold{n'}{\dual S} \subtypeo \dual T$ because an orphan-message-free
subtyping relation is still such even if we add pairs $(\unfold rV,Z)$ assuming $(V,Z)$ already in the 
relation. Having $\unfold {n'}{\dual S} = \dual{\unfold {n'}{S}} =
\context {\dual{A'}}{\Tbranchindex l{\dual{S_k}}{h}{L_k}}{k\in \{\!1,\ldots, m'\!\}}$ 
and 
$\dual T= \Tbranch{l}{\dual T}$, it is easy to see that, given an orphan-message-free 
subtyping relation $\mathcal {R}'$
such that $(\context {\dual{A'}}{\Tbranchindex l{\dual{S_k}}{h}{L_k}}{k\in \{\!1,\ldots, m'\!\}},\Tbranch{l}{\dual T})\in \mathcal {R}'$,
the relation obtained by enriching $\mathcal {R}'$
with the pairs $(\context {\dual{A''}}{\dual{S_k}_i}{k\in K \subseteq\{\!1,\ldots, m'\!\}}\!,\dual{T'}_i)$,
where $\mathcal A''$ and types $T'_i$, with $i \in I$, are such that
$(\context {\dual{A''}}{\Tbranchindex l{\dual{S_k}}{h}{L_k}}{k\in K \subseteq \{\!1,\ldots, m'\!\}}\!,$ $\Tbranch{l}{\dual {T'}}) \in \mathcal {R}'$, is still an orphan-message-free subtyping relation. 
Above we adopt an abuse of notation for input contexts:
$\context {\dual{B}}{W_k}{k\in K \subseteq \{1,\ldots, t\}}$ 
does not have holes numbered consistently from $1$ to $t$, but 
some numbers in $\{1,\ldots,t\}$ could be missing.
\end{enumerate}

\item
$T=\Tbranch{l}{T}$.\\
We have $\dual T= \Tselect{l}{\dual T}$. Having $\dual S \subtypeo \Tselect{l}{\dual T}$, by definition of $\subtypeo$,
in particular by $n$ applications of item $4.$ (with $n \geq 0$) and one application of item $2.$,
it follows that $\dual S = \Trec t_1. \dots \Trec t_n.\Tselectindex l{\dual S}jJ$, with $J\subseteq I$,
and $\unfold n{\dual S} = \Tselectindex l{\dual{S'}}jJ$ with 
$\dual{S'}_j \subtypeo \dual T_j$ for every $j \in J$.
Hence $S = \Trec t_1. \dots \Trec t_n.\Tbranchindex lSjJ$, then we can conclude
what requested, i.e., 
    $\unfold nS = \Tbranchindex l{S'}jJ$, $J\subseteq I$ and
    $\forall j\in J. (T_j,S'_j)\in\mathcal R$.
Notice that we have used the fact that $\unfold n{\dual S} = \dual{\unfold n{S}}$.     

\item
$T=\Trec t.T'$.\\
We first observe that $V \subtypeo \Trec t.Z$ implies
$V \subtypeo Z\{\Trec t.Z/\Tvar t\}$. This directly follows from the fact
that if $(V, \Trec t.Z)$ belongs to an orphan-message-free 
subtyping relation, then the same relation enriched with the pair $(V, Z\{\Trec t.Z/\Tvar t\})$
is still an orphan-message-free subtyping relation.
We now proceed by considering $\dual T=\Trec t.\dual{T'}$. As $\dual S \subtypeo \dual T$,
we have $\dual S \subtypeo \Trec t.\dual{T'}$. By the above observation we have 
$\dual S \subtypeo \dual{T'}\{\Trec t.\dual{T'}/\Tvar t\}$ that implies what requested, i.e.,
$(T'\{\Trec t.T'/\Tvar t\},S) \in \mathcal R$.
\qed
\end{enumerate}

\restateTHM{thm:orphanMessageEquiv}
\emph{Given two session types $T$ and $S$, we have $T \subtype S$ if and only if $T \subtypedc S$.}\\
\proof
Direct consequence of Lemmas \ref{lemma:easyOrphEquiv} and \ref{lemma:hardOrphEquiv}.
\qed

\restatePRO{prop:dualclosed}
\emph{    
The $\subtypetintout$ relation and the $\subtypesin \! \cup \! \subtypesout$ relation are asynchronous dual closed subtyping relations.
} \\
\proof
We first show that $\subtypetintout$ is an asynchronous dual closed subtyping relation.
We consider $\mathcal R = \{ (\dual{S},\dual{T}) \,|\, T \subtypetintout S\}$ and show that it is an asynchronous subtyping relation
when in Definition \ref{def:subtyping} we require $I=J_k$ in item 2.\  and $I=J$ in item 3.\ to hold.
This implies $\{ (\dual{S},\dual{T}) \,|\, T \subtypetintout S\} \subseteq \subtypetintout$, thus showing that $\subtypetintout$ is dual closed.
Given $(T,S) \in \mathcal R$, we show that $\dual{S} \subtypetintout \dual{T}$ implies items 1.-4.\ of Definition \ref{def:subtyping} (where we require $I=J_k$ in item 2.\ and $I=J$ in item 3.), apart from the no orphan message constraint of item 2., by case analysis on the structure of type $T$ exactly as in the proof of \mbox{Lemma \ref{lemma:hardOrphEquiv}}
(where $\subtypetintout$ is considered instead of $\subtype$ and all subset inclusions related to covariance/contravariance are replaced by subset equalities).
Concerning the no orphan message constraint of item 2., in the case 2.a of the proof of Lemma \ref{lemma:hardOrphEquiv} just an 
$[\,]^1$ input context arises (so it obviously holds); in the case 2.b,  instead, a generic input context $\mathcal A'$ arises: if
$\mathcal A' \neq [\,]^1$ then this means that $\dual S$, after some possible unfoldings, starts with an output and the constraint
is an immediate consequence of the fact that $\dual{S} \subtypetintout \dual{T}$ (as in the proof of Lemma \ref{lemma:easyOrphEquiv}).

We now show that $\subtypesin \cup \subtypesout$ is an asynchronous dual closed subtyping relation.
We use $T^{in}$ and $T^{out}$ to denote the set of single-in and single-out session types, respectively.
We have
$\subtypesin \cup \subtypesout =
   (\subtype \cap T^{in} \times T^{in}) \cup (\subtype \cap T^{out} \times T^{out}) =
   (\subtypedc \cap T^{in} \times T^{in}) \cup (\subtypedc \cap T^{out} \times T^{out})$, due to Theorem \ref{thm:orphanMessageEquiv}.
We now show that, for any $(T,S) \in \subtypesin \cup \subtypesout$,
all constraints considered by Definition \ref{def:dcsubtyping} hold.
We take $(T,S) \in \subtypedc \cap T^{in} \times T^{in}$, the other case $(T,S) \in \subtypedc \cap T^{out} \times T^{out}$ is dealt with symmetrically. Since $(T,S) \in \subtypedc$ we have that $(T,S)$ satisfies all constraints in items 1.-4.\ of Definition \ref{def:dcsubtyping}: we just have to additionally observe that, since all reached pairs belong to $\subtypedc$, they also obviously belong to $\subtypedc \cap T^{in} \times T^{in}$. 
Concerning the duality constraint, from $(T,S) \in \subtypedc$, we have $(\dual{S},\dual{T}) \in \subtypedc$, hence $(\dual{S},\dual{T}) \in (\subtypedc \cap T^{out} \times T^{out})$.
\qed

\restatePRO{prop:complicated}
\emph{    
The $\subtypesintout$ and $\subtypetinsout$ relations are such that: $T \,\subtypesintout\, S$ if and only if $\dual{S} \,\subtypetinsout\, \dual{T}$.
} \\
\proof
Concerning the only if part, we show $\{ (\dual{S},\dual{T}) \,|\, T \subtypesintout S\} \subseteq \subtypetinsout$ as follows.
We consider $\mathcal R = \{ (\dual{S},\dual{T}) \,|\, T \subtypesintout S\}$ and show that it is an asynchronous subtyping relation
when in Definition \ref{def:subtyping} we require $I=J$ in item 3. to hold and related types to be both single-out.
Given $(\dual{S},\dual{T}) \in \mathcal R$, we obviously have that $\dual{S}$ and $\dual{T}$ are both single-out and we show that
$T \subtypesintout S$ implies items 1.-4.\ of Definition \ref{def:subtyping} (where in item 3.\ we require $I=J$)
as in the proof of Proposition \ref{prop:dualclosed}. The only difference is that, when resorting to the case analysis in the proof of Lemma \ref{lemma:hardOrphEquiv} we consider $\subtypesintout$ instead of $\subtype$ and we replace all subset inclusions related to covariance/contravariance in item 3.\ and the subset inclusion $J \subseteq J_k$ in item 2.\ by equalities.
%FORSE SI POTREBBE RIMUOVERE $J \subseteq J_k$ DALLA DIM DI Lemma \ref{lemma:hardOrphEquiv} E RENDERE TUTTO SIMMETRICO

Concerning the if part, we show $\{ (\dual{S},\dual{T}) \,|\, T \subtypetinsout S\} \subseteq \subtypesintout$ in a completely symmetric way 
by observing that  
$\mathcal R = \{ (\dual{S},\dual{T}) \,|\, T \subtypetinsout S\}$ is an asynchronous subtyping relation
when in Definition \ref{def:subtyping} we require $I=J_k$ in item 2. to hold and related types to be both single-in.
In this case, when resorting to the case analysis in the proof of Lemma \ref{lemma:hardOrphEquiv} we consider $\subtypetinsout$ instead of $\subtype$ and we replace all subset inclusions related to covariance/contravariance in item 2., apart from $J \subseteq J_k$, by equalities.
\qed
%FORSE SI POTREBBE RIMUOVERE $J \subseteq J_k$ DALLA DIM DI Lemma \ref{lemma:hardOrphEquiv} E RENDERE TUTTO SIMMETRICO

 \section{Proofs of Section \ref{sec:algorithm}}\label{AppendixDecidible}

 \subsection{Proof of Proposition \ref{prop:semidecidable}}

Proposition \ref{prop:semidecidable} states that the procedure defined in Figure \ref{fig:algo}
is a semi-algorithm for checking whether $T \not\!\!\subtype S$.
The proof of the proposition is divided in two parts. The first one
shows that if it is not possible to reach
a judgement in which no rule can be applied,
i.e. there exist no $\Sigma', T', S'$ such that
$\emptyset \vdash T \subtypea S  \rderiv^* \Sigma' \vdash T' \subtypea S' \notrderiv$,
then it is possible to define an asynchronous subtyping
relation $\mathcal R$ such that $(T,S) \in \mathcal R$.
The second part shows that if $T \subtype S$ then the 
procedure either continues indefinitely or terminates successfully
by application of the $\textsf{Asmp}$ or $\textsf{End}$ rules.

\restatePRO{prop:semidecidable}
\emph{  
  Given the types $T$ and $S$, we have that there exist $\Sigma', T', S'$
  such that
 $T \not\!\!\subtype S$   if and only if
  $\emptyset \vdash T \subtypea S
  \rderiv^* \Sigma' \vdash T' \subtypea S' \notrderiv$.
}\\
\proof 
We prove the two implications separately.  

We start with the
{\em only if} part and proceed by contraposition, i.e,
we assume that it is not true that
$\exists \Sigma', T', S' \ldotp \; \emptyset \vdash T \subtypea S
\rderiv^* \Sigma' \vdash T' \subtypea S' \notrderiv$
and show that $T \subtype S$.

In order to do this we need to perform a preliminary observation:
under the assumption that it is not true that
$\exists \Sigma', T', S' \ldotp \; \emptyset \vdash T \subtypea S
\rderiv^* \Sigma' \vdash T' \subtypea S' \notrderiv$,
even if we remove rule $\textsf{Asmp}$
from the procedure it is still impossible to reach a judgement 
$\Sigma' \vdash T' \subtypea S' \notrderiv$, i.e. a judgement
on which no rule can be applied. This can be showed as follows.
Let $\rderiv_{noAsmp}$ be our decision procedure under the assumption
that $\textsf{Asmp}$ is not used.
Observe that the set of pairs $\Sigma$ in the judgements
is irrelevant for the decision procedure $\rderiv_{noAsmp}$
because $\textsf{Asmp}$ is the unique rule influenced by it.
Now, by contraposition, assume 
$\emptyset \vdash T \subtypea S
\rderiv_{noAsmp}^*\Sigma' \vdash T' \subtypea S' \notrderiv$.
%where no rule can be applied to $\Sigma' \vdash T' \subtypea S'$
%(neither $\textsf{Asmp}$).
Since the standard procedure $\rderiv^*$ cannot reach $\notrderiv$,
we must have that there exists an intermediary judgement 
$\Sigma'' \vdash T'' \subtypea S''$ where $\textsf{Asmp}$ is applied.
That is, there exists $\Sigma'' \vdash T'' \subtypea S''$ such that
$\emptyset \vdash T \subtypea S
\rderiv^* \Sigma'' \vdash T'' \subtypea S''$
(notice the use of the standard procedure),
$(T'',S'') \in \Sigma''$
and 
$ \Sigma'' \vdash T'' \subtypea S'' \rderiv_{noAsmp}^*
\Sigma' \vdash T' \subtypea S'$. 
%We have that $\Sigma'' \subseteq \Sigma'$
%as the environment only grows upon application
%of the rules.
Within the sequence of rule applications
$ \Sigma'' \vdash T'' \subtypea S'' \rderiv_{noAsmp}^*
\Sigma' \vdash T' \subtypea S'$ 
we consider the last judgement $ \Sigma''' \vdash T''' \subtypea S'''$ 
such that $(T''',S''')\in \Sigma''$
(such judgement exists as the first one $ \Sigma'' \vdash T'' \subtypea S''$
already has this property).
It is not restrictive to assume that in the sequence 
$ \Sigma''' \vdash T''' \subtypea S''' \rderiv_{noAsmp}^*
\Sigma' \vdash T' \subtypea S'$ there are no two judgements
$ \Sigma_1 \vdash T_1 \subtypea S_1$
and
$ \Sigma_2 \vdash T_2 \subtypea S_2$
with $T_1 = T_2$ and $S_1 = S_2$
(otherwise we can shorten the sequence 
$ \Sigma''' \vdash T''' \subtypea S''' \rderiv_{noAsmp}^*
\Sigma' \vdash T' \subtypea S'$
by removing the steps between the judgements 
$ \Sigma_1 \vdash T_1 \subtypea S_1$
and
$ \Sigma_2 \vdash T_1 \subtypea S_1$,\footnote{The sequence of rules 
applied to $ \Sigma_2 \vdash T_1 \subtypea S_1$ can be applied also to
$ \Sigma_1 \vdash T_1 \subtypea S_1$ because, as already observed, 
$\rderiv_{noAsmp}$ is not
sensitive to the differences between $\Sigma_1$ and $\Sigma_2$.}
obtaining a new one having the same properties
but without the second judgement
$ \Sigma_2 \vdash T_1 \subtypea S_1$).
Consider now, in the standard application of the 
procedure  $\emptyset \vdash T \subtypea S
\rderiv^* \Sigma'' \vdash T'' \subtypea S''$,
the intermediary judgement $\Sigma_i \vdash T''' \subtypea S'''$ that added 
$(T''',S''')$
to the environment.
We have that $(T''',S''') \not \in \Sigma_i$  (otherwise rule $\textsf{Asmp}$
was applied that does not introduce any new pair in $\Sigma_i$) 
and moreover $\Sigma_i \subset \Sigma''$ as the sets of type pairs grow monotonically
during the execution of the decision procedure $\rderiv^*$.
These last observations guarantee 
that %from this judgement
there exists a standard application of the procedure
$\emptyset \vdash T \subtypea S
\rderiv^* \Sigma_i \vdash T''' \subtypea S''' 
\rderiv^* \Sigma_i' \vdash T'''' \subtypea S''''\notrderiv$
simply by considering from $\Sigma_i \vdash T''' \subtypea S'''$
the same rules used in the sequence
$ \Sigma''' \vdash T''' \subtypea S''' \rderiv_{noAsmp}^* \Sigma' \vdash T' \subtypea S'$.
This holds because the pairs of types in the judgements traversed
by this sequence are not in $\Sigma''$
(due to the assumption on the judgement $ \Sigma''' \vdash T''' \subtypea S'''$) 
hence also not in $\Sigma_i$,
and moreover such pairs are all distinct 
(guaranteed by the assumption on the absence of two judgements
$ \Sigma_1 \vdash T_1 \subtypea S_1$
and
$ \Sigma_2 \vdash T_2 \subtypea S_2$
with $T_1 = T_2$ and $S_1 = S_2$).

Consider now the relation
${\mathcal R}=\{(T',S')\ |\ \exists \Sigma' \ldotp \Sigma' \vdash T'
\subtypea S' \in \mathcal{S}\}$ where $\mathcal S$ is the minimal set of judgements 
satisfying the following:
\begin{itemize}
\item
$\emptyset \vdash T \subtypea S \in \mathcal S$;
\item
if $\Sigma' \vdash T' \subtypea S' \in \mathcal S$ and
$\Sigma' \vdash T' \subtypea S' \rderiv \Sigma'' \vdash T'' \subtypea S''$,
without applying rule $\textsf{Asmp}$ or $\textsf{RecR}_2$, then
$\Sigma'' \vdash T'' \subtypea S'' \in \mathcal S$;
\item
if $\Sigma' \vdash T' \subtypea S' \in \mathcal S$ and
$\Sigma' \vdash T' \subtypea S' \rderiv \Sigma'' \vdash T'' \subtypea S''$
by applying $\textsf{RecR}_2$, then
$\Sigma'' \vdash T'' \subtypea \unfold{\depth(S')}{S'} \in \mathcal S$.
\end{itemize}
We observe that to each judgement $\Sigma' \vdash T'
\subtypea S' \in \mathcal{S}$ it is always possible to apply at least
one rule. In fact, if this is not possible, we would have
also $\emptyset \vdash T \subtypea S
\rderiv_{noAsmp}^*\Sigma'' \vdash T'' \subtypea S'' \notrderiv$
for a judgement $\Sigma'' \vdash T'' \subtypea S''$
with $T''=T'$ and $S''$ less unfolded than $S'$.
In fact, the unique difference between the judgements
in $\mathcal S$ and those reachable without adopting $\textsf{Asmp}$
is that those in $\mathcal S$ are more unfolded (see the difference
between $\outunfold{S}$ used in rule $\textsf{RecR}_2$ and 
$\unfold{\depth(S')}{S'}$ used in the definition of $\mathcal S$).

We finally show that ${\mathcal R}$ is an (orphan-message-free) subtyping relation
according to Definition \ref{def:subtyping}.
%hence $T \subtype_o S$ that, by Theorem \ref{thm:orphanMessageEquiv}, 
%implies also $T \subtype S$ . 
Let
$(T',S') \in {\mathcal R}$.
%, i.e. 
%$\exists \Sigma' \ldotp \emptyset \vdash T \subtypea S \rderiv^* \Sigma' \vdash T' \subtypea S'$.
Then 
%$\emptyset \vdash T \subtypea S \rderiv^* \Sigma' \vdash T' \subtypea
%S'$ 
$\Sigma' \vdash T' \subtypea S' \in \mathcal S$
and it is possible to apply at least one rule to
$\Sigma' \vdash T' \subtypea S'$.
% for the environment $\Sigma'$ such
%that
%$\emptyset \vdash T \subtypea S \rderiv^* \Sigma' \vdash T' \subtypea
%S'$.  
We proceed by cases on $T'$.
  \begin{itemize}

  \item \label{item:end} If $T'=\Tend$ then item $1.$ of Definition
    \ref{def:subtyping} for pair $(T',S')$ is shown by induction on
    $k=\mathsf{nrec}(S')$, i.e.\ the number of unguarded (not prefixed
    by some input or output) occurrences of recursions $\Trec t. S''$ in
    $S'$ for any $S'',\Tvar t$.
%NON CI POSSONO ESSERE MULTIPLE rec t PER LO STESSO t UNA DENTRO L'ALTRA!
  \begin{itemize}
  \item Base case $k=0$. The only rule applicable to
    $\Sigma' \vdash T' \subtypea S'$ is $\textsf{End}$, that immediately
    yields the desired pair of ${\mathcal R}$.
  \item Induction case $k>0$. The only rules applicable to
    $\Sigma' \vdash T' \subtypea S'$ are $\textsf{Asmp}$ and
    $\textsf{RecR}_1$.  In the case of $\textsf{Asmp}$ we have that
    $(T',S') \in \Sigma'$, hence there exists $\Sigma''$ with
    $(T',S') \notin \Sigma''$ such that
    $\Sigma'' \vdash T'
    \subtypea S' \in \mathcal S$.
    $\textsf{RecR}_1$ can be applied to
    $\Sigma'' \vdash T' \subtypea S'$.
    % IN REALTA' E' IMPOSSIBILE!! CHE DAL CENTRALE SI VADA AL FINALE
    % PERCHE' DOPO END L'ALGO SI FERMA!
    So for some $\Sigma'''$ ($=\Sigma'$ or $=\Sigma''$) we have that
    the procedure applies rule $\textsf{RecR}_1$ to
    $\Sigma''' \vdash T' \subtypea S'$. Hence
    $\Sigma''' \vdash T' \subtypea S' \rderiv \Sigma'''' \vdash T'
    \subtypea \unfold 1{S'}$. Since $\mathsf{nrec}(\unfold 1{S'})=k-1$, by induction
    hypothesis item $1.$ of Definition~\ref{def:subtyping} holds for
    pair $(T', \unfold 1{S'})$, hence it holds for pair $(T',S')$.
  \end{itemize}

\item \label{item:internal} If $T'=\Tselect{l}{T}$ then item $2.$ of
  Definition \ref{def:subtyping} for pair $(T',S')$ is shown as
  follows.
\begin{itemize}
\item If $\depth(S')=0$ then the only rule applicable
  to $\Sigma' \vdash T' \subtypea S'$ is $\textsf{Out}$, that
  immediately yields the desired pairs of ${\mathcal R}$.
\item If $\depth(S')\geq1$ then the only rules
  applicable to $\Sigma' \vdash T' \subtypea S'$ are $\textsf{Asmp}$
  and $\textsf{RecR}_2$.  In the case of $\textsf{Asmp}$ we have that
  $(T',S') \in \Sigma'$, hence there exists $\Sigma''$ with
  $(T',S') \notin \Sigma''$ such that
      $\Sigma'' \vdash T'
    \subtypea S' \in \mathcal S$.
%  $\emptyset \vdash T \subtypea S \rderiv^* \Sigma'' \vdash T'
%  \subtypea S' \rderiv^* \Sigma' \vdash T' \subtypea S'$
%  and rule 
  $\textsf{RecR}_2$ can be applied to
  $\Sigma'' \vdash T' \subtypea S'$.  So for some $\Sigma'''$
  ($=\Sigma'$ or $=\Sigma''$) we have that the procedure applies rule
  $\textsf{RecR}_2$ to $\Sigma''' \vdash T' \subtypea S'$. Hence
  $(T',\unfold{\depth(S')}{S'}) \in \mathcal R$.
%  $\Sigma''' \vdash T' \subtypea S' \rderiv \Sigma'''' \vdash T'
%  \subtypea \outunfold{S'} $.
%  ,
%  with $k = \depth(S,\emptyset)$.  
  Since
  $\depth(\unfold{\depth(S')}{S'})=0$, we end up in the previous
  case. Therefore item $2.$ of Definition \ref{def:subtyping} holds for
  pair $(T'\!,\!\!\; \unfold{\depth(S'\!\!\;)\!\!\;}{S'}\!\!\;)$, hence it holds for pair $(T',S')$.

\end{itemize}

\item \label{item:external} If $T'=\Tbranch{l}{T}$ then item $3.$ of
  Definition \ref{def:subtyping} for pair $(T',S')$ is shown by
  induction on $k=\mathsf{nrec}(S')$.
  \begin{itemize}
  \item Base case $k=0$. The only rule applicable to
    $\Sigma' \vdash T' \subtypea S'$ is $\textsf{In}$, that immediately
    yields the desired pairs of ${\mathcal R}$.
  \item Induction case $k>0$. The only rules applicable to
    $\Sigma' \vdash T' \subtypea S'$ are $\textsf{Asmp}$ and
    $\textsf{RecR}_1$.      In the case of $\textsf{Asmp}$ we have that
  $(T',S') \in \Sigma'$, hence there exists $\Sigma''$ with
  $(T',S') \notin \Sigma''$ such that
      $\Sigma'' \vdash T'
    \subtypea S' \in \mathcal S$.
        $\textsf{RecR}_1$ can be applied to
    $\Sigma'' \vdash T' \subtypea S'$.   
%    we have that
%    $(T',S') \in \Sigma'$, hence there exists $\Sigma''$ with
%    $(T',S') \notin \Sigma''$ such that
%    $\emptyset \vdash T \subtypea S \rderiv^* \Sigma'' \vdash T'
%    \subtypea S' \rderiv^* \Sigma' \vdash T' \subtypea S'$
%    and rule $\textsf{RecR}_1$ has been applied to
%    $\Sigma'' \vdash T' \subtypea S'$.  
    So for some $\Sigma'''$
    ($=\Sigma'$ or $=\Sigma''$) we have that the procedure applies
    rule $\textsf{RecR}_1$ to $\Sigma''' \vdash T' \subtypea S'$.
    Hence
    $\Sigma''' \vdash T' \subtypea S' \rderiv \Sigma'''' \vdash T'
    \subtypea \unfold 1{S'} $. Since $\mathsf{nrec}(\unfold 1{S'})=k-1$, by induction
    hypothesis item $3.$ of Definition~\ref{def:subtyping} holds for
    pair $(T', \unfold 1{S'})$, hence it holds for pair $(T',S')$.

  \end{itemize}

\item \label{item:rec} If $T'= \Trec t.{T'} $ then item $4.$ of
  Definition \ref{def:subtyping} for pair $(T',S')$ holds because the
  only rule applicable to $\Sigma' \vdash T' \subtypea S'$ is
  $\textsf{RecL}$ that immediately yields the desired pair of
  ${\mathcal R}$.

  \end{itemize}

  We now prove the {\em if} part and proceed by contraposition.
  We assume that $T \subtype S$ and show that 
  there exist no $\Sigma', T', S'$,
  such that
  $\emptyset \vdash T \subtypea S
  \rderiv^* \Sigma' \vdash T' \subtypea S' \notrderiv$.
%  If $T \subtype S$ then also $T \subtype_o S$ (by Theorem \ref{thm:orphanMessageEquiv}).
  So we can assume the existence of a relation ${\mathcal R}$ that is an
  (orphan-message-free) subtyping relation, according to Definition \ref{def:subtyping}, such that $(T,S) \in {\mathcal R}$.

  We say that
  $\Sigma \vdash T \subtypea S \rderiv_w \Sigma' \vdash T' \subtypea
  S'$
  if
  $\Sigma \vdash T \subtypea S \rderiv^* \Sigma' \vdash T' \subtypea
  S'$ and: the last rule applied is one of
  % $(Asmp)$,$(End)$,
  $\textsf{Out}$, $\textsf{In}$ or $\textsf{RecL}$ rules; while all
  previous ones are $\textsf{RecR}_1$ or $\textsf{RecR}_2$ rules.
  As another notation we use input-output-end contexts $\mathcal B$
  defined as the input contexts in Definition \ref{def:context}
  with the difference that also the output construct and $\Tend$ are
  part of the grammar in the definition.

  We start by showing that 
  $\exists \Sigma \ldotp \emptyset \vdash T \subtypea S \rderiv_w^*
  \Sigma \vdash T' \subtypea S'$
  implies $S' = \context{B}{S_k}{k\in\{1\dots m\}}$, $S_k =\Trec {t_k}.S'_k$, for some
  $\Tvar {t_k}$ and $S'_k$, and
  $\exists n_1,\dots,n_m \ldotp (T',$ $\context{B}{\unfold {n_k}{S_k}}{k\in\{1\dots m\}}) \in {\mathcal R}$. The
  proof is by induction on the length of such computation $\rderiv_w^*$.
  The base case is for a $0$ length computation: it
  yields $(T,S) \in {\mathcal R}$ which holds.  For the inductive case
  we assume it to hold for all computations of a length $k$ and we
  show it to holds for all computations of length $k+1$, by considering
  all judgements $\Sigma' \vdash T'' \subtypea S''$ such that
  $\Sigma \vdash T' \subtypea S' \rderiv_w \Sigma' \vdash T''
  \subtypea S''$.
  This is shown by first considering the case in which rule
  $\textsf{Asmp}$ applies to $\Sigma \vdash T' \subtypea S'$: in this
  case there is no such a judgement and there is nothing to prove.  Then
  we consider the case in which $T'=\Tend$ and
  $\Sigma \vdash \Tend \subtypea S' \rderiv^* \Sigma''' \vdash \Tend
  \subtypea \Tend$
  (by applying $\textsf{RecR}_1$ rules) and rule $\textsf{End}$
  applies to $\Sigma''' \vdash \Tend \subtypea \Tend$. Also in this
  case there is no such a judgement $\Sigma' \vdash T'' \subtypea S''$
  and there is nothing to prove.  Finally, we proceed by an immediate
  verification that judgements $\Sigma' \vdash T'' \subtypea S''$
  produced in remaining cases are required to be in $ {\mathcal R}$ by
  items $2.$, $3.$ and $4.$ of Definition~\ref{def:subtyping}:
  $T'=\Tselect{l}{T}$ ($\rderiv_w$ is a possibly empty sequence of
  $\textsf{RecR}_2$ applications followed by $\textsf{Out}$
  application), $T'=\Tbranch{l}{T}$ ($\rderiv_w$ is a possibly empty
  sequence of $\textsf{RecR}_1$ applications followed by $\textsf{In}$
  application) or $T'= \Trec t.{T'}$ ($\rderiv_w$ is simply
  $\textsf{RecL}$ application).

  We finally observe that, given a judgement
  $\Sigma \vdash T' \subtypea S'$ such that
  $S' = \context{B}{S_k}{k\in\{1\dots m\}}$, $S_k =\Trec {t_k}.S'_k$, for some
  $\Tvar {t_k}$ and $S'_k$, and
  $\exists \, n_1,\dots,n_m \, \ldotp (T',$ \linebreak $\context{B}{\unfold {n_k}{S_k}}{k\in\{1\dots m\}}) \in {\mathcal R}$
  we have:
  \begin{itemize}
  \item either rule $\textsf{Asmp}$ applies to
    $\Sigma \vdash T' \subtypea S'$, or

  \item $T'=\Tend$ and, by item $1.$ of Definition
    \ref{def:subtyping}, there exists $\Sigma'$ such that
    $\Sigma \vdash \Tend \subtypea S' \rderiv^* \Sigma' \vdash \Tend
    \subtypea \Tend$
    (by applying $\textsf{RecR}_1$ rules) and rule $\textsf{End}$ is
    the unique rule applicable to
    $\Sigma' \vdash \Tend \subtypea \Tend$, with $\textsf{RecR}_1$ being
    the unique rule applicable to intermediate judgements, or

  \item by items $2.$, $3.$ and $4.$ of Definition
    \ref{def:subtyping}, there exist $\Sigma', T'',S''$ such that
    $\Sigma \vdash T' \subtypea S' \rderiv_w^* \Sigma' \vdash T''
    \subtypea S''$, with each intermediate judgement
    having a unique applicable rule.
    In particular this holds for $T'=\Tselect{l}{T}$ ($\rderiv_w$ is a
    possibly empty sequence of $\textsf{RecR}_2$ applications followed
    by $\textsf{Out}$ application), $T'=\Tbranch{l}{T}$ ($\rderiv_w$
    is a possibly empty sequence of $\textsf{RecR}_1$ applications
    followed by $\textsf{In}$ application) or $T'= \Trec t.{T'}$
    ($\rderiv_w$ is simply $\textsf{RecL}$ application). 
    \qed
  \end{itemize}

\subsection{Proof of Theorem \ref{theo:kbounded}}

Theorem \ref{theo:kbounded} states that 
$\!\subtypeak\!$ indeed provides an algorithm for checking whether $T \not\!\!\subtype_{\mathsf k} S$, this is proved in the following
by also resorting to the proof of Proposition \ref{prop:semidecidable}.

\restateTHM{theo:kbounded}
\emph{    The algorithm for $\subtypeak$ always terminates and, given the types $T$
  and $S$, 
there exist $\Sigma', T', S'$
  such that
  $\emptyset \vdash T \subtypeak S
  \rderiv^* \Sigma' \vdash T' \subtypea S' \notrderiv$
  if and only if $T \not\!\!\subtype_{\mathsf k} S$.
} \\
\proof
We first observe that the decision algorithm for ${\mathsf k}$-bounded asynchronous subtyping
terminates. By contraposition, if the algorithm does not terminate, there exists
an infinite sequence $\Sigma \vdash T \subtypea S \rderiv \Sigma_1 \vdash T_1 \subtypea S_1
\rderiv^* \Sigma_i \vdash T_i \subtypea S_i \rderiv^*$. Along this infinite sequence
infinitely many distinct pairs $(T,S)$ will be added to $\Sigma$. As only finitely many
distinct terms can be reached as first element of the pairs, there will be infinitely many
distinct terms as second element. Such terms will have unbounded depth, but this is not
possible due to the constraint added to rule $\textsf{Out}$ that impose the use of ${\mathsf k}$-bounded
input contexts.

We now prove that, given the types $T$
  and $S$, 
there exist $\Sigma', T', S'$
  such that
  $\emptyset \vdash T \subtypeak S
  \rderiv^* \Sigma' \vdash T' \subtypea S' \notrderiv$
  if and only if $T \not\!\!\subtype_{\mathsf k} S$.

We start with the
{\em if} part and proceed by contraposition.
We assume that it is not true that
$\exists \Sigma', T', S' \ldotp \; \emptyset \vdash T \subtypeak S
\rderiv^* \Sigma' \vdash T' \subtypeak S' \notrderiv$
and we build a relation
${\mathcal R}$ that we show to be a ${\mathsf k}$-bounded Asynchronous Subtyping relation.
The relation ${\mathcal R}$ is built from the judgments 
$\Sigma'' \vdash T'' \subtypeak S''$ exactly as we did for the $\!\subtypea\!$ subtyping procedure
in the first part (the {\em if} part) of the proof of Proposition \ref{prop:semidecidable}.
In such a proof we show ${\mathcal R}$ to be an orphan-message-free subtyping relation, hence
we just have to show it to be ${\mathsf k}$-bounded. It is immediate to observe that, since when 
applying rule $\textsf{Out}$ to a judgment $\Sigma'' \vdash T'' \subtypeak S''$
we require the input context $\mathcal A$ to be ${\mathsf k}$-bounded, we may
include in ${\mathcal R}$ only pairs $(T'', S'')$ that satisfy the same constraint
in item $2$ of ${\mathsf k}$-bounded Asynchronous Subtyping relation definition (Definition \ref{def:kbounded}), because otherwise
we would have $\Sigma'' \vdash T'' \subtypeak S'' \rderiv^* \Sigma''' \vdash T''' \subtypeak S''' \notrderiv$
by possibly applying $\textsf{RecR}_1$/$\textsf{RecR}_2$ rules. Hence, as justified in Proposition \ref{prop:semidecidable} this would lead to violating the assumption that the algorithm does not reach an error.
The justification provided there still holds because judgments $\Sigma'' \vdash T'' \subtypeak S''_1$ and $\Sigma'' \vdash T'' \subtypeak S''_2$, with $S''_1$ and $S''_2$ that just differ for the level of internal unfoldings, behave equivalently with respect to errors due to ${\mathsf k}$-boundedness violations. This because the ${\mathsf k}$-boundedness of context $\mathcal A$  is established by the $\textsf{Out}$ rule after unfolding in $S''_1$/$S''_2$ all recursions occurring before the first output of every possible branch by means of the $\textsf{RecR}_1$/$\textsf{RecR}_2$ rules.

%We now prove the {\em only if} part and proceed by contraposition.
%Assume that there exists a relation ${\mathcal R}$ that is a
%k-bounded Asynchronous Subtyping relation such that $(T,S) \in {\mathcal R}$.

  We now prove the {\em only if} part and proceed by contraposition.
  We assume that $T \subtype_{\mathsf k} S$ and show that 
  there exist no $\Sigma', T', S'$,
  such that
 $\emptyset \vdash T \subtypeak S
  \rderiv^* \Sigma' \vdash T' \subtypeak S' \notrderiv$.
  If $T \subtype_{\mathsf k} S$ then also $T \subtype S$.
  So we can assume the existence of a relation ${\mathcal R}$ that is an
  orphan-message-free subtyping relation such that $(T,S) \in {\mathcal R}$.
We then use exactly the same proof as that of the second part (the {\em only if} part) of the proof
of Proposition \ref{prop:semidecidable} to establish a correspondance between judgements   $\Sigma'' \vdash T'' \subtypeak S''$,
such that
$\emptyset \vdash T \subtypeak S \rderiv_w^*
  \Sigma'' \vdash T'' \subtypeak S''$, and pairs in ${\mathcal R}$ (see the construction of the corresponding pair in the proof
of Proposition \ref{prop:semidecidable}).
Since ${\mathcal R}$ includes only pairs that satisfy the constraint
in item $2$ of ${\mathsf k}$-bounded Asynchronous Subtyping relation definition (Definition \ref{def:kbounded})
requiring context $\mathcal A$ to be ${\mathsf k}$-bounded; and since any judgment $\Sigma'' \vdash T'' \subtypeak S''$
such that $\emptyset \vdash T \subtypeak S \rderiv_w^* \Sigma'' \vdash T'' \subtypeak S''$
implies there is in ${\mathcal R}$ a corresponding pair $(T'',S''_1)$, with $S''_1$ differing from $S''$ just for the level of internal unfoldings, we have that reachable judgments $\Sigma'' \vdash T'' \subtypeak S''$ cannot be 
such that: $\Sigma'' \vdash T'' \subtypeak S'' \rderiv^* \Sigma''' \vdash T''' \subtypeak S'''$, by possibly applying $\textsf{RecR}_1$/$\textsf{RecR}_2$ rules,
and $\Sigma''' \vdash T''' \subtypeak S''' \notrderiv$ due to 
not satisfying the requirement about the input context $\mathcal A$ to be ${\mathsf k}$-bounded 
in the rule $\textsf{Out}$.
This because the difference in unfolding levels between $S''$ and $S''_1$ (inside judgment $\Sigma'' \vdash T'' \subtypeak S''$ and the corresponding pair $(T'', S''_1)$ in ${\mathcal R}$) is not significant: the ${\mathsf k}$-boundedness
of context $\mathcal A$  is established both in the rule $\textsf{Out}$ and in item $2$ of $\subtype_{\mathsf k}$ definition
after unfolding all recursions occurring before the first output of every possible branch.

This observation makes it possible to carry out the proof as in  Proposition~\ref{prop:semidecidable}, hence to show  that there exist no $\Sigma', T', S'$,
  such that
 $\emptyset \vdash T \subtypeak S
  \rderiv^* \Sigma' \vdash T' \subtypeak S' \notrderiv$.
\qed

\subsection{Proof of Proposition \ref{prop:outAntInfDecidable}}

We start by providing the proof of the first part of Proposition \ref{prop:outAntInfDecidable}, i.e. finiteness 
of $\reach T$ for single-out session types $T$, as a separate preliminary lemma, then we move to proving
the second part, i.e. decidability of $\antOutInf{T}$.

\begin{lemma}\label{lem:reachFinite}
Given a single-out session type $T$, $\reach T$ is finite.
\end{lemma}
\proof
We now define a finite set of session types $\finclose T$,
and then we prove that it satisfies all the constraints $1.,\dots,4.$
in Definition \ref{def:reach}.
Hence $\reach T \subseteq \finclose T$ by definition, from which finiteness of $\reach T$
follows.

It is not restrictive to assume that all the recursion variables of $T$
are distinct: let $\Tvar{x_1},\dots,\Tvar{x_n}$ be such variables.
We consider the rewriting variables $X_1,\dots,X_n$. 
Let $T_i$ be such that $\Trec {x_i}.T_i$ occurs in $T$;
let $T'$ be $T$ with $X_i$ that replaces $\Trec {x_i}.T_i$;
and similarly 
let $T'_i$ be $T_i$ with $X_j$ that replaces each occurrence of $\Trec {x_j}.T_j$ and $\Tvar{x_j}$.
We now consider the rewriting rules 
$X_i \rightarrow_i^1 T_i'$ and $X_i \rightarrow_i^2 \Tvar{x_i}$.
Given one of the above term $S$ containing rewriting variables, we denote with $\recclosure{S}$ the
session type obtained by repeated application of the rewriting rules in the following way:
if $X_i$ occurs inside a subterm $\Trec{x_i}.S'$ apply $\rightarrow_i^2$,
otherwise apply $\rightarrow_i^1$.
We now define another closure function on sets of terms $\mathcal S$: 
$\closure{\mathcal S}=\{S'|\mbox{$S'$ is a subterm of $S \in \mathcal S$}\}$.
Consider finally $\finclose{T}=\{\recclosure{S}|S \in \closure{\{T',T_1',\dots,T_n'\}}\}$.
We have that $\finclose{T}$ is finite and it satisfies all the constraints $1.,\dots,4.$
in Definition \ref{def:reach}. 
\qed

We now report some definitions and preliminary results used in the proof of Proposition~\ref{prop:outAntInfDecidable} concerning the second part about decidability of $\antOutInf{T}$ for single-out session types $T$.
We introduce the relation $\antEq{T}$: intuitively, $(T',T'') \in \antEq{T}$ if $T'$ and
$T''$ are terms in $\reach{T}$ capable of anticipating the same infinite sequence of
outputs. For instance, assuming that the following
$T' = \Trec t.\oplus\{l:\& \{l_1:  \oplus\{l':\Tvar t\}, l_2:  \oplus\{l':\Tvar t\}\}\}$
and 
$T'' = \Trec t.\& \{ l_1: \oplus \{l: \oplus\{l': \Tvar t\} \} \}$
belong to $\reach{T}$, for some session type $T$,
we have that $(T',T'') \in \antEq{T}$ because they can anticipate the sequence of
outputs $l$ and $l'$, indefinitely.

\begin{definition}\label{def:antEqT}
Let $T$ be a single-out session type.  A relation $\mathcal R$ over $\reach{T}$ is an
$\antEq{T}$ relation if $(T',T'') \in \mathcal R$ implies: there exist $l, \mathcal{A'}, \mathcal{A''}$ such that $\outunfold{T'} = 
\context {A'}{\Tselectsingle{l}{T'_i}}{i \in \{1,\ldots, n\}}$ and
$\outunfold{T''} = 
\context {A''}{\Tselectsingle{l}{T''_j}}{j \in \{1,\ldots, m\}}$, with $(T'_i,T''_j) \in \mathcal R$ for all $i \in \{1,\ldots, n\}$ and $j \in \{1,\ldots, m\}$.
%  with (W,S) in R per ogni W,S presi dall’unione dei T1_i e T2_j
%
We say that $T' \antEq{T} T''$ if there is an $\antEq{T}$ relation  $\mathcal R$ such that $(T',T'') \in \mathcal R$.
\end{definition}
%dim che unione di due antEq relations è ancora una antEq relation?

Notice that $\antEq{T}$ itself is an $\antEq{T}$ relation because, obviously, the union of two $\antEq{T}$ relations is an $\antEq{T}$ relation and $\reach{T}$ is finite. Moreover notice that, given a term $T' \in \reach{T}$, all terms $T'_i$ (with $i \in \{1,\ldots, n\}$) for which $\outunfold{T'} = \context {A'}{\Tselectsingle{l}{T'_i}}{i \in \{1,\ldots, n\}}$ are always 
such that $T'_i \in \reach{T}$ as well (because $\outunfold{T'}$ never unfolds recursions occurring inside terms $T'_i$).
Finally, notice that $\antEq{T}$ is %(effectively) 
decidable in that it is a relation over $\reach{T}$, which is a finite set.
%VA BENE ``EFFECTIVELY''???

\begin{definition}
$\antSet{T}$ is the field of $\antEq{T}$, that is the set of session types $T' \in \reach{T}$ such that there exists $T''$ with $(T',T'') \in \antEq{T}$ or $(T'',T') \in \antEq{T}$.
\end{definition}

\begin{lemma}\label{lemma:anteq}
$\antEq{T}$ is an equivalence relation on $\antSet{T}$.
\end{lemma}
\proof
The reflexive, symmetric and transitive closure of an $\antEq{T}$ relation is an $\antEq{T}$ relation, hence this holds true
for $\antEq{T}$ as well.
\qed

%\begin{definition}
%We say that $T' \extAntEq{T} T''$ if there exist $l, \mathcal{A'}, \mathcal{A''}$ such that 
%$\outunfold{T'} = 
%\context {A'}{\Tselectsingle{l}{T'_i}}{i \in \{1,\ldots, n\}}$ and
%$\outunfold{T''} = 
%\context {A''}{\Tselectsingle{l}{T''_j}}{j \in \{1,\ldots, m\}}$, with $T'_i \antEq{T} T''_j$ for all $i \in \{1,\ldots, n\}$ and $j \in \{1,\ldots, m\}$.
%
%Moreover, $\extAntSet{T}$ is the field of $\extAntEq{T}$.
%\end{definition}
%
%Notice that, all terms $T'_i$, with $i \in \{1,\ldots, n\}$ and $T''_j$, with $j \in \{1,\ldots, m\}$, are in $\antSet{T} \subseteq \reach{T}$. Moreover, notice that $\extAntEq{T}$ is obviously an equivalence relation on $\extAntSet{T}$.

\begin{lemma}\label{lemma:soundness1}
Let $T' \in \reach{T}$. We have that $\antOutInf{T'}$ if and only if $T' \in \antSet{T}$.
\end{lemma}
\proof
We prove the two implications separately, starting from the $\emph{if}$ part,
e.g. by assuming $T' \in \antSet{T}$. By Lemma \ref{lemma:anteq} we have $T' \antEq{T} T'$.
We now prove by induction on $m$ that for every $m$ there exists $l_{i_1} \cdots l_{i_m}$
such that $\antOut{T'}{l_{i_1}\cdots l_{i_{m}}}$ is defined.
If $m=1$ it is sufficient to consider $l_{i_1}=l$ where
$\outunfold{T'} = 
\context {A'}{\Tselectsingle{l}{T'_i}}{i \in \{1,\ldots, n\}}$
(with $\mathcal {A'}$ and $T'_i$ that exist by Definition \ref{def:antEqT}).
Consider now that $T''=\antOut{T'}{l_{i_1}\cdots l_{i_{m-1}}}$ is defined.
By Definition \ref{def:outAnt}, we have 
$T''=\context A{T_k}{k}$ with $\outunfold {\antOut{T}{l_{i_1}\cdots l_{i_{m-2}}}} =
               \context A{\Tselectsingle{l_{i_{m-1}}}{T_k}}{k}$.
As $T' \antEq{T} T'$, we can apply $m-1$ times Definition \ref{def:antEqT}
to conclude that $T_i \antEq{T} T_j$, for
every $i,j \in {1\dots k}$. This guarantees the existence of 
the input contexts $\mathcal{A}^k$, session types $T^k_r$, and
label $l$ such that such that $\outunfold{T_k}= \mathcal{A}^k[{\Tselectsingle{l}{T^k_r}}]^{r}$.
This implies that it is possible to define $\antOut{T''}{l}$ hence also
$\antOut{T'}{l_{i_1}\cdots l_{i_{m}}}$ by taking $l_{i_{m}} = l$.

We now move to the $\emph{only if}$ part assuming that 
%We know that 
there exists an infinite label sequence $l_{i_1} \cdots l_{i_n} \cdots$ such that, 
%CAMBIARE DEF IN MAIN PAPER!!!!?? (ANCHE LA i NON SEMBRA SERVIRE A NULLA!)
%for every $n \geq 0$ there exists $\gamma$, with $|\gamma|=n$, such that 
%$\antOut{T'}{\gamma}$ is defined.
for every $n$,  $\antOut{T'}{l_{i_1}\cdots l_{i_{n}}}$
is defined.  
Let $\mathcal R$ be the minimal relation such that $(T',T') \! \in \! \mathcal R$ and:
$\!\outunfold {\antOut{T'}{l_{i_1} \!\cdots l_{i_{n-1}}}}$ $ =
               \context A{\Tselectsingle{l_{i_{n}}}{T_k}}{k \in \{1 \dots m_n\}}$, for any $n \geq 1$,
implies  $\forall i,j \in \{1 \dots m_n\} \ldotp (T_i, T_j) \in \mathcal R$.
We now show that $\mathcal R$ above is an $\antEq{T}$ relation.
Considered any $(T'',T''')$ in $\mathcal R$, we have that there exists $h$, with $h \geq 1$, such that, for some 
$\mathcal A'$, $\mathcal A''$, we have:
 $\outunfold{T''} = 
\context {A'}{\Tselectsingle{l_{i_{h}}}{T'_i}}{i \in \{1,\ldots, m'\}}$ and
$\outunfold{T'''} = 
\context {A''}{\Tselectsingle{l_{i_{h}}}{T''_j}}{j \in \{1,\ldots, m''\}}$, with $(T'_i,T''_j) \in \mathcal R$ for all $i \in \{1,\ldots, m'\}$ and $j \in \{1,\ldots, m''\}$.
This holds, according to the definition of $\mathcal R$: for $(T'',T''')=(T',T')$ by taking $h=1$ and by observing that pairs $(T'_i,T''_j) \in \mathcal R$ because they are added to $\mathcal R$ in the case $n=1$; for any $(T'',T''')$ added to $\mathcal R$ in the case $n$, by taking $h=n+1$ and by observing that pairs $(T'_i,T''_j) \in \mathcal R$ because they are among the pairs that are added to $\mathcal R$ in the case $n+1$.
%Definisco \calR tale che per ogni \gamma t.c. ant(T',\gamma) è definito
%\calR mette in relazione ogni coppia (W1,W2) con W1 e W2 che riempiono 
%buchi di ant(T',\gamma). Tale \calr è una antEq_T in quanto data (W1,W2) in \calR,
%ho che possono anticipare la stessa l (quella di ant(T',\gamma l)) ed inoltre 
%i termini dopo tale anticipazione saranno in relazione per definizione di \calR.
%Ho quindi anche che T' \in antSet_T per anticipazione del primo elemento di \gamma.
%%(in def di \calR considero anche caso \gamma = \epsilon)
\qed

\restatePRO{prop:outAntInfDecidable}
\emph{Given a single-out session type $T$, $\reach{T}$ is finite and it is 
decidable whether $\antOutInf{T}$.} \\
\proof
Direct consequence of Lemmas \ref{lem:reachFinite}, Lemma \ref{lemma:soundness1}
and the finiteness of $\antSet{T}$.
\qed

\subsection{Proof of Theorem \ref{theo:termination}}

Theorem \ref{theo:termination} states that, for single-out session types $T$ and $S$, the algorithm provided 
by $T \subtypet S$ indeed terminates. The proof is based on characterizing terms $S''$ such that 
$(T'',S'') \in \Sigma'$ for any judgment $\Sigma' \vdash T' \subtypet S'$ reached by the algorithm.
In particular, we show that any such term $S''$ can be obtained by anticipating a sequence of output labels $\gamma$
in a term $R$ belonging to the finite set $\reach S$.  This preliminary result is proved by means of the following lemma and corollary.

Lemma \ref{lem:tuttiAntOut} states that given a type $S'$
at the right hand side of a judgement
reachable during the execution of our algorithm, i.e.
$\emptyset \vdash T \subtypet S
\rightarrow^* \Sigma' \vdash T' \subtypet S'$
for some initial types $T$ and $S$, the type $S'$ (and the types in $\reach {S'}$)
can be obtained from the initial type $S$ (or one of the types in $\reach {S}$)
by means of anticipation of a sequence of output labels, i.e. 
for all $Q \in \reach {S'}$
there exist $R \in \reach S$ and a sequence of labels $\gamma$ such that
$Q = \antOut{R}{\gamma}$. There is only one case in which this property is not
guaranteed to hold, namely, when the last applied rule in $\emptyset \vdash T \subtypet S
\rightarrow^* \Sigma' \vdash T' \subtypet S'$ is $\textsf{RecR}_2$.
As a counter-example, consider for instance
$$
\begin{array}{lll}
\emptyset & \vdash & \Trec t.\oplus \{l:\oplus \{l: \& \{l':\Tvar t\}\}\} \subtypet 
\Trec t.\& \{l': \oplus \{l:\Tvar t\}\}
\rightarrow^* \\
\Sigma' & \vdash & \oplus \{l: \& \{l':\Trec t.\oplus \{l:\oplus \{l: \& \{l':\Tvar t\}\}\}\}\}  \subtypet\\
& &  \& \{l': \& \{l': \oplus \{l: \Trec t.\& \{l': \oplus \{l:\Tvar t\}\}\}\}\}
\end{array}
$$
where $\rightarrow^*$ denotes application of the sequence of rules 
$\textsf{RecL}$, $\textsf{RecR}_2$, $\textsf{Out}$ and
$\textsf{RecR}_2$.
We have that the last type 
$\& \{l': \& \{l': \oplus \{l: \Trec t.\& \{l': \oplus \{l:\Tvar t\}\}\}\}\}$
is different from $\antOut{R}{\gamma}$, for every $R \in \reach {\Trec t.\& \{l': \oplus \{l:\Tvar t\}\}}$
and every sequence of labels $\gamma$. 

\begin{lemma}\label{lem:tuttiAntOut}
Consider two single-out session types $T$ and $S$.
Given a judgement
$\Sigma' \vdash T' \subtypet S'$ such that $\emptyset \vdash T \subtypet S
\rightarrow^* \Sigma' \vdash T' \subtypet S'$, in such a way that the final
rule applied is not $\textsf{RecR}_2$, we have that for all $Q \in \reach {S'}$
there exist $R \in \reach S$ and a sequence of labels $\gamma$ such that
$Q = \antOut{R}{\gamma}$.
\end{lemma}
\proof
By induction on the length of the sequence of rule applications 
$\emptyset \vdash T \subtypet S \rightarrow^* \Sigma' \vdash T' \subtypet S'$.
In the base case we have $S'=S$. Consider now $Q\in \reach{S'}$.
Obviously $Q = \antOut{Q}{\epsilon}$ with $Q \in \reach{S}$ because
$\reach{S}=\reach{S'}$.

In the inductive case we proceed by case analysis
on the last rule application 
$\Sigma'' \vdash T'' \subtypet S'' \rightarrow 
 \Sigma' \vdash T' \subtypet S'$. We have two possible cases:
\begin{itemize}
\item
We can apply the induction hypotheses on the judgement $\Sigma'' \vdash T'' \subtypet S''$.
Hence for all $Q'' \in \reach {S''}$
there exist $R \in \reach S$ and a sequence of labels $\gamma$ such that
$Q'' = \antOut{R}{\gamma}$. Consider now $Q \in \reach {S'}$.
We proceed by cases on the applied rule.

For the rules  $\textsf{In}$, $\textsf{RecR}_1$ and $\textsf{Out}$
with $\mathcal A= [\,]^1$ we have that $S' \in \reach{S''}$ hence
also $Q \in \reach {S''}$ because if $S' \in \reach{S''}$ then
$\reach{S'} \subseteq \reach{S''}$ by definition of $\reach{\_}$.

If the rule is $\textsf{Out}$ with $\mathcal A\neq [\,]^1$
we have that $S' = \antOut{R}{\gamma \cdot l}$ with $R \in \reach{S}$ and $\gamma$
such that $S''=\antOut{R}{\gamma}$ and $l$ is the label of the anticipated
output. We limit our analysis to the case in which
$Q \not \in \reach {S''}$ (in the other cases we can 
proceed as above). This happens if $Q$ is obtained by applying
rule $2.$ of Definition \ref{def:reach} to remove some but not all
the inputs in front of one of the output anticipated in $S''$.
Consider now the term $V$ being like $Q$ but without the 
$l$ output anticipation. Formally, $V$ is defined as follows. Denoted $S''$ with 
$\context {A}{ \Tselectindex l{S_k}{j}{J_k} }{k}$ we know that
%$\forall k \ldotp 
for all $k$
there exists a $j_k$  such that $l_{j_k} = l $.
This means that $S'$ is 
$\context {A}{ S_{k j_k}}{k}$. Hence, being $Q$ reachable from $S'$ by consuming some inputs of the 
input context $\mathcal A$ only, we have that there exists $\mathcal A'$ such that $Q$ is 
$\context {A'}{ S'_{h j'_h}}{h}$, where, 
considered the hole $k$ corresponding to the hole $h$,
%for each $h$ there is a corresponding index $k$ such 
we have that $j'_h = j_k$ and
$S'_{h j'_h}=S_{k j_k}$. Therefore, the previously mentioned term $V$ is $\context {A'}{ 
\Tselectindex l{S'_h}{j}{J_h}}{h}$, where, considered $k$ corresponding to $h$, we have that $J_h = J_k$.
%\in \{1,\ldots, n\}
We conclude by observing that $V \in \reach{S''}$, hence
there exist $R' \in \reach{S}$ and $\gamma'$ such that $V=\antOut{R'}{\gamma'}$.
But $Q=\antOut{R'}{\gamma'\cdot l}$, hence proving the thesis.

\item We cannot apply the induction hypotheses on the judgement $\Sigma'' \vdash T'' \subtypet S''$
because the rule used to obtain $\Sigma'' \vdash T'' \subtypet S''$ is $\textsf{RecR}_2$.
As $\textsf{RecR}_2$ cannot be applied in sequence, it is surely possible
to apply the induction hypothesis on the previous judgement $\Sigma''' \vdash T''' \subtypet S'''$
such that $\Sigma''' \vdash T''' \subtypet S''' \rightarrow 
 \Sigma'' \vdash T'' \subtypet S''$.
Then we have that for all $Q''' \in \reach{S'''}$
we have $Q''' = \antOut{R}{\gamma}$ with $R \in \reach{S}$
and a sequence of labels $\gamma$.
We also have that the rule applied in 
$\Sigma'' \vdash T'' \subtypet S'' \rightarrow 
 \Sigma' \vdash T' \subtypet S'$ is $\textsf{Out}$, which is the 
only rule that can applied after $\textsf{RecR}_2$. Let $l$ be the
label of the output involved in the application of the $\textsf{Out}$ rule.
Consider now $Q \in \reach {S'}$.
% and we show that $Q = \antOut{R}{\gamma}$
%for some $R \in \reach S$. 
We consider two possible cases:
\begin{itemize}
\item $Q$ is obtained from $S'$ by consuming inputs present
in the input context $\mathcal A$ used in the last
application of the rule $\textsf{Out}$. Consider now $Q'''$ obtained
from $S'''$ by consuming the same inputs and performing
the needed unfoldings. Obviously $Q''' \in \reach{S'''}$: hence, by 
induction hypothesis, $Q''' = \antOut{R}{\gamma}$ with $R \in \reach{S}$.
We have $Q = \antOut{R}{\gamma\cdot l}$ hence proving the thesis.

\item $Q$ is obtained from $S'$ by consuming strictly more than a sequence
of inputs present in the input context $\mathcal A$ used in the last
application of the rule $\textsf{Out}$. This means that $Q \in \reach{W}$
where $W$ is a term starting with an output that populates one of the holes
of $\mathcal A$ in $S''$. But the terms starting with an output that can 
occur in $S''$, assuming
$\emptyset \vdash T \subtypet S \rightarrow^* \Sigma'' \vdash T'' \subtypet S''$,
are already in $\reach S$. In fact the rules do not perform transformations
under outputs, excluding those strictly performed by top level unfoldings.
Hence $W \in \reach{S}$, which implies $Q \in \reach S$ from which the thesis
trivially follows (because $Q = \antOut{Q}{\epsilon}$).  
\qed
\end{itemize}
\end{itemize}

\begin{corollary}\label{Gigio}
Consider two single-out session types $T$ and $S$.
Given a judgement
$\Sigma' \vdash T' \subtypet S'$ such that $\emptyset \vdash T \subtypet S
\rightarrow^* \Sigma' \vdash T' \subtypet S'$
and a pair $(T'',S'') \in \Sigma'$, we have that $S''=\antOut{R}{\gamma}$
for some $R \in \reach S$ and a sequence of labels $\gamma$.
\end{corollary}
\proof
Let $(T'',S'') \in \Sigma'$ and
consider the sequence of rule applications $\emptyset \vdash T \subtypet S
\rightarrow^* \Sigma'' \vdash T'' \subtypet S''$
that preceeds the application of the rule that introduces $(T'',S'')$ in $\Sigma'$.
Such rule must be one of $\textsf{RecL}$, $\textsf{RecR}_1$ or $\textsf{RecR}_2$:
hence on the judgement $\Sigma'' \vdash T'' \subtypet S''$ it is possible
to apply one of these three rules.
Since after the application of a rule $\textsf{RecR}_2$ the uniqe applicable rule is 
$\textsf{Out}$, we have the guarantee that the last rule in $\emptyset \vdash T \subtypet S
\rightarrow^* \Sigma'' \vdash T'' \subtypet S''$ is not $\textsf{RecR}_2$.
%Let $(T'',S'') \in \Sigma'$. This pair has been introduced by application 
%of one of the rules $\textsf{RecL}$, $\textsf{RecR}_1$ or $\textsf{RecR}_2$.
%But before the application of these rules it is not possible to apply rule 
%$\textsf{RecR}_2$, because after such rule only $\textsf{Out}$ can be applied.
%So the pair $(T'',S'')$ corresponds to a sequence of rule applications
%$\emptyset \vdash T \subtypet S
%\rightarrow^* \Sigma'' \vdash T'' \subtypet S''$ in which
%$\textsf{RecR}_2$ is not the last applied rule.
Hence it is possible to apply Lemma \ref{lem:tuttiAntOut}, from which
the thesis directly follows.
% from  Lemma \ref{lem:tuttiAntOut}.
\qed

%\proof
%Let $(T'',S'') \in \Sigma'$. This pair has been introduced by application 
%of one of the rules $\textsf{RecL}$, $\textsf{RecR}_1$ or $\textsf{RecR}_2$.
%But before the application of these rules it is not possible to apply rule 
%$\textsf{RecR}_2$, because after such rule only $\textsf{Out}$ can be applied.
%So the pair $(T'',S'')$ corresponds to a sequence of rule applications
%$\emptyset \vdash T \subtypet S
%\rightarrow^* \Sigma'' \vdash T'' \subtypet S''$ in which
%$\textsf{RecR}_2$ is not the last applied rule.
%The thesis directly follows from  Lemma \ref{lem:tuttiAntOut}.
%\qed

\medskip

\restateTHM{theo:termination}
\emph{Given two single-out session types $T$ and $S$, the algorithm applied to the 
initial judgement $\emptyset \vdash T \subtypet S$ terminates.} \\
\proof
Assume by contraposition that there exists single-out session types $T$ and $S$
such that the algorithm applied to the 
initial judgement $\emptyset \vdash T \subtypet S$ does not terminate.
This means that there exists an infinite sequence of rule applications 
$\emptyset \vdash T \subtypet S \rightarrow \Sigma_1 \vdash T_1 \subtypet S_1
\rightarrow^* \Sigma_i \vdash T_i \subtypet S_i \rightarrow^*$.
Within this infinite sequence, there are infinitely many applications
of the unfolding rules $\textsf{RecL}$, $\textsf{RecR}_1$ or $\textsf{RecR}_2$,
that implies the existence of infinitely many distinct pairs $(T_j,S_j)$ that are
introduced in the environment (assuming that $j$ ranges over the instances of application
of such rules). All these pairs are distinct, otherwise the precedence of the 
$\textsf{Asmp}$ rule would have blocked the algorithm.
It is obvious that the distinct r.h.s. $T_j$ are 
finitely many, because every $T_j \in \reach{T}$, which is a finite set.
On the contrary, the distinct $S_j$ are infinitely many, but Corollary \ref{Gigio}
guarantees that for each of them, there exists $S'_j \in \reach{S}$ and a sequence 
of labels $\gamma_j$ such that $S_j = \antOut{S'_j}{\gamma_j}$.
%Within this infinite sequence of pairs $(T_j,S_j)$ d

Due to the finiteness of the possible $T_j$ and $S'_j$, there exists
$T''$ and $S''$ such that there exists an infinite subsequence
of $(T_{j_1},S_{j_1}),(T_{j_2},S_{j_2}),\dots,(T_{j_k},S_{j_k}),\dots$
such that $T_{j_i}=T''$ and $S_{j_i}=\antOut{S''}{\gamma_{j_i}}$.
It is not restrictive to consider $j_h<j_{h+1}$ for every $h$.
The presence of infinitely many distinct $\gamma_{j_i}$ 
for which $\antOut{S''}{\gamma_{j_i}}$ is defined, guarantees $\antOutInf{S''}$.
Moreover, this guarantees also 
the possibility to define an infinite subsequence 
$(T_{j_{l_1}},S_{j_{l_1}}),(T_{j_{l_2}},S_{j_{l_2}}),\dots,$
$(T_{j_{l_k}},S_{j_{l_k}}),\dots$
such that $|\gamma_{j_{l_i}}\!|\!<\!|\gamma_{j_{l_{i+1}}}\!|$.
We now consider the leaf sets $\leafset{S_{j_{l_i}}\!}$. These sets are
defined on a finite domain because the subterms of such types starting
with a recursive definition or an output, and preceded by inputs only, 
are taken from $\reach S$. This because the algorithm does not 
apply transformations under recursive definitions or outputs, 
excluding the effect of the standard top level unfolding of previous
recursive definitions, which is considered in the definition of $\reach S$.
Hence there are only finitely many distinct $\leafset{S_{j_{l_i}}}$,
that guarantees the existence of $v<w$ such that 
$\leafset{S_{j_{l_v}}}=\leafset{S_{j_{l_w}}}$.
Consider now the judgement $\Sigma_{j_{l_w}} \vdash T_{j_{l_w}} \subtypet S_{j_{l_w}}$.
We know that $(T_{j_{l_v}},S_{j_{l_v}}) \in \Sigma_{j_w}$, $T_{j_{l_v}}=T_{j_{l_w}}$, 
$S_{j_{l_v}}=\antOut{S''}{\gamma_{j_{l_v}}}$, $S_{j_{l_w}}=\antOut{S''}{\gamma_{j_{l_w}}}$,
$S'' \in \reach{S}$, and $|\gamma_{j_{l_v}}|<|\gamma_{j_{l_v}}|$. Hence it is possible 
to apply to such judgement the rule $\textsf{Asmp}_2$.
As $\textsf{Asmp}_2$ has priority, it should be applied on this judgement
thus blocking the sequence of rule applications. But this contradicts the initial
assumption of non termination of the algorithm.
\qed

% %\section{Dio c'\'e. Scemo! :)}
 %\begin{theorem}\label{theo:soundness}
%\end{theorem}

\subsection{Proof of Theorem \ref{theo:soundness}}

The soundness Theorem \ref{theo:soundness} states that the $\subtypet$ algorithm reaches $\notrderiv$
if and only if the $\subtypea$ procedure does so. This is proved in the following by resorting to some preliminary definitions and results.

In Definition \ref{def:antEqT} we have defined the relation $\antEq{T}$
among types that have the same infinite
sequence of outputs that can be anticipated. But, in order
to have a decidable relation, we had to limit to types belonging to the
set $\reach{T}$. Now, we define a more general relation $\extAntEq{T}$
applicable to types having (once unfolded) the following shape: any possible input context 
with holes filled with single outputs having a continuation belonging to 
$\reach{T}$. This extension of the $\antEq{T}$
is necessary because the execution of the subtyping algorithm can generate
new terms (as a consequence of output anticipations) having this specific 
shape.

\begin{definition}
Let $T'$, $T''$ be single-out session types.
We say that $T' \!\!\; \extAntEq{T} T''\!$ if there exist $l, \mathcal{A'}\!, \mathcal{A''}\!$ such that 
$\outunfold{T'} \!=\! 
\context {A'}{\Tselectsingle{l}{T'_i}}{i \in \{\!1,\ldots, n\!\!\; \}}$ and
$\outunfold{T''} \!= \!
\context {A''}{\Tselectsingle{l}{T''_j}}{j \in \{\!1,\ldots, m\!\!\; \}}$, with $T'_i \antEq{T} T''_j$ for all $i \in \{1,\ldots, n\}$ and $j \in \{1,\ldots, m\}$.

Moreover, $\extAntSet{T}$ is the field of $\extAntEq{T}$.
\end{definition}

Notice that, all terms $T'_i$, with $i \in \{1,\ldots, n\}$ and $T''_j$, with $j \in \{1,\ldots, m\}$, are in $\antSet{T} \subseteq \reach{T}$. Moreover, notice that $\extAntEq{T}$ is obviously an equivalence relation on $\extAntSet{T}$.

\begin{lemma}\label{lemma:soundness2}
Let $T' \!\in\! \antSet{T}$ and $T''\!=\!\antOut{T'}{\gamma}$ for some $\gamma$. We have that $T'' \in \extAntSet{T}$.
\end{lemma}
\proof
We have to show that there exist $l, \mathcal{A}$ for which we have 
$\outunfold{\antOut{T'\!}{\!\!\;\gamma}}$ 
$ = \context {A}{\Tselectsingle{l}{T_i}}{i \in \{1,\ldots, m\}}$, with $T_i \antEq{T} T_j$ for all $i,j \in \{1,\ldots, m\}$.
We denote $\gamma l = l_{i_1} \cdots l_{i_h}$, with $h \geq 1$. 
For any $n$, with $1 \leq n \leq h$, considered $\mathcal{A'}$ and terms $T_k$ with $k \in \{1 \dots m_n\}$ such that
$\outunfold {\antOut{T'}{l_{i_1}\cdots l_{i_{n-1}}}} =
               \context {A'}{\Tselectsingle{l_{i_{n}}}{T_k}}{k \in \{1 \dots m_n\}}$,
we have that $\forall i,j \in \{1 \dots m_n\} \ldotp T_i \antEq{T} T_j$.
This is easily shown by induction on $n$, applying the definition of $\antEq{T}$ (the base case is directly derived from
$T' \antEq{T} T'$). The case $n=h$ yields the desired result.
%Anticipare \gamma coincide con fare la stessa quantità di anticipazioni da tutte le foglie di 
%smartunfold(T') che sono inizialmente tutte pairwise in antEq_T per riflessività su antSet_T. 
%Le foglie ottenute dopo aver anticipato \gamma (quindi le foglie di T''), sono pairwise in 
%antEq_T per Proposition sopra.
%% IN REALTA' SI DEVE FARE PER INDUZIONE SULLA LUNGHEZZA DI GAMMA E LA PROVA 
%%SI FA CONSIDERANDO IL FATTO CHE LE POSSIBILI RELAZIONI FRA BUCHI SONO LE POSSIBILI
%%RELAZIONI FRA BUCHI DI BUCHI
\qed

\begin{lemma}\label{lemma:soundness3}
Let $T'\!,T'' \!\!\!\; \in \!\!\!\; \extAntSet{T}\!\!$ and $\leafset{T'}\!=\!\leafset{T''}$. 
We have that $T' \extAntEq{T} T''$.
\end{lemma}
\proof 
It is easy to see that $\leafset{T'}\!=\!\leafset{T''}$ implies $\leafset{\outunfold{T'}}\!=\!\leafset{\outunfold{T''}}$. This because $\outunfold{}$ causes a leaf $T'''$ belonging to both $\leafset{T'}$ and $\leafset{T''}$ to yield the same new set of leaves $\leafset{T'''}$ in both $T'$ and $T''$.
By definition of $\extAntSet{T}$ we have that exist $l', \mathcal{A'}$ such that 
$\outunfold{T'} = 
\context {A'}{\Tselectsingle{l'}{T'_i}}{i \in \{1,\ldots, n\}}$, with $T'_i \antEq{T} T'_j$ for all $i,j \in \{1,\ldots, n\}$.
Similarly, there exist $l'', \mathcal{A''}$ such that 
$\outunfold{T''} = 
\context {A''}{\Tselectsingle{l''}{T''_j}}{j \in \{1,\ldots, m\}}$, with $T''_i \antEq{T} T''_j$ for all $i,j \!\in\! \{1,\ldots, m\}$.
From the fact that $\leafset{\outunfold{T'}}$ $=\leafset{\outunfold{T''}}$ we have that $l'=l''$ and that: for all 
$T'_i$, with $i \in \{1,\ldots, n\}$, there exists $T''_j$, with $j \in \{1,\ldots, m\}$, such that $T'_i = T''_j$; and, vice versa,
for all $T''_j$, with $j \in \{1,\ldots, m\}$, there exists $T'_i$, with $i \in \{1,\ldots, n\}$, such that $T''_j = T'_i$.
Therefore we conclude that $T' \extAntEq{T} T''$.
%Ho che leafSet(smartunfold(T')) = leafSet(smartunfold(T'')).
%Per definizione di extAntSet_T ho che per ogni coppia di foglie dello smartunfold hanno la 
%stessa 
%etichetta l e le continuazioni sono tutte in antEq_T. Avendo T' ed T'' le stesse foglie, questa 
%proprietà vale sull’unione delle loro foglie (che comunque è uguale all’insieme delle foglie 
%di uno) 
%quindi banalmente 
%T' extAntEq_T T''.
\qed

In order to prove Theorem \ref{theo:soundness} we also need to consider a simplified subtyping procedure, whose judgments are denoted by $\subtypesa\!$, and the notions of {\it IO steps}, {\it blocking judgments} and {\it blocking paths}.

\myparagraph{Simplified Subtyping Procedure.}
%We now consider a {\it new subtyping procedure}, 
We here denote by $\subtypesa$ the judgements of the subtyping procedure that
is defined exactly as our procedure (defined in Section \ref{subsec:subproc} and based on applications of the rules therein over judgments of the form $\Sigma \vdash T \subtypea S$) with the only difference that the $\textsf{Asmp}$ rule is removed
(i.e. the subtyping procedure whose transitions were denoted by $\rderiv_{noAsmp}$ in the proof of Proposition \ref{prop:semidecidable}). Since, in the absence of the $\textsf{Asmp}$ rule
the content of environment $\Sigma$ is never accessed for reading, it has no actual effect on the procedure (on rule applications) and can be removed as well, together with updates on such environment made by the rules. As a consequence we will denote $\subtypesa$ judgments just by $\vdash T \subtypesa S$ for some $T$ and $S$.
Here, differently from the $\rderiv_{noAsmp}$ notation used in the proof of  Proposition \ref{prop:semidecidable}, since we adopt a new notation for judgements, we will simply use:
$\vdash T \subtypesa S \rderiv \;\vdash T' \subtypesa S'$
to denote that the latter can be obtained from the former by one rule
application. Finally, as usual, $\vdash T \subtypesa S \notrderiv$ denotes
that there is no rule that can be applied to the judgement
$\vdash T \subtypesa S$.

\begin{definition}
A \emph{blocking judgment} $\vdash T \subtypesa S$, denoted by  $\vdash T \subtypesa S \blocking$, is 
a judgment such that, for some $T', S'$ we have: $\vdash T \subtypesa S \rderiv^* \;\vdash T' \subtypesa S' \notrderiv$ by applying rules $\textsf{RecL}$, $\textsf{RecR}_1$ and $\textsf{RecR}_2$ only.
%A \emph{terminating judgment} $\Sigma \vdash T \subtypea S$, denoted by  $\Sigma \vdash T \subtypea S \terminating$, is 
%a judgment such that, for some $\Sigma', T', S'$ we have: $\Sigma \vdash T \subtypea S \rderiv^* \Sigma' \vdash T' \subtypea S' \rderiv$ by applying rules $\textsf RecL$, $\textsf RecR1$ and $\textsf RecR2$ only.
\end{definition}

\begin{definition}
An \emph{IO step} $a$, denoted by $\iostep{a}$, with $a \in \{ \&_l, \oplus_l \mid l \in L \}$
is a sequence of $\subtypesa$ rule applications $\rderiv^*$ such that the last applied rule is an $\textsf{In}$  (in the case $a=\&_l$, where $l$ is the input label singling out which of the rule premises we consider),
or an $\textsf{Out}$ rule (in the case $a=\oplus_l$, where $l$ is the output label singling out which of the rule premises we consider) and all other rule applications concern $\textsf{RecL}$, $\textsf{RecR}_1$ and $\textsf{RecR}_2$ rules only.
\end{definition}

\begin{definition}
$a_1 \dots a_n$, with $n \geq 0$, is a \emph{blocking path} for judgment $\vdash T \subtypesa S$ if there exist $T', S'$ such that $\vdash T \subtypesa S \iostep{a_1} \dots \iostep{a_n} \;\vdash T' \subtypesa S' \blocking$ (where $T'=T$ and $S'=S$ in the case $n=0$).
%$a_1 \dots a_n$, with $n \geq 0$, is a \emph{terminating path} for judgment $\Sigma \vdash T \subtypea S$ if there exist $\Sigma', T', S'$ such that $\Sigma \vdash T \subtypea S \iostep{a_1} \dots \iostep{a_n} \Sigma' \vdash T' \subtypea S' \terminating$ (where $\Sigma'=\Sigma$, $T'=T$ and $S'=S$ in the case $n=0$).
\end{definition}

\begin{lemma}\label{lemma:soundnessmain}
Let $S \!\in\! \reach{Z}$ and $\, \vdash T \subtypesa \antOut{S}{\gamma}$, $\vdash T \subtypesa\!\!$ $\antOut{S}{\beta}$ be such that: $|\gamma| \!<\! |\beta|$ and $\antOut{S}{\beta} \extAntEq{Z} \antOut{S}{\gamma}$.
If $a_1\dots a_n$, with $n \geq 0$, is a blocking path for $\, \vdash T \subtypesa \antOut{S}{\beta}$ then
there exists an $m$ long prefix of $a_1 \dots a_n$, with $0 \leq m \leq n$, that is a blocking path for $\vdash T \subtypesa \antOut{S}{\gamma}$.
\end{lemma}
\proof
%It is obvious in the case $|\gamma| = |\beta|$.
%We now prove that it holds in the case $|\gamma| < |\beta|$ 

The proof is by induction on $n \geq 0$.

We start by proving the base case $n = 0$. 
That is $\vdash T \subtypesa \antOut{S}{\gamma} \blocking$, i.e. for some $T', S'$ we have: $\vdash T \subtypesa \antOut{S}{\gamma} \rderiv^* \;\vdash T' \subtypesa S' \notrderiv$ by applying rules $\textsf{RecL}$, $\textsf {RecR}_1$ and $\textsf {RecR}_2$ only.

We first observe that $\vdash T \subtypesa \antOut{S}{\gamma} \iostep{a}$ is not possible for any $a \in \{ \&_l, \oplus_l \mid l \in L \}$. This because: if we had $\vdash T \subtypesa \antOut{S}{\gamma} \iostep{\&_l}$ for some $l \in L$, then $\antOut{S}{\beta} = \Tbranch{l}{T}$ with $l = l_i$ for some $i \in I$, hence we would have that also $\vdash T \subtypesa \antOut{S}{\beta} \iostep{\&_l}$; and 
if we had $\vdash T \subtypesa \antOut{S}{\gamma} \iostep{\oplus_l}$ for some $l \in L$, then,
since $\antOut{S}{\beta} \extAntEq{Z}\!$ $\antOut{S}{\gamma}$, we would have that also $\vdash T \subtypesa \antOut{S}{\beta} \iostep{\oplus_l}$.

Therefore, given that it is not possible that $\vdash T \subtypesa \antOut{S}{\gamma} \rderiv^* \;\vdash \Tend  \subtypesa \Tend$ by applying rules $\textsf{RecL}$, $\textsf{RecR}_1$ and $\textsf{RecR}_2$ only (because otherwise  $\antOut{S}{\beta}$ would not be defined), we conclude  $\vdash T \subtypesa \antOut{S}{\gamma} \blocking$ (notice that the number of times
a  $\textsf{RecL}$, $\textsf{RecR}_1$ or $\textsf{RecR}_2$ is applicable to a judgment is finite because we do not have unguarded recursion and $\textsf {RecR}_2$ cannot be consecutively applied for more than one time).

We now consider the induction case for blocking path $a_1 \dots a_n$ of length $n \geq 1$.

We first consider the case $a_1=\&_l$ for some $l \in L$.
Given that $\antOut{S}{\beta}$ is defined and that $\vdash T \subtypesa \antOut{S}{\beta} \iostep{\&_l}$, we deduce
that $\antOut{S}{\gamma}$ is: either $\Tselectsingle{l'}{T'}$ (possibly preceded by some recursion operators),  for some $l',T'$; 
%IN REALTA' l' SARA IL PRIMO OUTPUT NELLA DIFFERENZA TRA \beta e \gamma
or $\Tbranch{l}{T}$ (possibly preceded by some recursion operators), for some terms $T_i$
and labels $l_i$ such that $l = l_i$ for some $i \in I$.
In the first case we have $\vdash T \subtypesa \antOut{S}{\gamma} \blocking$, hence the the lemma trivially holds; in the second case
we have $\vdash T \subtypesa \antOut{S}{\gamma} \iostep{\&_l}$ and we proceed with the proof.
We have that there exist $T'$, $S'$, $\sigma$ such that
$\vdash T \subtypesa \antOut{S}{\gamma} \iostep{\&_l} T' \subtypesa \antOut{S'}{\gamma'}$
and
$\vdash T \subtypesa \antOut{S}{\beta} \iostep{\&_l} T' \subtypesa \antOut{S'}{\beta'}$, with $\gamma= \sigma \gamma'$ and $\beta= \sigma \beta'$. In particular $S'$ is obtained from $S$ by removing all its initial (single-)outputs (and intertwined recursions, that are unfolded) until the first input $\Tbranch{l}{T}$ is reached, which is also removed, thus yielding $S' = T_i$ for the $i \in I$ such that $l = l_i$. This corresponds, in the definition of $\reach{Z}$ (Definition \ref{def:reach}), to repeatedly applying, starting from $S \in \reach{Z}$, rules $3$ and $4$ and finally rule $2$, thus yielding $S' \in \reach{Z}$. 
Notice that $\sigma$ is the sequence of labels of the initial outputs that were removed during this procedure
and that, obviously, $|\gamma’| < |\beta’|$. 

Now, in order to be able to apply the induction hypothesis we have also to show that $\antOut{S}{\beta'} \extAntEq{Z} \antOut{S}{\gamma'}$. 
We observe that $\antOut{S}{\gamma'}$ $\extAntEq{Z} \antOut{S}{\gamma}$. This holds because
$\antOut{S}{\gamma}$ is a $\Tbranch{l}{T}$ term, with $l = l_i$ for some $i \in I$, possibly preceded by some recursion operators, and from the following observations: obviously, for any $t,T''$, it holds $\Trec t.T'' \extAntEq{Z} T''\{\Trec t.T''/\Tvar t\}$; and $\leafset{T_i} \subseteq \leafset{\Tbranch{l}{T}}$. In the same way, we have $\antOut{S}{\beta'} \extAntEq{Z} \antOut{S}{\beta}$.

It is therefore possible to apply the induction hypothesis to $T' \!\!\!\;\subtypesa\! \antOut{S'\!}{\gamma'}$ and
$T' \subtypesa \antOut{S'}{\beta'}$ that possesses the shorter blocking path $a_2\dots a_n$.

Finally, we consider the case $a_1\!=\!\oplus_l$ for some $l \!\in\! L$.
Since $\vdash T \!\subtypesa\! \antOut{S}{\beta}$ $\iostep{\oplus_l}$ and
$\antOut{S}{\beta} \extAntEq{Z} \antOut{S}{\gamma}$, we have that also $\vdash T \subtypesa\!\!\!$ $\antOut{S}{\gamma} \iostep{\oplus_l}$.
In particular, we have that there exists $T'$ such that
$\vdash T \subtypesa \antOut{S}{\gamma} \iostep{\oplus_l} T' \subtypesa \antOut{S}{\gamma l}$
and
$\vdash T \subtypesa \antOut{S}{\beta} \iostep{\oplus_l} T'$ $\subtypesa \antOut{S}{\beta l}$, where, obviously, $|\gamma l| < |\beta l|$.
Moreover, since $\antOut{S}{\beta}$ $\extAntEq{Z} \antOut{S}{\gamma}$ it is immediate to show (by applying the definitions of $\mathsf{antOut}$, $\extAntEq{}$ and $\antEq{}$) that also
$\antOut{S}{\beta l} \extAntEq{Z} \antOut{S}{\gamma l}$.

It is therefore possible to apply the induction hypothesis to $T' \!\!\!\;\subtypesa\! \antOut{S}{\!\!\;\gamma l}$ and
$T' \subtypesa \antOut{S}{\beta l}$ that possesses the shorter blocking path $a_2\dots a_n$.
\qed\\

\restateTHM{theo:soundness} 
\emph{Given two single-out session types $T$ and $S$, we have that 
%for every $\Sigma', T', S'$,
there exist $\Sigma', T', S'$
such that
$\emptyset \vdash T \subtypea S
  \rderiv^* \Sigma' \vdash T' \subtypea S' \notrderiv$
if and only if
there exist $\Sigma'', T'', S''$
such that
$\emptyset \vdash T \subtypet S
  \rderiv^* \Sigma'' \vdash T'' \subtypet S'' \notrderiv$. }

\noindent
\proof
We consider the two implications separately starting from the
\emph{if} part.  
Assume that
$\emptyset \vdash T \subtypet S \rderiv^* \Sigma'' \vdash T''
\subtypet S'' \notrderiv$.
In this sequence of rule applications, the new rule $\textsf{Asmp}2$
is never used otherwise the sequence would terminate
successfully by applying such a rule. Hence, by applying the same
sequence of rules, we have
$\emptyset \vdash T \subtypea S \rderiv^* \Sigma' \vdash T' \subtypea
S'$
with $T''=T'$, $S''=S'$ and $\Sigma''=\Sigma'$.
We have that $\Sigma' \vdash T' \subtypea S' \notrderiv$, otherwise if
a rule could be applied to this judgement, the same rule could be
applied also to $\Sigma'' \vdash T'' \subtypet S''$ thus contradicting
the assumption $\Sigma'' \vdash T'' \subtypet S'' \notrderiv$.

We now move to the \emph{only if} part. Assume that
$\emptyset \vdash T \subtypea S \rderiv^* \Sigma' \vdash T' \subtypea S' \notrderiv$ and that,
by contradiction, $\emptyset \vdash T \subtypet S
  \rderiv^* \Sigma'' \vdash T'' \subtypet S'' \notrderiv$ does not hold.

From $\emptyset \vdash T \subtypea S \rderiv^* \Sigma' \vdash T'
\subtypea S' \notrderiv$ (since in this sequence of rule applications the $\textsf{Asmp}$ rule
is never used, otherwise the sequence would  terminate
successfully by applying such a rule), by applying the same
sequence of rules, we have
$\vdash T \subtypesa S \rderiv^* \;\vdash T' \subtypesa
S'$.

We now observe that, since we assumed (by contradiction) that we do not get the error when using the $\subtypet$ procedure,
there must exist at least a triple $\Sigma''', T''', S'''$ such that:
$\emptyset \vdash T \subtypet S
  \rderiv^* \Sigma''' \vdash T''' \subtypet S'''$ (and correspondingly 
$\vdash T \subtypesa S
  \rderiv^* \;\vdash T''' \subtypesa S'''$ because the $\textsf{Asmp}$ and $\textsf{Asmp}2$ rules, that would have led to successful termination, cannot have been applied),
$\Sigma''' \vdash T''' \subtypet S''' $ successfully terminates by applying the $\textsf{Asmp}$ or
$\textsf{Asmp}2$ rule, and $\vdash T''' \subtypesa S'''$ has a blocking path.

Let us now consider one of such triples $\Sigma''', T''', S'''$ (possessing the above stated properties) that has a blocking path of minimal length, i.e. there is no other $\Sigma''', T''', S'''$ triple
of the kind above such that $\vdash T''' \subtypesa S'''$ has a shorter blocking path. Let $a_1 \dots a_n$ be such a path.
%Notice that a triple of this kind exists because of the existence of the above triple $\Sigma''', T''', S'''$.
Since the $\textsf{Asmp}$ or $\textsf{Asmp}2$ rule is applied to $\Sigma''' \vdash T''' \subtypet S'''$, we have $S''' = \antOut{S}{\beta}$ (in the case of $\textsf{Asmp}$ this is obtained by Corollary \ref{Gigio}).

We now consider $\gamma$ such that  $(T''',\antOut{S}{\gamma}) \in \Sigma'''$ %, with $|gamma| < |beta|$% 
was used in the premise of $\textsf{Asmp}$ or $\textsf{Asmp}2$ rule: $\gamma= \beta$ in the case of the
$\textsf{Asmp}$ rule, $|\gamma| < |\beta|$ in the case of the $\textsf{Asmp}2$ rule.
Moreover, let us also consider $\Sigma_\gamma$ to be the environment such that  
$\emptyset \vdash T \subtypet S \rderiv^* \Sigma_\gamma \vdash T''' \subtypet \antOut{S}{\gamma}$, where
$\Sigma_\gamma \vdash T''' \subtypet \antOut{S}{\gamma}$ is the judgment to which the rule that caused $(T''',\antOut{S}{\gamma})$ to be inserted in the environment was applied.

We now observe that there exists a $m$ long prefix of $a_1 \dots a_n$, with $0 \leq m \leq n$, that is a blocking path for $\vdash T''' \subtypesa \antOut{S}{\gamma}$.
This is obvious in the case $\gamma = \beta$; it is due to Lemma \ref{lemma:soundnessmain} in the case $|\gamma| < |\beta|$: we obtain $\antOut{S}{\beta} \extAntEq{Z} \antOut{S}{\gamma}$ as needed by such a Lemma from the statements in the premise of rule $\textsf{Asmp}2$ and by applying Lemmas \ref{lemma:soundness1}, \ref{lemma:soundness2} and \ref{lemma:soundness3}.

Since we assumed (by contradiction) that $\emptyset \vdash T \subtypet S
\rderiv^* \Sigma'' \vdash T'' \subtypet S'' \notrderiv$ does not hold, this would be possible only if
%either there is a terminating path for $\Sigma'''' \vdash T'''' \subtypea \antOut{S}{\gamma}$ and the same 
%sequence of rule applications can be performed also from $\Sigma'''' \vdash T'''' \subtypet \antOut{S}{\gamma}$; or
there existed a triple $\Sigma'''', T'''', S''''$ such that: there is a sequence of rule applications $\Sigma_\gamma \vdash T''' \subtypet \antOut{S}{\gamma} \rderiv^* \Sigma'''' \vdash T'''' \subtypet S''''$ that is a prefix of the sequence of rule applications of the blocking path for $\vdash T''' \subtypesa \antOut{S}{\gamma}$; and $\Sigma'''' \vdash T'''' \subtypet S''''$ successfully terminates by applying the $\textsf{Asmp}$ or $\textsf{Asmp}2$ rule. Notice that such a sequence  $\Sigma_\gamma \vdash T''' \subtypet \antOut{S}{\gamma} \rderiv^* \Sigma'''' \vdash T'''' \subtypet S''''$ should necessarily include the application of, at least, an $\textsf{In}$ rule (causing the algorithm to branch), because otherwise (given that $\Sigma_\gamma \vdash T''' \subtypet \antOut{S}{\gamma} \rderiv^* \Sigma''' \vdash T''' \subtypet \antOut{S}{\beta}$) we could not have that $\Sigma'''' \vdash T'''' \subtypet S''''$ successfully terminates by applying the $\textsf{Asmp}$ or $\textsf{Asmp}2$ rule.

However the existence of such a triple $\Sigma'''', T'''', S''''$ is not possible, because $\vdash T'''' \subtypesa S''''$ would have
a $k$ long blocking path with $k < n$ (being such a path strictly shorter than that of $\vdash T''' \subtypesa \antOut{S}{\gamma}$), thus violating the minimality assumption about the blocking path length of the $\Sigma''', T''', S'''$ triple.
\qed

 \section{Proofs of Section \ref{sec:undecidability}}
 \subsection{Proof of Lemma \ref{lem:boundedUnd} and Theorem \ref{thm:undecidableBounded}}

We here prove Lemma \ref{lem:boundedUnd} and Theorem \ref{thm:undecidableBounded} that show undecidability of $\subtypebound$ 
by reduction from the bounded non 
termination problem.

\restateLEM{lem:boundedUnd}
\emph{Given a queue machine $M$ and an input $x$, it is undecidable whether $M$ does not terminate and is bound
  on $x$.}

\noindent  
\proof
We first prove that boundedness is undecidable.
If, by contraposition, boundedness was decidable, 
termination could be decided by first checking boundedness, and then perform
a finite state analysis of the queue machine behaviour. More precisely, 
termination on bounded queue machines
can be decided by forward exploration of the reachable configurations until
a terminating configuration is found, or a cycle is detected by reaching an
already visited configuration.

We now conclude by observing that given a queue machine $M$ and the input $x$,
it is not possible to decide whether \emph{$M$ does not terminate and is bound on $x$}.
Assume by contraposition one could decide the above property of queue machines.
Then boundedness could be decided as follows: transform $M$ in a new machine $M'$
that behaves like $M$ plus an additional special symbol $\#$ which is 
enqueued every time it is dequeued; boundedness of $M$ on input $x$ can be
decided by checking the above property on $M'$ and input $\#x$ (in fact
$M'$ never terminates and is bound on $\#x$ if and only if $M$ is bound on $x$). 
\qed

\medskip

\restateTHM{thm:undecidableBounded}
\emph{
  Given a queue machine $M= (Q , \Sigma , \Gamma , \$ , s , \delta )$ and
  an input string $x$, 
  we have that $\semT{s}{\emptyset} \subtypebound \semS{x\$}$ if and only if
  $M$ does not terminate and is bound on $x$.}

\noindent
\proof
We need a preliminary result: given $(q,\gamma) \rightarrow _{M} (q',\gamma')$,
if $\semT{q}{\emptyset} \subtype \semS{\gamma}$ then
we also have that
$\semT{q'}{\emptyset} \subtype \semS{\gamma'}$.
In fact, assuming $\gamma=C_1\cdots C_m$ and $\delta(q,C_1)=(q',B^{C_1}_1\cdots B^{C_1}_{n_{C_1}})$,
we have $\gamma'=C_2\cdots C_m B^{C_1}_1\cdots B^{C_1}_{n_{C_1}}$.
Having $\semT{q}{\emptyset} \subtype \semS{\gamma}$,
%there exists an asynchronous subtyping relation ${\mathcal R}$
%s.t. $(\semT{q}{\emptyset},S) \in {\mathcal R}$.
%By application of 
by one application of item 4. of Definition \ref{subtype}, 
one application of item 3., and $n_{C_1}$ applications of item 2.,
we can conclude that $\semT{q'}{\emptyset} \subtype \semS{\gamma'}$.
%$
%\big(\semT{q'}{\emptyset}                      
%\ ,\
%S'
%\big) \in  {\mathcal R}
%$.

%We need a preliminary result: given $(q,\gamma) \rightarrow _{M} (q',\gamma')$,
%if an asynchronous subtyping relation
%includes the pair $(\BsemT{q}{\emptyset}, \BsemS{\gamma})$ then
%it contains also the pair
%$(\BsemT{q'}{\emptyset}, \BsemS{\gamma'})$.
%%
%In fact, assuming $\gamma=C_1\cdots C_m$ and $\delta(q,C_1)=(q',B^{C_1}_1\cdots B^{C_1}_{n_{C_1}})$,
%we have $\gamma'=C_2\cdots C_m B^{C_1}_1\cdots B^{C_1}_{n_{C_1}}$.
%Consider now an asynchronous subtyping relation ${\mathcal R}$
%s.t. $(\BsemT{q}{\emptyset}, \BsemS{\gamma}) \in {\mathcal R}$.
%By application of item 4. of Definition \ref{subtype}, 
%one application of item 2., and $n_A$ applications of item 4.,
%we can conclude that also $(\BsemT{q'}{\emptyset}, \BsemS{\gamma'}) \in {\mathcal R}$.

We now observe that if $M$ is not bound on $x$ we have that it is not
possible to have $\semT{s}{\emptyset} \subtypebound \semS{x\$}$.
Assume by contraposition that $\semT{s}{\emptyset} \subtypebound \semS{x\$}$. 
From the previous preliminary result, we have that also
$\semT{q'}{\emptyset} \subtypebound \semS{\gamma'}$
for each reachable configuration $(q',\gamma')$. But due to unboundedness of $M$ on $x$
we have that, for every $k$, there is an enqueue
operation that is executed when the queue is longer than $k$.
Assume this happens when the configuration $(q',\gamma')$
performs its computation action.
%there is no limit to the length of $\gamma'$. This means that the session types
%$\semS{\gamma'}$ have input contexts of unbounded length.
%But as $M$ is single-consuming, we have that $\semT{q'}{\emptyset}$ has at most two inputs
%before the first output. Hence, i
In order to relate
$\semT{q'}{\emptyset}$ and $\semS{\gamma'}$, we need
a relation that contains pairs with the l.h.s. starting with an output
and the r.h.s. with an input context of depth greater than $k$. But this cannot
hold if we fix a maximal depth smaller than $k$ to the input context.

Now we observe that $\semT{s}{\emptyset} \subtypebound \semS{x\$}$
if and only if $M$ does not terminate and is bound on $x$.
Following the \emph{(Only if part)} of the proof of Theorem 3.1 \cite{BCZ16} 
stating the undecidability of $\selsubtype$,
we prove that if $\semT{s}{\emptyset} \subtypebound \semS{x\$}$
then $M$ does not terminate. Moreover, we also have that $M$ is bound on $x$
in the light of the previous observation. 

Consider now that $M$ does not
terminate. As in the \emph{(If part)} of the same
proof mentioned above, we define
$\mathcal C=\{(q_i,\gamma_i)\ |\ (s,x\$)=(q_0,\gamma_0) \rightarrow _{M} (q_1,\gamma_1) \rightarrow _{M}
\cdots \rightarrow _{M} (q_i,\gamma_i), i \geq 0\}$
and the following relation $\mathcal R$ on types:
$$
\begin{array}{ll}
\mathcal{R} & =
\\
\{ 
&
\big(\ \semT{q}{\emptyset}, \semS{C_1 \cdots C_m}\ \big),\\
&
\big(\ %\unfold 1{\semT{q}{\emptyset}}
%\unfold 1{\semT{q}{\emptyset}},\semS{C_1 \cdots C_m}\ \big),
\Tbranchset{A}{
          \Tselectsingle{B^A_1}{\cdots
          \Tselectsingle{B^A_{n_A}}{\semT{q'}{\emptyset}
          }
          }
          }                  
          {\Gamma}\ ,\ \semS{C_1 \cdots C_m}\ \big),
\\
&
\big(\Tselectsingle{B^{C_1}_1}{
         \Tselectsingle{B^{C_1}_2}{\cdots
           \Tselectsingle{B^{C_1}_{n_{C_1}}}{\semT{q'}{\emptyset}
                              }
                            }
                          }
\ ,\
%,\\
%&\ \
 \Tbranchsingle{C_2}{\cdots
           \Tbranchsingle{C_m}
             {Z}
                              }
\ \big),                                                        

\\

&
\big(      
         \Tselectsingle{B^{C_1}_2}{\cdots
           \Tselectsingle{B^{C_1}_{n_{C_1}}}{\semT{q'}{\emptyset}
                              }
                            }                          
\ ,\
%,\\
%&\ \
 \Tbranchsingle{C_2}{\cdots
           \Tbranchsingle{C_m}{
           \Tbranchsingle{B^{C_1}_1}{  
             {Z}
             					    }
                              }
                              }
\ \big),                                                        

\\

& \cdots

\\
&
\big(\      
         \semT{q'}{\emptyset}
                                                      
\ ,\
%,\\
%&\ \
 \Tbranchsingle{C_2}{\cdots
           \Tbranchsingle{C_m}{
           \Tbranchsingle{B^{C_1}_1}{  \cdots
             {           \Tbranchsingle{B^{C_1}_{n_{C_1}}}{  
             {Z}
             					    }}
             					    }
                              }
                              }
\ \big)

\\

| & (q,C_1 \cdots C_m) \in \mathcal C, \delta(q,C_1)=(q',B^{C_1}_1\cdots B^{C_1}_{n_{C_1}}), \\
&
Z = {\Trec{t}.\Tselectset{A}{\Tbranchsingle{A}{\Tvar{t}}}{\Gamma}}
\ \ \}
\end{array}
$$
Following the proof of Theorem 3.1 \cite{BCZ16} 
we show that this relation is an asynchronous subtyping relation. Moreover
boundedness of $M$ on $x$ guarantees boundedness on the length of the
reachable queue contents $C_1 \cdots C_m$,
that implies boundedness of the depth of the input contexts of the
r.h.s. of all the pairs in $\mathcal{R}$. This proves that
$\semT{s}{\emptyset} \subtypebound \semS{x\$}$.
\qed

%%%%%%%%%%%%%%%%%

\subsection{Proof of Theorem \ref{thm:construction}}\label{app:scqm}

Theorem \ref{thm:construction} states that, given a single consuming queue machine $M$ and an input $x$, termination of $M$ on $x$ is undecidable. The theorem is proved by resorting to Turing completeness of queue machines. In order to do this we preliminarily provide an encoding $\encodeSC{M}$ from a queue machine $M$ into a single-consuming queue machine and 
two lemmas that guarantee that, given a queue machine $M$ and an input $x$,
$M$ terminates on $x$ if and only if the single-consuming queue machine $\encodeSC{M}$ terminates on $x$.

\begin{definition}\label{def:singleConsunimgEncoding}
  Let $M=(\{q_1,..,q_n\} , \Sigma , \Gamma , \$ , s , \delta )$ be a queue machine and
  let $\#$ be a special character not in $\Gamma$. We 
  denote with $\encodeSC{M}$ the following single-consuming queue machine
  $(\{q_1,..,q_n,q_1',..,q_n'\} , \Sigma , \Gamma\cup\{\#\} , \$ , s ,
  \delta' )$ with $\delta'$ defined as follows:
  \begin{itemize}
  \item  $\delta'(q_i,a)=(q_j',\epsilon)$ if $\delta(q_i,a)=(q_j,\epsilon)$

  \item $\delta'(q_i,a)=(q_j,\gamma)$ if $\delta(q_i,a)=(q_j,\gamma)$ with $\gamma\neq \epsilon$

  \item $\delta'(q_i,\#)=(q_i',\epsilon)$
  
  \item  $\delta'(q_i',a)=(q_j,\#)$ if $\delta(q_i,a)=(q_j,\epsilon)$

  \item $\delta'(q_i',a)=(q_j,\gamma)$ if $\delta(q_i,a)=(q_j,\gamma)$ with $\gamma\neq \epsilon$

  \item $\delta'(q_i',\#)=(q_i,\#)$
  
%  \item $\delta'(q_i',\#)=q_i',\epsilon$
%    
%  \item $\delta'(q_i'��,\#)=q_i',\#$
%
  \end{itemize}
Given a configuration $(q,\gamma)$ of $\encodeSC{M}$,
we denote with $\unencodeSC{(q,\gamma)}$ the configuration
$(z,\beta)$ where $z=q$, if $q \in \{q_1,..,q_n\}$, or $z=q_i$, if $q=q_i'$,
while $\beta$ is obtained from $\gamma$ by removing each instance of the special
symbol $\#$.
\end{definition}

\begin{example}
We now comment the construction $\encodeSC{M}$
that, given a queue machine $M$, returns a single-consuming queue machine.
As an example, consider $M = (\{q_1,q_2\} , \{a\} , \{a,\$\} , \$ , q_1 , \delta )$
with $\delta$ such that $\delta(q_1,a)=(q_2,\epsilon)$, $\delta(q_1,\$)=(q_1,\$)$,
$\delta(q_2,a)=(q_2,a)$, and $\delta(q_2,\$)=(q_2,\epsilon)$. 
This machine accepts the input string $"a"$ by consuming in sequence $a$ 
and then $\$$.
Consider now $\encodeSC{M} = (\{q_1,q_2,q'_1,q'_2\} , \{a\} , \{a,\$,\#\} , \$ , q_1 , \delta' )$
with $\delta'$ such that $\delta'(q_1,a)=(q'_2,\epsilon)$, $\delta'(q_1,\$)=(q_1,\$)$,
$\delta'(q_1,\#)=(q_1',\epsilon)$, 
$\delta'(q_2,a)=(q_2,a)$, $\delta'(q_2,\$)=(q_2',\epsilon)$,
$\delta'(q_2,\#)=(q_2',\epsilon)$,
$\delta'(q'_1,a)=(q_2,\#)$, $\delta'(q'_1,\$)=(q_1,\$)$,
$\delta'(q'_1,\#)=(q_1,\#)$,
$\delta'(q'_2,a)=(q_2,a)$, $\delta'(q'_2,\$)=(q_2',\#)$, and
$\delta'(q'_2,\#)=(q_2,\#)$. 
This new queue machine also accepts the input string $"a"$
but it does simply consume $a$ and $\$$ in sequence, but when $\$$
is dequeued the special symbol $\#$ is enqueued (which is subsequently
consumed thus emptying the queue). Notice that the queue machine $\encodeSC{M}$
cannot consume two symbol in sequence because after the first one
is consumed, it enters in one of the primed state $q_i'$
that always enqueue some symbol.  
\end{example}

\begin{lemma}\label{lem:firstSCencoding}
Let $M=(Q , \Sigma , \Gamma , \$ , s , \delta )$ be a queue machine and let $x\in \Sigma^*$.
If $(s,x\$) \rightarrow_{M}^{*} (q,\gamma)$ then there exists a configuration $(q',\gamma')$
such that $(s,x\$) \rightarrow_{\encodeSC{M}}^* (q',\gamma')$ with $\unencodeSC{(q',\gamma')}=(q,\gamma)$. 
\end{lemma}
\proof
By induction on the number of steps in the sequence $(s,x\$) \rightarrow_{M}^{*} (q,\gamma)$.
The base case is trivial. In the inductive case we perform a case analysis. The unique 
non trivial case is when the configuration reached by $\encodeSC{M}$ according
to the inductive hypothesis has the queue starting with the special symbol
$\#$. In this case, $\encodeSC{M}$ must perform more transitions, first to consume all 
the instances of $\#$ in front of the queue and then to mimick the new transition of $M$.
\qed

\begin{lemma}\label{lem:secondSCencoding}
Let $M=(Q , \Sigma , \Gamma , \$ , s , \delta )$ be a queue machine and let $x\in \Sigma^*$.
If $(s,x\$) \rightarrow_{\encodeSC{M}}^{*} (q,\gamma)$ then
$(s,x\$) \rightarrow_{M}^* \unencodeSC{(q,\gamma)}$.
\end{lemma}
\proof
By induction on the number of steps in the sequence $(s,x\$) \rightarrow_{\encodeSC{M}}^{*} (q,\gamma)$.
The base case is trivial. In the inductive case we perform a case analysis. The unique 
non trivial case is when $\gamma$ starts with the special symbol
$\#$. In this case, $M$ does not perform any new transition as if
$(q',\gamma')$ is the new configuration we have that $\unencodeSC{(q,\gamma)}=\unencodeSC{(q',\gamma')}$.
\qed

\medskip

\restateTHM{thm:construction}
\emph{Given a single consuming queue machine $M$ and an input $x$, the termination of $M$ on $x$
  is undecidable.}
  
\noindent  
\proof
The thesis directly follows from the Turing completeness of queue machines, and the
two above Lemmas that guarantee that given a queue machine $M$ and an input $x$,
$M$ terminates on $x$ if and only if the single-consuming queue machine $\encodeSC{M}$ terminates on $x$.
This is guaranteed by the fact that if $\encodeSC{M}$ reaches a configuration with the queue containing
only instances of $\#$, it is guaranteed to eventually terminate by emptying the queue.
\qed

\subsection{Proof of Theorem \ref{thm:undecidableNoBranching}}

Theorem \ref{thm:undecidableNoBranching} 
states that a single-consuming queue machine does not terminate if and only if the types obtained by the encoding of Figure~\ref{fig:encoding2} are in the $\subtypetintout$
relation.
The proof is done by separately showing, as preliminary lemmas, 
both implications (one in each direction) to hold. 

Concerning such lemmas and their proof, we need to introduce some preliminary notation.
%\myparagraph{Notation.}
Given a sequence of queue symbols $\gamma$, we denote with
$\BsemS{\gamma}_u$ the set of session types that can be
obtained from $\BsemS{\gamma}$ by independently replacing each occurrence, inside it, of the term $T''$
defined in Figure \ref{fig:encoding2} with $\antOut{T''}{l_{i_1} \dots l_{i_{n}}}$,
for some sequence of labels $l_{i_1} \dots l_{i_{n}}$ with $n \geq 0$ (distinguished label sequences can be considered for replacing different occurrences of $T''$ inside $\BsemS{\gamma}$).
%, and by applying to the obtained term $\unfold n{\_}$ for some $n \geq 0$.
Observe that $\BsemS{\gamma}_u$ is well defined because $T''$ can anticipate 
every possible sequence of outputs.
Moreover, for simplicity, we will consider the asynchronous subtyping relation
$\subtype$ instead of $\subtypetintout$. Nevertheless, we will apply such
relation on types that have all their choices labeled on the
same set of labels, hence the two relations obviously coincide on
such types.

\begin{lemma}\label{lem:easyUndecidable}
  Given a single-consuming queue machine $M= (Q , \Sigma ,$ $\Gamma , \$ , s , \delta )$ and
  an input string $x\in \Sigma^*$, if $\BsemT{s}{\emptyset} \subtype \BsemS{x\$}$ then $M$ does not terminate on $x$.
\end{lemma}
\proof  
We need a preliminary result: given $(q,\gamma) \rightarrow _{M} (q',\gamma')$
and a term $S \in \BsemS{\gamma}_u$,
if $\BsemT{q}{\emptyset} \subtype S$ then
there exists $S' \in \BsemS{\gamma'}_u$ such that
$\BsemT{q'}{\emptyset} \subtype S'$.
In fact, assuming $\gamma=C_1\cdots C_m$ and $\delta(q,C_1)=(q',B^{C_1}_1\cdots B^{C_1}_{n_{C_1}})$,
we have $\gamma'=C_2\cdots C_m B^{C_1}_1\cdots B^{C_1}_{n_{C_1}}$.
Consider now $S\in \BsemS{\gamma}_u$.
Having $\BsemT{q}{\emptyset} \subtype S$,
%there exists an asynchronous subtyping relation ${\mathcal R}$
%s.t. $(\semT{q}{\emptyset},S) \in {\mathcal R}$.
%By application of 
by one application of item 4. of Definition \ref{subtype}, 
one application of item 3., and $n_A$ applications of item 2.,
we can conclude that there exists $S' \in \BsemS{\gamma'}_u$
such that $\BsemT{q'}{\emptyset} \subtype S'$.
%$
%\big(\semT{q'}{\emptyset}                      
%\ ,\
%S'
%\big) \in  {\mathcal R}
%$.

We now prove the thesis by showing that assuming that $M$ accepts $x$ we have
$\BsemT{s}{\emptyset} \not\!\!\subtype\ \BsemS{x\$}$.  
By definition of queue machines, we have that: $M$ accepts $x$ implies
$(s,x\$)\rightarrow _{M}^* (q,\epsilon)$.  Assume now, by
contraposition, that 
$\BsemT{s}{\emptyset} \subtype \BsemS{x\$}$.  As
$(s,x\$)\rightarrow _{M}^* (q,\epsilon)$, by repeated application of
the above preliminary result we have that exists $S' \in \BsemS{\epsilon}_u$
such that
$\BsemT{q}{\emptyset} \subtype S'$. But this cannot
hold because $\BsemT{q}{\emptyset}$ is a recursive definition that upon
unfolding begins with an input that implies (according to
items 4. and 3. of Definition \ref{def:subtyping}) that also $S'$
(once unfolded) starts with an input.  But this
is false, in that, by definition of the queue encoding 
$\BsemS{\epsilon} =
\Trec t \oplus\Big\{A : \& \Big( \big\{A:\Tvar t\big\} \labelunion \big\{A':T''\big\}_{A'\in \Gamma\setminus\{A\}}\Big)
                           \Big\}_{A \in \Gamma}$.
\qed

\begin{lemma}\label{lem:hardUndecidable}
  Given a single-consuming queue machine $M= (Q , \Sigma ,$ $\Gamma , \$ , s , \delta )$ and
  an input string $x\in \Sigma^*$, if $M$ does not terminate on $x$ then $\BsemT{s}{\emptyset} \subtype \BsemS{x\$}$.
\end{lemma}
\proof  
Assuming that $M$ does not accept $x$ we show that
$\BsemT{s}{\emptyset} \subtype \BsemS{x\$}$.
%We show how to define a subtyping
%relation $\mathcal R$ such that $(T,S) \in \mathcal R$.
When a queue machine does not accept an input,
the corresponding computation never ends. In our case, this means that
there is an infinite
sequence $(s,x\$)=(q_0,\gamma_0) \rightarrow _{M} (q_1,\gamma_1) \rightarrow _{M}
\cdots \rightarrow _{M} (q_i,\gamma_i) \rightarrow _{M} \cdots$.
Let $\mathcal C$ be the set of reachable configurations, i.e.
$\mathcal C=\{(q_i,\gamma_i)\ |\ i \geq 0\}$. 
We now define a relation $\mathcal R$ on types, where 
$T'$ and $T''$ are as in Figure \ref{fig:encoding2},
$T_0 = \oplus\{A: T''\}_{A \in \Gamma}$ and
$T_n = \&\{A: T_{n-1}\}_{A \in \Gamma}$:

$$
\!\!\!
\begin{array}{lll}
\mathcal{R} & =\!\!
\\
& \{ 
& 
\big(\ \BsemT{q}{\emptyset}\ ,\ S_{C_1 \cdots C_m}\ \big),
%\\
%&
\big(\ 
  \Tbranchset{A}{\semTcont{B^A_1\cdots B^A_{n_A}}_{q'}^{\emptyset}
          }                  
          {\Gamma}\ ,\ S_{C_1 \cdots C_m}\ \big),
\\
& &
\big(\ 
{\semTcont{B^{C_1}_1\cdots B^{C_1}_{n_{C_1}}}_{q'}^{\emptyset}}\ ,\
S_{C_2 \cdots C_m}                              
\ \big),                                                        
%\\
%&
\big(\ 
{\semTcont{B^{C_1}_2\cdots B^{C_1}_{n_{C_1}}}_{q'}^{\emptyset}}\ ,\
S_{C_2 \cdots C_m B^{C_1}_1}                              
\ \big),
\\
& & \cdots
\\
& &
\big(\ {\semTcont{B^{C_1}_{n_{C_1}}}_{q'}^{\emptyset}}\ ,\ S_{C_2 \cdots C_mB^{C_1}_1\cdots B^{C_1}_{n_{C_1}-1}}\ \big)
\\
& | & (q,C_1 \cdots C_m) \in \mathcal C, \delta(q,C_1)=(q',B^{C_1}_1\cdots B^{C_1}_{n_{C_1}}), 
S_\gamma \in \BsemS{\gamma}_u
\ \ \}
\\ 
\bigcup 
\\
& \{ 
&
\big(\ \BsemT{q}{\emptyset}\ ,\ T_n\ \big),
%\\
%&
\big(\ 
  \Tbranchset{A}{\semTcont{B^A_1\cdots B^A_{n_A}}_{q'}^{\emptyset}
          }                  
          {\Gamma}\ ,\ T_n\ \big),
\\
& &
\big(\ 
{\semTcont{B^{C_1}_1\cdots B^{C_1}_{n_{C_1}}}_{q'}^{\emptyset}}\ ,\
T_m                              
\ \big),                                                        
%\\
%&
\big(\ 
{\semTcont{B^{C_1}_2\cdots B^{C_1}_{n_{C_1}}}_{q'}^{\emptyset}}\ ,\
T_m                              
\ \big),
\\
& & \cdots
\\
& &
\big(\ {\semTcont{B^{C_1}_{n_{C_1}}}_{q'}^{\emptyset}}, T_m\ \big)
\\
& | & (q,C_1 \cdots C_m) \in \mathcal C, \delta(q,C_1)=(q',B^{C_1}_1\cdots B^{C_1}_{n_{C_1}}), \\
& &
\mbox{if $\exists q'',C$ s.t. $\delta(q,C)=(q'',\epsilon)$ then $n \geq 2$ else $n \geq 1$},   m \geq 0
\ \ \}
\\ 
\bigcup 
\\
& \{ 
&
\big(
T'
\ ,\
T_n
\big),
%\\
%&
\big(
\&\big\{A_1\!: \oplus\{A_2\!: T'\}_{A_2 \in \Gamma} \big\}_{A_1 \in \Gamma}
\ ,\
T_n
\big),
%\\
%&
\big(
\oplus\{A_2\!: T'\}_{A_2 \in \Gamma} 
\ ,\
T_m
\big)
\\
& | &
n\geq 1, m \geq 0
\ \ \}
\\
\bigcup
\\
& 
\{ 
&
\big(
T'
\ ,\
S_{\gamma}
\big),
%\\
%&
\big(
\&\big\{A_1\!: \oplus\{A_2\!: T'\}_{A_2 \in \Gamma} \big\}_{A_1 \in \Gamma}
\ ,\
S_{\gamma}
\big),
%\\
%&
\big(
\oplus\{A_2\!: T'\}_{A_2 \in \Gamma} 
\ ,\
S_{\gamma}
\big)
\\
& | &
\gamma \in \Gamma^*, S_\gamma \in \BsemS{\gamma}_u
\ \ \}
\end{array}
$$
We have that the above $\mathcal{R}$
is an asynchronous subtyping relation because each of the pairs satisfies the conditions in Definition
\ref{subtype} thanks to the presence of other pairs in $\mathcal{R}$.
We can conclude observing that $(s,x\$) \in
{\mathcal C}$ implies that $( \BsemT{q}{\emptyset}, \BsemS{x\$})$ 
belongs to the above asynchronous subtyping relation ${\mathcal R}$, hence
$\BsemT{q}{\emptyset} \subtype \BsemS{x\$}$.  
\qed

%%%%%%%%%%%%%%%%%%%%%%%

\medskip

\restateTHM{thm:undecidableNoBranching}
\emph{    Given a single consuming queue machine $M= (Q , \Sigma , \Gamma , \$ , s , \delta )$ and
  an input string $x\in \Sigma^*$, we have $\BsemT{s}{\emptyset} \subtypetintout \BsemS{x\$}$ 
  if and only if $M$ does not terminate on $x$.}\\
\proof
Direct consequence of Lemmas \ref{lem:easyUndecidable} and \ref{lem:hardUndecidable}.
\qed

\end{document}